\documentclass[10pt]{article}

\usepackage[margin=1in]{geometry}

\usepackage{helvet}

\usepackage{amsmath,bm}
\usepackage{graphicx,pdflscape}
\usepackage{float}
\usepackage[toc,page]{appendix}
\usepackage{simpler-wick}
\usepackage{color,subfigure}
\usepackage{cite}
\usepackage[normalem]{ulem}
\usepackage{hyperref}
\hypersetup{
    colorlinks,
    citecolor=red,
    filecolor=cyan,
    linkcolor=blue,
    urlcolor=red 
}
\usepackage[capitalise]{cleveref}

\usepackage{xparse}
\ExplSyntaxOn
\NewDocumentCommand{\mref}{m}{\quinn_mref:n {#1}}
\seq_new:N \l_quinn_mref_seq
\cs_new:Npn \quinn_mref:n #1
 {
  \seq_set_split:Nnn \l_quinn_mref_seq { , } { #1 }
  \seq_pop_right:NN \l_quinn_mref_seq \l_tmpa_tl
  ( 
  \seq_map_inline:Nn \l_quinn_mref_seq
    { \ref{##1},\nobreakspace } 
  \exp_args:NV \ref \l_tmpa_tl 
  ) 
 }
\ExplSyntaxOff

\newcommand{\disp}[1]{Eq.~\mref{#1}}
\newcommand{\figdisp}[1]{Fig.~\mref{#1}}
\newcommand{\tbldisp}[1]{Table~\mref{#1}}
\newcommand{\secdisp}[1]{Section-\mref{#1}}
\newcommand{\appdisp}[1]{Appendix-\mref{#1}}
\newcommand{\refdisp}[1]{Ref.~\cite{#1}}
\newcommand{\disprange}[2]{Eq.~(\ref{#1}--\ref{#2})}

\newcommand{\beq}{\begin{eqnarray}}
\newcommand{\eeq}{\end{eqnarray}}
\newcommand{\barray}{\begin{eqnarray}}
\newcommand{\earray}{\end{eqnarray}}

\newcommand{\lessim} {\ {\raise-.5ex\hbox{$\buildrel<\over\sim$}}\ }
\newcommand{\gssim}{\ {\raise-.5ex\hbox{$\buildrel>\over\sim$}}\ }

\newcommand{\si}{\sigma}
\newcommand{\sib}{\bar{\sigma}}
\newcommand{\tJ}{\ $t$-$J$ \ }
\newcommand{\tJV}{\ $t$-$J$-$V_C$ \ }

\newcommand{\nn}{\nonumber}

\newcommand{\w}{\omega}
\renewcommand{\Re}{\mathrm{Re}}
\renewcommand{\Im}{\mathrm{Im}}

\renewcommand{\emph}{\textit}

\newcommand{\iden}{ {\bf 1}}

\newcommand{\sr}{\textcolor{black}}

\newcommand{\bra}{\langle\langle}
\newcommand{\ket}{\rangle\rangle}
\newcommand{\half}{\frac{1}{2}}
\newcommand{\chem}{{\bm \mu}}
\newcommand{\A}{{\cal A}}
\newcommand{\G}{{\cal G}}

\newcommand{\GH}{{\bf g}}
\newcommand{\GHI}{\GH^{-1}}

\newcommand{\wt}{\widetilde}

\newcommand*{\bigchi}{\mbox{\large$\chi$}}
\newcommand*{\hatchi}{\mbox{\large$\widetilde{\chi}$}}
\newcommand{\vq}{{\vec{q}}}
\newcommand{\vk}{{\vec{k}}}

\graphicspath{{Figures-Reduced/}}

\begin{document}


\title{Dielectric response of electrons with strong local  correlations and long-ranged Coulomb interactions   }
\author{ B Sriram Shastry\footnote{sriram@physics.ucsc.edu} \, and Michael Arciniaga\\
\small \em Physics Department, University of California, Santa Cruz, CA, 95064 \\}
\date{\today}

\maketitle

\vspace{1 in} 

\abstract{   
    Motivated by recent experiments, we append long ranged Coulomb interactions to dominant strong local correlations and study the resulting \tJV model for the 2-dimensional cuprate materials. This model includes the effect of short ranged Hubbard-Gutzwiller-Kanamori type correlations {and} long ranged Coulomb interactions on tight binding electrons. 
We calculate the $ \{\vq,\w\}$ dependent charge density fluctuations in this model using the extremely correlated Fermi liquid theory characterized by quasiparticles with very small weight $Z$. We develop a novel set of formulae to represent the dynamical charge susceptibility and the dielectric function, using a version of the charge-current continuity equation for a band system, valid for arbitrary $\vq$. Combining these ingredients, we present results for the irreducible dynamical charge susceptibility $\wt{\chi}_{\rho\rho}(\vq,\w)$, (longitudinal) dielectric function $\varepsilon(\vq,\w)$, current susceptibility $\wt{\chi}_{J J}(\vq,\w)$, conductivity $\sigma(\vq,\w)$, and the plasma frequency for any $\vq$. We also present calculations for the first moment of the structure function and discuss a characteristic energy scale $\Omega_p(\vq)$  which locates a peak in $\Im \, \wt{\chi}_{\rho\rho}(\vq,\w)$.
}

\newpage



\section{Introduction \label{Intro}}

    The role of strong local correlations and their interplay with long ranged Coulomb interactions, is an important problem in condensed matter physics. In the context of the metal insulator (Mott-Hubbard) transition of a Hubbard-Gutzwiller-Kanamori type model of strong correlations with added long ranged Coulomb interactions, early work \cite{Rice-Brinkman,Brinkman-Rice,Vollhardt} emphasized that this combination of the two types of interactions, quite generally leads to a metal with poor screening. These works noted that strong local correlations enhance the effective mass of electrons near a Mott transition, with $m^*/m$$\sim$$1/(1-U/U_c)$ at half filling $n=1$ with $U\gg t$\cite{FN1} and $U_c$ is the putative critical interaction strength discussed in \cite{Brinkman-Rice}. \sr{Closer to the considerations of this paper, away from $n=1$ a  reduction of the compressibility ($\chi_{comp}=\frac{V}{N^2} \frac{d N}{ d \mu}$  \disp{compressibility-2}) occurs for $U\gg t$ in the Gutzwiller  theory \cite{Vollhardt,Brinkman-Rice}. As emphasized by Vollhardt \cite{Vollhardt}, an enhancement of effective mass $m^*/m$$\sim\frac{1}{(1-n)}$, is offset by an even greater enhancement of an appropriate Landau Fermi liquid parameter.}
 These combine to give  a net suppression of compressibility. In turn this suppresses  the screening constant $q_s$, which is related to the compressibility by a sum-rule \cite{Nozieres} (see \disp{compressibility-2,static-screening-3d,static-screening-2d} below):
\beq
q^2_s=\frac{4\pi  q_e^2}{a_0^3 N_s} \frac{dN}{d\mu}\to 0. \label{screening-length}
\eeq
 The screening length defined through $\lambda_s =2\pi/q_s$  increases, and hence the
 metal  has progressively poorer screening properties as we move close to the insulator. More recent theoretical work \cite{recent-screening-1,recent-screening-2} has focussed on the dynamical aspects of screening, within the program of unifying band structure methods with dynamically screened Coulomb interaction and short ranged correlations. The latter are usually treated within the dynamical mean field theory \cite{recent-screening-1,recent-screening-2}.

    An immediate motivation for the present work comes from a set of experiments using the recently developed tool of momentum resolved electron energy loss spectroscopy (M-EELS) \cite{Mills,MEELS-1,MEELS-2,MEELS-3}. This technique gives a direct readout of the structure function $S(\vq,\omega)$ or equivalently the dielectric function $\varepsilon(\vq,\omega)$, for a broad range of momentum transfer $\vq$ and energy transfer $\omega$. The initial application of this technique has provided high resolution data on the structure function for the archetypical strongly correlated cuprate superconducting material $Bi_{2.5}Sr_{1.9}CaCu_2O_{8+x}$ ($BSSCO$), for two samples with $T_c=91$K and $T_c=50$K respectively. In the normal state, the data looks very different from what one might expect for a conventional weakly correlated Fermi liquid, e.g., one describable by the random-phase approximation (RPA). Sharp features arising from long lived quasiparticles in that theory are rounded off to broad peaks, and the spectrum has surprisingly long frequency tails. Understanding the data seems to require reducing the quasiparticle domination in charge response functions, as argued in \refdisp{Varma,MFL}.

    In this work we extend the {\em extremely correlated Fermi liquid theory} (ECFL)\cite{ECFL}, by adding the long ranged component of the Coulomb interaction. We thus calculate the charge dynamics of the \tJV model \disp{model}, which is a generalization of the \tJ model obtained by adding to it a long-ranged Coulomb interaction $V_C$. For this model we calculate the $\{\vq,\omega\}$ dependent dielectric function $\varepsilon\{\vq,\omega\}$ and the charge and current susceptibilities. 

    The ECFL theory was developed to describe the very large $U$ Hubbard model, or equivalently the short ranged \tJ interaction \cite{ECFL}. It therefore deals with the propagation and interaction of Gutzwiller projected electrons, obeying non-canonical anticommutators \disp{non-canonical}, within a tight binding model. The ECFL theory is characterized by a small but non-zero quasiparticle weight $Z \ll 1$ \cite{ECFL}, and is therefore suitable for describing the above experiments. This generalized ECFL calculation provides a microscopic theory of charge fluctuations in a metal, with fragile quasiparticles. In \figdisp{spectrum} the resulting single electron spectral function from ECFL in two dimensions is displayed with typical values of the model parameters. The role of strong correlations in suppressing the quasiparticle weight  from the free electron value, i.e. $Z_{k_F}\ll 1$ is seen here. The closely related momentum distribution function  in \figdisp{occ-num} illustrates this suppression, through the reduced (Migdal) discontinuity  at $k_F$. The suppression of the compressibility in Refs. \cite{Vollhardt,Brinkman-Rice} mentioned above,   is also  obtained in the ECFL theory, as illustrated in \figdisp{fig-compressibility}. The ECFL theory gives  a  set of results for the  wave vector dependent static susceptibility,  the first frequency moment of the structure function, and the plasma dispersion \figdisp{static,chiBubWWstatic,sumrule-dispersion}.

    The theory of the interacting 2-d electron system presented here differs significantly from established theories designed in the contexts of semiconductor inversion layers, surfaces of metals and more recently for graphene \cite{2deg1,2deg2,2deg3,2deg4}. In the current study, the dominant interaction is the short ranged Coulomb repulsion on the scale of a single atom, i.e., the Gutzwiller-Hubbard correlation. If one starts from weakly-interacting electrons within a perturbative scheme, it is very difficult to build in the strong local correlations, since the perturbation parameter is the largest energy scale! We start instead with non-canonical Gutzwiller projected electrons $\widetilde{C}_{i \sigma}$ \disp{model,non-canonical}, and then introduce long ranged Coulomb interactions, giving the \tJV model. In this treatment the physics of the Mott-Hubbard insulator at half filling is obtained naturally, in view of the inbuilt Gutzwiller projection.


\subsection{Highlights of new formulas}

    The calculations on this \tJV model use the {\em extremely correlated Fermi liquid} (ECFL) theory \cite{ECFL} for the \tJ model. The highly correlated single electron Green's function of this theory $\G(\vec{k},\omega)$  is computed using a systematic expansion in a parameter $\lambda \in[0,1]$, explained below in Sec.(\ref{meaning-of-lambda}). 
We use the results reported in our recent work     
 to      ${\cal O}(\lambda^2)$ \cite{ECFL-2d,ECFL-2d-Mai-a,ECFL-2d-Mai-b}, in 2-d.
    
     This theory produces an electron liquid with a very small, but non-zero quasiparticle weight at the Fermi momentum  $Z_{k_F}$- often abbreviated in this paper as  $Z$. It therefore has a fair {\em a-priori} possibility of reproducing the broad backgrounds seen in experiments. We also note that the ECFL theory provides a quantitative set of results for resistivity of cuprates for the single layer compounds \cite{ECFL-Resistivity} in fair agreement with a large body of data. It also provides a set of results for the inelastic non-resonant Raman scattering in different channels for the \tJ model from the fluctuations of the kinetic energy components\cite{ECFL-Raman}, that give a fair account of Raman scattering experiments\cite{Girsh,Sugai}. 

    In order to calculate the fluctuations of the charge  density, one needs information beyond that contained in $\G(\vec{k},\omega)$. We require the two particle response rather than the single particle Green's functions. Generalizing the ECFL calculations in that direction is a non-trivial task. Therefore we are  obliged to make approximations using the correlated single particle Green's functions. 

    This work extends the general formalism in two important directions described in the next paragraph. These extensions  enable the formulation of  suitable approximations using the available Green's functions $\G(\vec{k},\omega)$. We describe these two extensions, and record  their  location in this paper. Some readers might find these extensions of potential use in problems other than the one considered here. Other readers interested in the concrete applications made here, can use this roadmap to skip certain  sections and appendices. 

    The first formal result is \disp{new-rep,Psi-chi-inverse}. This formula is valid for any density response function that admits a high frequency moment expansion in powers of $\omega^2$ \disp{moment-expansion-1,moment-expansion-2}. It expresses the $\{\vq,\omega\}$ dependent irreducible susceptibility in terms of (i) its  static limit,  (ii) the leading high frequency moment and (iii) the complex self energy $\Psi(\vq,\omega)$ for this object. This self energy $\Psi(\vq,\omega)$   has not been discussed in literature, as far as we are aware. It is obtained following a Luttinger type analysis of the susceptibility \cite{Luttinger}, by reorganizing the moment expansion formulas.

    The next formal result is the derivation of an important pair of alternate formulas \disp{dielectric-constant-1,dielectric-constant-2} for the dielectric function valid for all $\vec{q},\omega$.  While \disp{dielectric-constant-1} is a familiar expression in terms of the density operator, \disp{dielectric-constant-2} is new and involves the $W$ operator, which is the divergence of the lattice current operator as seen in \disp{W1,W2}. These formulas are modeled after analogous formulas due to Nozi\`eres in \refdisp{Nozieres}, valid for the (continuum) homogeneous electron gas. In the latter context, Nozi\`eres uses diagrammatic perturbation theory and regroups  terms  so that the conservation of charge is reflected in the relationship between appropriate correlation functions- thus finally leading to his twin formulas.

    The two alternate formulas \disp{dielectric-1, scr-5} for the {\em inverse} dielectric constant 
 are relatively more   straightforward, and
 follow from the continuity equation. These involve the reducible correlation functions $\bigchi_{\rho\rho}$, and provide  the starting point for obtaining the Nozi\`eres type formulas, which are analogous relations for the {\em irreducible} susceptibility $\hatchi_{\rho\rho}$. The connection between the reducible i.e. $\bigchi_{\rho\rho}$, and irreducible i.e. $\hatchi_{\rho\rho}$ susceptibilities is straightforward when the electrons are canonical. The relationship is expressed using Feynman diagrams, which encode perturbation theory compactly and elegantly, as shown in text books \cite{Nozieres,Rajagopal}. However for the \tJV model, we are dealing with non-canonical electrons, and hence the identification and extraction of irreducible pieces needs to be accomplished without the use of vertex functions,  or of manipulating sums of Feynman diagrams. The needed analysis is carried out in \secdisp{Sec-5}. The method employed by us decomposes the charge source, i.e. an auxiliary external potential used to generate the Greens functions  into a  part containing a Hartree type term from the remainder as  described in \secdisp{Sec-5} and \appdisp{App-B,App-C}. As stated, this leads to the final formulas \disp{dielectric-constant-1,dielectric-constant-2}, with a central result being the identity \disp{scr-6}, relating the (irreducible) charge and W-type correlations functions. The W-type response functions involve the W-type vertex \disp{W2,W1}, these contain the full set of hopping parameters in the tight binding model, and crucially for our purposes, enable us to address the $\vec{q},\omega$ dependence of the charge response over the {\em entire} Brillouin zone (BZ). It is also evident that by taking the long wavelength limit $\lim_{q\to 0}$, we recover the homogeneous electron gas relations originally written by Nozi\`eres\cite{Nozieres}

    Combining the formal expressions \disp{new-rep,Psi-chi-inverse} for the charge and W-type susceptibilites with the twin Nozi\`eres type relations \disp{dielectric-constant-1,dielectric-constant-2} enables us to make useful approximations for the charge response. We use the ECFL single particle Green's functions $\G$ to perform the explicit calculations, and thereby obtain two independent bubble susceptibilities \disp{Inter-1,Inter-2}. These are the basic computations from ECFL. Using them in \disp{chiA,chiB}, we get two alternate estimates of the irreducible charge susceptibility $\hatchi_{\rho\rho}(\vq,\omega)$, and there from the dielectric constant by using \disp{dielectric-constant-1}. If we were to use exact (instead of bubble) susceptibilities, these two results would coincide, by virtue of the exact result \disp{scr-6}. Since the approximations for the bubble calculations are not exact, these two estimates differ from each other in general. In fact these provide two complementary approximations, valid in different regimes $\omega\to0$ and $|\omega| \gg t$ ($t$ is the hopping parameter). We then combine expressions \disp{new-rep,Psi-chi-inverse}, guided by considerations of validity at low and high $\omega$ as summarized in \appdisp{App-X} and \appdisp{App-Limits}. We finally  arrive at  alternate approximations \disp{appxI,appxII}. These two approximations are overall similar in most features. They only differ at very small $\vq,\omega$ where quasiparticle excitations that are missing in \disp{appxI}, but are present in \disp{appxII}, cause some differences. Another novel result presented here is the identification of an important characteristic energy scale $\Omega_p(\vq)$. This scale locates a peak in $\Im \, \hatchi_{\rho\rho}(\vq,\omega)$ \disp{im-chi-expression,Omega-p}, and is also expressible as a specific moment of the $\Im \, \hatchi_{\rho\rho}(\vq,\omega)$ in \disp{emergentOmega-2,emergentOmega-3}. We present results for this scale and show that it is quite low at small $\vq$. 


\subsection{The plan of the paper}

    We define the \tJV model below in \secdisp{Sec-Model}, and summarize the method used to calculate the charge response. The calculation uses the ECFL theory to calculate the electron Green's function $\G$ to a certain approximation (termed as ${\cal O}(\lambda^2)$), which has been described in detail in our recent publications \cite{ECFL-2d,ECFL-2d-Mai-a,ECFL-2d-Mai-b}. To make this work self contained, we summarize the scheme and the equations used to compute $\G$ in \appdisp{App-ECFL}. 

    In \appdisp{App-A} we recall the formal definitions of the susceptibility and the structure function for  describing the charge response. \secdisp{Sec-4} summarizes the definitions of charge $\bigchi_{\rho\rho}$ and ``current-type'' susceptibilities $\bigchi_{WW}$, and their cross susceptibilities $\bigchi_{\rho W}, \bigchi_{W \rho }$, for electrons in a narrow band, and their mutual relationship from the conservation law of charge.

     In \appdisp{App-B} we define the electronic Green's function $\G$, its equation of motion generated conveniently by external potentials, which include a charge and a current source, and express the susceptibilities in terms of variational derivatives of the Green's functions, with respect to the external potentials. 

    Instead we present the necessary formal results here, directly using the susceptibilities. The strategy used is to redefine the external potential by absorbing a Hartree type term into it, as described in \appdisp{App-C}.

    We define in \appdisp{App-C} the irreducible susceptibilities $\hatchi_{\mu \nu}$ in terms of the reducible ones. The irreducible susceptibilities are calculated by taking functional derivatives of the Green's function $\G$. The details of the formalism are provided in \appdisp{App-C}. The dielectric function satisfies a {\em linear} relationship \disp{dielectric-constant-1} with it, in contrast to the non-linear relation with the reducible susceptibility \disp{dielectric-1}. In \appdisp{App-C} we show that the conservation laws connect the screened, or irreducible susceptibilities with results that parallel those for canonical electrons.

    In \secdisp{Sec-5} we  express the susceptibilities in terms of their screened, or irreducible pieces $\hatchi_{\rho,\rho}, \hatchi_{\rho,W}$, $\hatchi_{W,\rho}, \hatchi_{W,W}$. We find a useful and important pair of formulas \disp{dielectric-constant-1,dielectric-constant-2}. These relations, obtained for tight-binding non-canonical electrons, are completely analogous to the results of \refdisp{Nozieres}, who worked with canonical electrons in the continuum, i.e., for the homogeneous electron gas. In these exact formulas, the dielectric function at arbitrary $(\vq,\omega)$ is expressed in alternate forms involving two different pairs of correlation functions. These alternate forms work better in complementary regions of $\omega$ and allow us to make useful approximations, as explored in \appdisp{sec3.5} and in \appdisp{sec3.6}.

    The frequency sum-rules for the susceptibility play an important role in our theory and are summarized in \appdisp{App-X}. The limiting values of the dielectric constant at low and high $\omega$ are noted in \appdisp{App-Limits}.

    In \secdisp{sec3.7} the formulae that approximates the dielectric function is presented and applications of the methodology to the computation of the dielectric function is described. In \secdisp{sec5} we discuss the results and present some conclusions.



\section{The \tJV  model \label{Sec-Model}}

    The \tJV Hamiltonian studied here is 
\beq
H&=&H_{\mbox{t}}+H_{\mbox{J}} \label{tJ}+V_C \label{model}  \\
H_{\mbox t}&=& -\sum_{ij \si} t_{ij} \wt{C}^\dagger_{i \si} \wt{C}_{j \si} - \chem \sum_i n_i   \\
H_{\mbox{J}}&=& \half \sum_{ij} J_{ij} (\vec{S}_i.\vec{S}_j - \frac{n_i n_j}{4})  \\
V_C&=&\frac{1}{2}\sum_{i\neq j} V_{ij} n_i n_j, \;\;\;V_{ij}=\frac{1}{\varepsilon_\infty} \frac{q_e^2}{ |\vec{r}_i-\vec{r}_j|} \label{VCoulomb}
\eeq
with the electronic charge $q_e=-|e|$, the density operator $n_i = \sum_\si \wt{C}^\dagger_{i \si}\wt{C}_{i \si}$, and spin density operator $S_i^\alpha =\half \sum_{\si \si'} \wt{C}^\dagger_{i \si} \tau^\alpha_{\si \si'}\wt{C}_{i \si'}$, $\tau^\alpha$ is a Pauli matrix, and the Coulomb potential is denoted by $V_{ij}$. The hopping parameters $-t_{ij}= \frac{1}{N_s} \sum_{ij} e^{i \vec{k} (\vec{r}_i-\vec{r}_j)} \varepsilon_k$ are Fourier components of the band energy $\varepsilon_k$, $N_s$ is the number of sites in the crystal. \sr{{[\bf Q-3]} Here we have  add the  long ranged Coulomb term to the familiar \tJ model. The  well studied \tJ model  is obtained from the large U limit of the Hubbard model, by performing an expansion in $t/U$, followed by the neglect of certain short ranged three body terms of the order $t^2/U$  arise in this transformation \cite{tJ-origin}. }
We will study both 3 and 2 dimensional (layered) strongly correlated electron systems, where the Fourier components of $V$ is given in 3-d, assuming a simple cubic cell of side $a_0$ by
\beq
V(\vq)=\frac{1}{N_s a_0^3 \varepsilon_\infty } \frac{4\pi q_e^2}{|\vec{q}|^2} \mbox{   (3-d)}, \label{Coulomb-3d}
\eeq
and in 2-d by
\beq
V(\vq) = \frac{1}{N_s a_0^2 \varepsilon_\infty } \frac{2\pi q_e^2}{ |\vec{q}|} \mbox{   (2-d)}.
\label{Coulomb-2d}
\eeq
To simplify notation we will set $\hbar=1$ and the lattice constant $a_0=1$ in most part below. Here $\varepsilon_\infty$ is the static dielectric constant due to screening by mobile charges other than the ones described by $H_t$, if any are present. Here the correlated Fermi destruction operator $\wt{C}_i$ is found from the plain (i.e. canonical or unprojected) operators $c_i$, by sandwiching it between two Gutzwiller projection operators $\wt{C}_{i \si}=P_G C_{i \si} P_G$. Let us note that these Fermions satisfy a non-canonical set of anticommutation relations
\beq
 \{ \wt{C}_{i \si_i}, \wt{C}_{j \si_j}^\dagger \}&=&\delta_{ij} \left(\delta_{\si_i \si_j} - \si_i \si_j \wt{C}_{i \sib_i}^\dagger  \wt{C}_{i \sib_j} \right), \mbox{and  }\nn \\  \{ \wt{C}_{i \si_i}, \wt{C}_{j \si_j} \}&=&0.
 \label{non-canonical}
\eeq
The physical meaning of this sandwiching process is that the Fermi operators act within the subspace where projector $P_G$ enforces single occupancy at each site. This model generalizes the well studied \tJ model by adding the long ranged Coulomb interaction term, and we will study the effect of the added term in determining the fluctuations of the charge density, the dielectric function and related structure function. We initially keep the dimensionality of the electronic system general so that the results apply to 3-dimensions, and later consider the case of 2-dimensional stacking of the electronic system,  for modeling cuprate superconductors.

The \tJV  model used here neglects multi-band aspects of the Coulomb interaction, and focusses on the extremely correlated single band containing the Fermi surface (FS). It throws out inter-band transition matrix elements of the Coulomb interaction and only retains intraband terms. A rough account of the other bands is taken, by rescaling the Coulomb interaction by an infinite frequency dielectric constant $\varepsilon_\infty$ as in \disp{VCoulomb}. This rescaling represents the cumulative effect of the ``fast'' (i.e. high energy) electrons on the ``slow'' (low energy) correlated electrons described by our model. This type of reasoning suggests that as long as the excitation energies do not exceed the inter-band energies, the single band model employed here should be quite reliable.

    In applying the results of these calculations to real systems, it must be kept in mind that the \tJV model is only a `low energy' abstraction of the narrow band containing the Fermi energy, which is further embedded in a continuum of bands extending to very high energies. Thus, in an experimental situation, curtailing the frequency integration in \disp{moment-expansion-1} up to a cutoff frequency $\Omega \sim 1,2$ eV is expected to capture the `low energy' model, with strong correlations built into the results. On the other hand by extending the integral to higher energies, one gets rid of the correlations and  the results should reveal the bare electron scales.

 \subsection{Comments on the novel features of our methodology \label{meaning-of-lambda}}
\sr{ The solution presented here using the ECFL formalism has some unique features that need an introduction. The main innovation consists of introducing  a parameter $\lambda$, lying between $[0,1]$ in the theory. One simple way is to generalize \disp{non-canonical} to 
\beq
 \{ \wt{C}_{i \si_i}, \wt{C}_{j \si_j}^\dagger \}&=&\delta_{ij} \left(\delta_{\si_i \si_j} - \lambda \si_i \si_j \wt{C}_{i \sib_i}^\dagger  \wt{C}_{i \sib_j} \right), \mbox{and  }\nn \\  \{ \wt{C}_{i \si_i}, \wt{C}_{j \si_j} \}&=&0,
 \label{non-canonical-lambda}
\eeq
so that  $\lambda=0$ gives us standard Fermions, whereas $\lambda=1$ gives us the non-canonical Fermions with Gutzwiller projection. As explained in \cite{ECFL} this procedure has a parallel in the expansion of spin algebra in terms of Bosons using the parameter $\frac{1}{2S}$, which plays a role similar to that of $\lambda$.
Another and equivalent   method of introducing $\lambda$ is through the Schwinger-Tomonaga  equations of motion \cite{ECFL}. Collecting terms of
a given order in $\lambda$ for the self energy type objects provides a   systematic  solution of the exact Schwinger-Tomonaga equations for the Greens functions of the \tJ or the \tJV model.
More physically we may consider $\lambda$ as representing a fraction of double occupancy, with $\lambda=1$ corresponding to their complete elimination. }

\sr{The theory leads to a novel form of the Greens functions in terms of a pair of self energies, as given in Appendix (\ref{App-ECFL}). For a more complete description the reader may consult \cite{ECFL}.}


\section{Reducible susceptibilities and Conservation laws \label{Sec-4}}
    In this section we outline the relationship between two reducible (dynamical) susceptibilities $\bigchi_{\rho\rho}$ and $\bigchi_{WW}$ for interacting electrons on a lattice, which follows from the conservation of charge. The basic definition of the susceptibility $\chi_{AB}$ for any pair of operators is given in \disp{susceptibility,Matsubara}, the local operators $\rho_m= q_e \sum_\si \wt{C}_{m \si}^\dagger \wt{C}_{m \si}$ correspond to the charge density of electrons at site $m$ and $W$ to the divergence of the lattice current defined in \disp{W1,W2} below. These susceptibilities and their easily derived relationship is valid at all $(\vq,\omega)$, and is then generalized to an almost identical relationship between irreducible susceptibilities below. This generalization is technically non-trivial, and is one of the main formal results of this work. Since it is likely to be of interest to specialists, we have separated out the derivation to appendices, and keep the main text relatively free of these details.
 
    The charge conservation laws follow from the basic observation that both $H_J$ and $V_C$ in the Hamiltonian commute with the local charge density $\rho_m$, thereby only $H_t$ governs its equation of motion. We find the commutator of $\rho_m$ can be expressed by an exact relation involving a Hermitian operator $W_m$
\beq
 [H,\rho_m]= -i W_m, \;\;\mbox{where}\;\; W_m = i q_e \sum_{n \si}t_{mn} \left( \wt{C}_{m \si}^\dagger \wt{C}_{n \si}- \wt{C}_{n \si}^\dagger \wt{C}_{m \si} \right). \label{W1}
\eeq
Defining its Fourier component $W_q$ through
\beq
 W_m&=& \frac{1}{N_s} \sum_q e^{i \vec{q}.\vec{r}_m} W_\vq,  \nn \\
 W_{\vq}& =& i q_e \sum_{k \si} \; (\varepsilon_\vk-\varepsilon_{\vk+\vq})  \; \wt{C}^\dagger_{k \si} \wt{C}_{\vk+\vq \si}, \;\; W^\dagger_\vq=W_{-\vq}, \label{W2}
\eeq
the conservation law for charge can be rewritten as
\beq
[H,\rho_\vq]= -i W_\vq. \label{conservation} 
\eeq
We may think of the W-variable as the lattice counterpart of the divergence of the current $\vec{\nabla}.\vec{J}$ from the following considerations. While \disp{conservation} is valid for arbitrary $q$, in the long wavelength  limit $q\to 0$, we note that 
\beq
\lim_{q\to0} \; W_q \to  -i  \vec{q}. \vec{J}_\vq, \label{OPWtoJ}
\eeq
where the electrical current operator $\vec{J}_\vq= q_e \sum_{k \si}  \left(\vec{\nabla}\varepsilon_k\right)  \wt{C}^\dagger_{\vk \si} \wt{C}_{\vk+\vq \si}$. Hence \disp{conservation} becomes the  familiar continuity equation
\beq
\left( [H,\rho_\vq]+ \vec{q}.\vec{J}_\vq \right)\big|_{q\to0}=0
\eeq
With this remark it is clear that \disp{conservation} can be taken as the condition for conservation of charge at arbitrary wavelengths. 
 
    This leads us to consider in addition to the charge susceptibility, the three W-susceptibilities
\beq
\bigchi_{W W}(\vq,\tau)&\equiv& \bigchi_{W_\vq W_{-\vq}}(\tau); \nn \\
 \bigchi_{\rho W}(\vq,\tau) &\equiv& \bigchi_{\rho_\vq W_{-\vq}}(\tau); \nn \\   \bigchi_{W \rho}(\vq,\tau) &\equiv& \bigchi_{W_\vq \rho_{-\vq}}(\tau). \label{chis}
\eeq
Note here that the location of $W$ in the subscript determines the sign of the attached wave vector. 

    For completeness we note that the optical conductivity is written in terms of a current-susceptibility (see \disp{resistivity}). The unscreened current-current susceptibility can be written in the same fashion as \disp{chis}
\beq
 \bigchi_{JJ}(\vq,\tau)\equiv \bigchi_{J_q J_{-q}}(\tau). \label{chiJJ}
\eeq
Using \disp{OPWtoJ} we can relate this to $\bigchi_{W W}$  for small $\vq$
\beq
\mbox{For} \; {|\vq| a_0 \ll 1 }, \; \;\; \bigchi_{W W}(\vq,\tau) \to |\vq|^2 \bigchi_{JJ}(\vq,\tau). \label{chiWtoJ}
\eeq 
The screened current-current susceptibility satisfies an analogous relation discussed later in \disp{WtoJ}. 

    In \disp{chis-1} and related equations we use the same symbol to represent the real space versions of the susceptibilities. It should be straightforward to distinguish between the two usages from their contexts. Let us first note the relationships between these and the charge susceptibility. From \disp{Matsubara} we note that $\bigchi_{\rho \rho}(\vq,\tau) = \langle T_\tau \rho_\vq(\tau) \rho_{-\vq}(0)\rangle$ and therefore on taking successive  $\tau$-derivatives we get
\beq
\frac{d}{d\tau} \bigchi_{\rho \rho}(\vq,\tau)&=& (-i)   \bigchi_{W \rho}(\vq,\tau) \label{scr-1}
\eeq 
where we used the vanishing of the equal time commutator $[\rho_q,\rho_{-q}]$. Taking a further derivative we find
\beq
\frac{d}{d\tau} \bigchi_{W \rho}(\vq,\tau)&=&\delta(\tau) \langle[W_\vq,\rho_{-\vq}]\rangle-\bigchi_{W [H,\rho]}(\vq,\tau)\nn\\
&=&- i N_s \kappa(\vq) \delta(\tau) + i \bigchi_{WW}(\vq,\tau), \label{scr-2}
\eeq
and taking $\vec{q}$ along the $x$ axis
\beq
\kappa(\vq) = \frac{2 q_e^2}{N_s}\sum_{k \si} \left( \varepsilon_{\vk+\vq}-\varepsilon_{\vk}  \right) \langle  \wt{C}^\dagger_{\vk \si} \wt{C}_{\vk \si} 
 \rangle. \label{kappa}
\eeq
For general non-parabolic bands 
\beq\lim_{q\to 0} \kappa(\vq) =  |\vq|^2 {\cal T} \label{tau-1}  \label{low-q-kappa} \eeq
where the variable ${\cal T}$ (equal to   the stress tensor per site $\frac{1}{N_s}\langle \tau^{xx}\rangle $ in \cite{Shastry-ROPP}),  is given by
\beq
{\cal T}= \frac{q_e^2}{N_s} \sum_{k \si} \left( \frac{d^2 \varepsilon_{\vk}}{d k_x^2}  \right) \langle  \wt{C}^\dagger_{\vk \si} \wt{C}_{\vk \si} 
 \rangle, \label{kappa-smallq}
\eeq

It can be seen that ${\cal T}$ is related to the f-sumrule for the optical conductivity
\beq
\int_{-\infty}^\infty \frac{d \omega}{\pi} \; \Re \, \sigma(\omega) = {\cal T}. \label{f-sumrule}
\eeq

    When parabolic bands $\varepsilon_\vk= k^2/(2m)$ are used, we find at all $\vec{q}$ the simple result 
\beq
{\cal T} = \left(\frac{nq_e^2}{m}\right)  , \label{kappa-parabolic}
\eeq 
where $n=N/N_s$ is the electron density \cite{dimension-kappa}. Combining \disp{scr-1,scr-2}, we find
\beq
\frac{d^2}{d\tau^2} \bigchi_{\rho \rho}(\vq,\tau)= - \delta(\tau) N_s \kappa(\vq) + \bigchi_{WW}(\vq,\tau),\label{scr-3}
\eeq

    Multiplying both sides by $e^{i \Omega_\nu \tau}$ and integrating over $\tau$ as in \disp{Matsubara} we find
\beq
\bigchi_{\rho \rho}(\vq,i\Omega_\nu)= \frac{1}{\Omega_\nu^2} \left(N_s \kappa(\vq) - \bigchi_{WW}(\vq,i \Omega_\nu) \right). \label{scr-4}
\eeq
The large $\Omega$ behaviour is determined by the first term, since $\bigchi$ vanishes there, and leads to the important plasma sum-rule discussed below in  \disp{plasma-limit,plasmons-1,plasmons-2,plasmons-3,plasmons-4}.

    Analogous relations can be derived for real frequencies using the definitions in \disp{susceptibility}. We write \disp{scr-1} and \disp{scr-2} directly in $\omega$ space as
\beq
(\omega)\bigchi_{\rho \rho}(\vq,\omega)&=& i \bigchi_{W \rho}(\vq,\omega)= -i\bigchi_{\rho W}(\vq,\omega) \label{der-1} \\
(\omega)\bigchi_{W \rho}(\vq,\omega)&=& i N_s \kappa(\vq) -i\bigchi_{W W}(\vq,\omega) \label{der-2} 
\eeq 
where $\kappa$ is defined in \disp{kappa}. It is clear that these relations in  $\omega$ can be obtained from \disp{scr-4} by analytically continuing the Matsubara frequency $i\Omega_\nu \to \omega+ i 0^+$. Combining these we get
\beq
\bigchi_{\rho \rho}(\vq,\omega)= -\frac{1}{\omega^2} \left(N_s \kappa(\vq) - \bigchi_{WW}(\vq,\omega) \right). \label{scr-5}
\eeq
which is analytically continued version of \disp{scr-4} for real frequencies.

    We note the relationship between the reducible susceptibility $\bigchi_{\rho\rho}$ and the dielectric function $\varepsilon(\vq,\omega)$
\beq
\frac{1}{\varepsilon(\vq,\omega)}=1- \frac{V(\vq)}{ q_e^2} \bigchi_{\rho \rho}(\vq,\omega).  \label{dielectric-1}
\eeq
This is easily established \cite{Nozieres}  from linear response theory. From \disp{scr-5} we note that we can compute $\varepsilon(\vq,\omega)$ directly from $\bigchi_{\rho \rho}(\vq,\omega)$, or alternately from $\bigchi_{WW}(\vq,\omega)$. When done exactly, these alternate formulas must of course coincide, but they offer important possibilities for approximations that we shall pursue below.


\section{ Nozi\`eres type  expressions for $\varepsilon(\vq,\omega)$ using two  irreducible susceptibilities   \label{Sec-5}}

We turn to the irreducible susceptibilities $\hatchi_{\rho \rho}$ and $\hatchi_{WW}$, which are more convenient since they already contain to a large extent the effects of the long ranged part of the Coulomb interaction. In the electron gas problem these susceptibilities can be rigorously defined diagrammatically by using screened vertex functions \cite{Nozieres}. We can easily generalize the treatment in Nozi\`eres to conventional electrons in a tight binding model. This corresponds to \disp{model} without the $H_J$ and with conventional electrons replacing the Gutzwiller projected electron operators $\wt{C}_{j \si}$. With Gutzwiller projection the entire calculation is non-trivial since the definition of vertex functions is beset with technical difficulties described elsewhere \cite{ECQL,ECFL}. {In \appdisp{App-C} we present a workaround, avoiding the use of vertex functions entirely and instead using the relationship between correlation functions directly.} The final relationships between the two sets of susceptibilities, valid for a tight binding band of non-canonical electrons at arbitrary $\vq,\omega$, are exactly the same as that for conventional electrons. 

    We denote the pair of subscripts $\{\rho,W \}$ by a symbol $\mu$ (or $\nu$), and introduce the irreducible susceptibilities $\hatchi_{\mu \nu}(\vq,\omega)$. Rules for calculating the reducible and irreducible susceptibilities from taking functional derivatives of the Green's functions are provided in the \appdisp{App-B} and \appdisp{App-C}. The relationships between the irreducible and the reducible susceptibilities are compactly given by (see \disp{hatunhat-a})
\beq
&&\bigchi_{\mu \nu}(q)= \hatchi_{\mu \nu}(q) - \frac{1}{q_e^2} V(\vq)  \hatchi_{\mu \rho}(q) \bigchi_{\rho \nu}(q). \label{hatunhat}
\eeq
This can be solved for all the components and displays the screened nature of the resulting susceptibilities. The density-density response $\bigchi_{\rho \rho}$ is simplest since all terms on the right have the same subscripts. Gathering terms $\bigchi_{\mu \nu}(q)$ on the left, we find
\beq
&&\bigchi_{\rho \rho}(q)= \frac{\hatchi_{\rho \rho}(q)}{  1+ \frac{1}{q_e^2} V(\vq) \hatchi_{\rho \rho}(\vec{q}, \omega)}. \label{relations-0} 
\eeq
Using \disp{dielectric-1}, dielectric function is given in terms of the irreducible susceptibility by
\beq
\varepsilon(q) \equiv \varepsilon(\vec{q}, \omega) = 1+ \frac{1}{q_e^2} V(\vq) \hatchi_{\rho \rho}(\vec{q}, \omega), \label{dielectric-constant-1}
\eeq
with the Coulomb potential given by \disp{Coulomb-3d,Coulomb-2d}. Proceeding similarly we find the other three susceptibilities in terms of their screened counterparts. With $q=(\vq,\omega)$ the relationships between the four susceptibilities are given by
\beq
&&\bigchi_{\rho \rho}(q)= \frac{\hatchi_{\rho \rho}(q)}{  \varepsilon({q})}\label{relations-01} \\
&& \bigchi_{\rho W}(q)= \frac{\hatchi_{\rho W}(q)}{ \varepsilon(q) },\label{relations-1} \\
&&  \bigchi_{W \rho }(q)= \frac{\hatchi_{W \rho }(q)}{ \varepsilon(q) },\;\;\; \label{relations-2} \\
&&\bigchi_{W W}(q)= \hatchi_{W W}(q)- \frac{V(\vq)}{q_e^2 \varepsilon(q)}\hatchi_{W \rho}(q)\; \hatchi_{\rho W}(q). \label{relations-3}
\eeq
It is worth noting the connection between these results and the equations presented by Nozi\`eres \cite{Nozieres} for the homogeneous electron gas --- denoted by a prefix ``N''. The vertex $W$ (see \disp{W1}) replaces the (longitudinal) current vertex $(-i) \frac{\vq .\vec{k}}{m}$ in \refdisp{Nozieres}, who chooses $\vq$ along the z (or 3) axis and denotes  $\frac{k_z}{m}$ by ``3''. Our pair of operators map as $\{\rho \to 4,W \to 3\}$ to those of Nozi\`eres. Our susceptibilities $\bigchi_{\mu,\nu}$ are $\frac{i}{2\pi \Omega} \times S_{\alpha,\beta}$ of Nozi\`eres. Our dielectric function in \disp{dielectric-constant-1} corresponds to his Eq.~(N-6.170), our \disp{relations-0,relations-1,relations-2,relations-3} correspond to Eq.~(N-6.168).

We next study the charge conservation laws for the screened susceptibilities $\hatchi_{\mu \nu}$,  combining the conservation relations  \disp{der-1,der-2} for the unscreened susceptibilities and the relations (\disp{relations-0,relations-1,relations-2,relations-3}). Now using $\hatchi_{\rho\rho}= \bigchi_{\rho\rho}\times \varepsilon$ and \disp{relations-1} we write
\beq
(\omega) \hatchi_{\rho \rho}(\vq,\omega)= i \hatchi_{W \rho}(\vq,\omega) = -i \hatchi_{ \rho W}(\vq,\omega) . \label{der-3}
\eeq
For the next step we rearrange \disp{scr-5} as 
$$N_s \kappa(\vq)= \bigchi_{WW}(\vq,\omega)- (\omega^2)\bigchi_{\rho\rho}(\vq,\omega),$$ 
and substitute the screening equations \disp{hatunhat}, \disp{relations-1}, and \disp{relations-2} for the right hand side. This yields
\beq
N_s \kappa(\vq)
&=& \hatchi_{WW}(\vq,\omega)- (\omega^2)\hatchi_{\rho\rho}(\vq,\omega)\nn\\
&&+\frac{V(\vq)}{q_e^2} \left((\omega)^2 \bigchi_{\rho \rho}(\vq,\omega)\hatchi_{\rho \rho}(\vq,\omega) -\hatchi_{W\rho}(\vq,\omega)\bigchi_{\rho W}(\vq,\omega)\right). \nn\\ \label{der-4}
\eeq
We now use the conservation laws \disp{der-1} $(\omega)\bigchi_{\rho \rho} =-i \bigchi_{\rho W }$, and
\disp{der-3} $(\omega) \hatchi_{\rho \rho}= i \hatchi_{ W  \rho}$. This shows that the second  term in \disp{der-4} vanishes identically! We thus find the exact result
\beq
\hatchi_{\rho \rho}(\vq,\omega)= \frac{1}{\omega^2} \left( \hatchi_{WW}(\vq,\omega)-N_s \kappa(\vq)  \right), \label{scr-6}
\eeq
as the screened version of \disp{scr-5}. At large $|\omega| \gg t$, since $\hatchi_{WW}(\vq,\omega)\to 0$, we find the important asymptotic behaviour for the real part
\beq
\lim_{\omega \gg t} \hatchi_{\rho \rho}(\vq,\omega) =  -\frac{N_s \kappa(\vq)}{\omega^2} .\label{asymptotic-plasmon}
\eeq
For any generic $\vec{q}$ we must obtain a finite static limit of $\hatchi_{\rho \rho}$, which requires an exact cancellation between the two terms in the bracket, i.e.
\beq
\hatchi_{WW}(\vq,0)= N_s \kappa(\vq), \label{constraint-chiWW}
\eeq
and therefore can alternately  write
\beq
\hatchi_{\rho \rho}(\vq,\omega)= \frac{N_s \kappa(\vq)}{\omega^2} \left( \hatchi_{WW}(\vq,\omega)/ \hatchi_{WW}(\vq,0)-1\right), \label{scr-7}
\eeq
Combining \disp{scr-6} we get an expression for $\varepsilon(\vq,\omega)$, alternate to \disp{dielectric-constant-1}
\beq
\varepsilon(\vec{q}, \omega) = 1+ \frac{1 }{q_e^2 \omega^2} V(\vq)\left( \hatchi_{W W}(\vec{q}, \omega)- N_s \kappa(\vq) \right). \label{dielectric-constant-2}
\eeq
The expressions \disp{dielectric-constant-1,dielectric-constant-2} are the twin Nozi\`eres formulas referred to in the introduction. The formal derivation shows that if the two expressions are evaluated exactly, then they must coincide. Approximations are not guaranteed to retain their equivalence. In certain classes of approximate calculations they do agree. For example the standard random phase approximation (RPA) uses the non-interacting Green's functions $G_0$, and the vertex is the bare one. The two susceptibilities are found from the bubble diagrams \cite{Nozieres}
\beq
\hatchi_{\rho \rho}^{(0)}(\vq,i \Omega_\nu)&=& -  q_e^2 \sum_{k \si} G_0(k) G_0(k+q)= 2 q_e^2 \sum_{\vec{k}} \frac{f_\vk-f_{\vk+\vq}}{\varepsilon_{\vk+\vq}-\varepsilon_{\vk}- i \Omega_\nu},\;\;\;\;\;\; \label{RPA-1} \\
\hatchi_{W W}^{(0)}(\vq,i \Omega_\nu)&=& -  q_e^2 \sum_{k\si} G_0(k) G_0(k+q)(\varepsilon_\vk-\varepsilon_{\vk+\vq})^2\nn \\
&=& 2 q_e^2 \sum_{\vec{k}} \frac{f_\vk-f_{\vk+\vq}}{\varepsilon_{\vk+\vq}-\varepsilon_{\vk}- i \Omega_\nu} (\varepsilon_\vk-\varepsilon_{\vk+\vq})^2\label{RPA-2}
\eeq
In this case the validity of \disp{scr-7} can be shown by multiplying \disp{RPA-1} by $(i\Omega_\nu)^2$, followed by the use partial fractions. This process reduces it to \disp{RPA-2} plus a term equivalent to $N_s q_e^2 \tau(q)$. 

    In the case of canonical electrons, we can define vertex functions suitably, and make approximations for the vertex as well as the Green's functions in a consistent way \cite{Nambu,Nozieres} so that the Ward-Takahashi identities are satisfied. Such approximations guarantee the equivalence of the approximate versions of \disp{dielectric-constant-1,dielectric-constant-2}. The RPA described above is an example of such an approximation, this scheme trivially satisfies the Ward-Takahashi identities.


\section{Formulas for  the Approximate Dielectric Function\label{sec3.7} }

    The main problem of interest in this work is the \tJV model. Here the short ranged Coulomb interactions lead to a Mott-Hubbard type insulating state at half filling, and doping such a state with holes leads to a metallic state of a very unusual \sr{nature, characterized} with a small quasiparticle weight. Adding long ranged Coulomb interactions to this state poses a considerable difficulty. While we are able to  obtain a fairly sophisticated single electron Green's function $\G$ from the ECFL theory\cite{ECFL}, the two particle response functions are currently unreliable. This is a difficult task even for the simpler case of canonical electrons, and has led to a variety of beyond-RPA type approximations \cite{Hubbard-Singwi}. For Gutzwiller projected electrons, it is indeed a formidable task. In this situation, the availability of the two alternate formulas \disp{dielectric-constant-1,dielectric-constant-2} is very helpful. We can compute the susceptibilities $\hatchi_{\rho \rho}(\vec{q}, \omega)$ and $\hatchi_{WW}(\vec{q}, \omega)$ at all $\{\vq,\w\}$, using only the above $\G$ within a bubble scheme $\G \G$ as described below in \disp{Inter-1,Inter-2}. Being approximate, these two estimates differ in general, but provide complementary perspective on the dielectric response at various $\vq,\w$. By comparing these estimates with known (exact) limiting behaviour of the  susceptibility detailed in \appdisp{App-Limits}, we can ascertain their respective regimes of validity. This provides us with the possibility of combining the two formulas, to obtain an approximate answer whose broad characteristics are known beforehand.

\subsection{Formula for irreducible susceptibility \label{new-rep-old} in terms of a  self-energy $\Psi(\vq,\omega)$}

    We begin with a novel representation for the susceptibility using the freedom to define suitable generalized self-energies of Green's functions, as discussed in \cite{Anatomy,Mori,Dupuis}. We start from the high frequency moment expansion \disp{moment-expansion-2}, in inverse powers of $\omega^2$ as discussed in \appdisp{moments-irreducible-chi}. This series can be formally rewritten in a continued fraction representation following Mori \cite{Mori,Dupuis} as
\beq
\frac{1}{q_e^2 N_s} \hatchi_{\rho\rho}(\vq,\omega) &=&\frac{\beta_1(\vq)}{\omega^2 - \alpha_1(\vq) - \Sigma^{(0)}_\chi(\vq,\omega)} \label{first-selfenergy} \\
\Sigma^{(0)}_\chi(\vq,\omega)&=&\frac{\beta_2(\vq)}{\omega^2 - \alpha_2(\vq) - \Sigma^{(1)}_\chi(\vq,\omega)} \\
& \vdots& \nn
\eeq
where $\beta_1 = - \widetilde{\omega}^{(1)}_{\vq}$ is the negative of the first moment of frequency \disp{equality-1,mu-dispersion}, and $\Sigma^{(m)}_\chi(\vq,\omega)$ with $m=0,1, \ldots$ represent the successive ``self-energies''. They are characterized by the property that for $\omega \gg t$ they behave as $\Sigma^{(m)}_\chi(\vq,\omega) \sim \frac{\beta_{m+2}}{\omega^2}$, and thus vanish. The coefficients $\alpha_m,\beta_m$ are functions of $\vq$ and can be found in principle, in terms of the frequency moments. It is more convenient for our purpose to rewrite \disp{first-selfenergy} in by eliminating $\alpha_1$ in favor of the static limit of $\Sigma^{(0)}_\chi$, and using $\beta_1= - \widetilde{\omega}^{(1)}_{\vq}$. This leads to
\beq
 \frac{1}{q_e^2 N_s} \hatchi_{\rho\rho}(\vq,\omega)=\left( \frac{q_e^2 N_s}{ \hatchi_{\rho\rho}(\vq,0)} - \frac{\omega^2}{\widetilde{\omega}^{(1)}(\vq)} + \frac{1}{\widetilde{\omega}^{(1)}(\vq)}\left(\Sigma^{(0)}_\chi(\vq,\omega)- \Sigma^{(0)}_\chi(\vq,0)  \right) \right)^{-1}. \label{new-rep-0}
\eeq
We can simplify the notation by defining a new self-energy type function 
\beq
\Psi(\vq,\omega)= \frac{1}{\widetilde{\omega}^{(1)}(\vq)}\left(\Sigma^{(0)}_\chi(\vq,0)- \Sigma^{(0)}_\chi(\vq,\omega)  \right), \label{Psi-Sigma}
\eeq
with $\widetilde{\omega}^{(1)}(\vq)$ detailed in \disp{equality-1,moment-2}. The irreducible susceptibility is now given by 
\beq
 \frac{1}{q_e^2 N_s} \hatchi_{\rho\rho}(\vq,\omega)=\left( \frac{q_e^2 N_s}{ \hatchi_{\rho\rho}(\vq,0)} - \frac{\omega^2}{\widetilde{\omega}^{(1)}(\vq)} - \Psi(\vq,\omega) \right)^{-1}. \label{new-rep}
\eeq
This self-energy $\Psi$ can be found from $\hatchi_{\rho\rho}(\vq,\omega)$, if the latter is known, by inversion of \disp{new-rep}, and can be expressed formally in terms of the higher moments $\widetilde{\omega}^{(2j+1)}(\vq)$ using \disp{moment-expansion-2} \cite{Anatomy,Mori,Dupuis}. The self energy vanishes in the static limit by construction
\beq 
\Psi(\vq,\omega)\vert_{\omega\to0} =0, \label{vanishing-Psi}
\eeq 
and has a finite high frequency limit (from the first term in \disp{Psi-Sigma}).

    We note that from the Lehmann representation of $\hatchi_{\rho\rho}(\vq,\omega)$ that the $\Sigma^{(m)}$ in complex $\omega$ are analytic everywhere except the real axis. This implies that all singularities are located on the real axis, and hence these can be further represented in the form
\beq
\Sigma^{(0)}_\chi(\vq,\omega)= -\frac{1}{\pi} \int_{-\infty}^\infty \, d\nu \, \frac{\Im \,\Sigma^{(0)}_\chi(\vq,\nu)}{\omega- \nu + i \eta} 
\eeq
where $\eta=0^+$. Using the fact that $\Sigma^{(0)}_\chi(\vq,0)$ and $\hatchi_{\rho\rho}(\vq,0)$ are real, it follows from \disp{new-rep} that
 \beq
 \Im \, \Psi(\vq,\omega)= - q_e^2 N_s  \Im \,\hatchi^{-1}_{\rho\rho}(\vq,\omega). \label{connectPsiChi}
 \eeq
Using the analyticity of $\Psi$ in the upper half complex $\omega$ plane, together with \disp{vanishing-Psi}, we obtain an expression for $\Psi$ in terms of the imaginary part of the inverse susceptibility
\beq
 \Psi(\vq,\omega) = (q_e^2 N_s) \left( \frac{1}{\pi} \int_{-\infty}^\infty \, d\nu \, \frac{\Im \,\hatchi^{-1}_{\rho\rho}(\vq,\nu)}{\omega- \nu + i \eta}+ \frac{1}{\pi} \int_{-\infty}^\infty \, d\nu \, \frac{\Im \,\hatchi^{-1}_{\rho\rho}(\vq,\nu)}{ \nu } \right). \label{Psi-chi-inverse}
\eeq
Here the second term is expected to be finite due to the odd-ness in frequency of $\Im \, \hatchi^{-1}_{\rho\rho}(\vq,\nu)$. It follows from \disp{connectPsiChi} that $\Im \, \Psi(\vq,\omega)$ is odd in $\omega$ while \disp{Psi-chi-inverse} says that $\Re \, \Psi(\vq,\omega)$ is even in $\omega$.
 
    In summary the susceptibility $\hatchi_{\rho\rho}(\vq,\nu)$ is determined in \disp{new-rep} by the self energy $\Psi(\vq,\omega)$ satisfying \disp{vanishing-Psi} and \disp{Psi-chi-inverse}, together with two functions of $\vq$ only:
 (a) the static susceptibility $\hatchi_{\rho\rho}(\vq,0)$ and (b) the moment $\widetilde{\omega}^{(1)}(\vq)$ (with dimensions of frequency). The latter is calculable for all $\vq$ in terms of equal time correlations from \disp{moment-2}.

    Separating $\Psi=\Psi' + i\Psi''$, we can write the complex susceptibility \disp{new-rep} conveniently as
\beq
\frac{1}{q_e^2 N_s} \hatchi_{\rho\rho}(\vq,\omega)=\left(\frac{1}{\wt{\w}^{(1)}(\vq)} \{\Omega^2(\vq,\w)-\w^2\} - i  \Psi''(\vq,\omega) \right)^{-1}, \label{new-rep-2}
\eeq
and hence
\beq
\frac{1}{q_e^2 N_s} \Im \, \hatchi_{\rho\rho}(\vq,\omega)= \frac{ [ \widetilde{\omega}^{(1)}(\vq) ]^2 \Psi''(\vq,\omega) }{[ \widetilde{\omega}^{(1)}(\vq)  \Psi''(\vq,\omega)]^2+ \{\Omega^2(\vq,\omega)-\omega^2\}^2 }. \label{im-chi-expression}
\eeq
In these expressions the characteristic energy scale $\Omega$ is given by
\beq
 \Omega^2(\vq,\omega)&=&  \widetilde{\omega}^{(1)}(\vq)  \left( \frac{d \mu}{ d n} \gamma(\vq) - \Psi'(\vq,\omega)\right) \label{Omega} \\
 \gamma(\vq) &=& \frac{\hatchi_{\rho\rho}(0,0)}{\hatchi_{\rho\rho}(\vq,0)}, \;\; \gamma(0)=1, \label{gamma}
\eeq
and we made use of the exact result \disp{compress-1} to express the static limit of the susceptibility in terms of the thermodynamic variable $\frac{d \mu}{ d n}$. Recall that the compressibility $\chi_{comp}=\frac{1}{2 n(0)}\frac{d n}{d \mu} \, \chi^{(non)}_{comp}$, where $n(0)$ is the density of states per site per spin, and hence this representation also satisfies the compressibility sum-rule \disp{compress-1}.

    From \disp{im-chi-expression} we see that  $\Im \, \hatchi_{\rho\rho}(\vq,\omega)$ 
    is expected to have peaks. The peak frequency is termed as  $\Omega_p(\vq)$, and  identified with  $\omega_0$,   the positive  root  of
     \beq \omega_0^2= \Omega^2(\vq,\omega_0), \; \; \mbox{ i.e. } \;\; \Omega_p(\vq)= \omega_0. \label{Omega-p-exact} \eeq 
     The  root is approximately located at the energy scale $\Omega(\vq,0)$, i.e.
     \beq
\Omega_p(\vq) \sim \Omega(\vq,0) = \sqrt{ \widetilde{\omega}^{(1)}(\vq) \, \frac{d \mu}{ d n} \gamma(\vq)} \label{Omega-p}.
\eeq 
We display alternate  versions of this expression in \disp{emergentOmega,emergentOmega-2}.
The width of the peak is given by
\beq
\Gamma_p(\vq) = \sqrt{ \widetilde{\omega}^{(1)}(\vq) \Psi''(\vq,\Omega_p(\vq))} \label{Gamma-p}.
\eeq
As we explicitly see later,   the approximation \disp{Omega-p} for  $\Omega_p(\vq)$ at low $\vq$, is larger than  the  exact  peak frequency $\omega_0$ in \disp{Omega-p-exact} by a factor of $\sim$2, the discrepancy arising from the substantial  breadth of the peak, $\Gamma_p\gssim \Omega_p$.
In terms of these variable  we can approximately write 
\beq
\frac{1}{q_e^2 N_s} \Im \, \hatchi_{\rho\rho}(\vq,\omega)\sim \frac{  \widetilde{\omega}^{(1)}(\vq)  \Gamma^2_p(\vq) }{ \Gamma^4_p(\vq)+ \{\Omega_p^2(\vq)-\omega^2\}^2 }. \label{im-chi-2}
\eeq

    The representation \disp{new-rep} also exactly satisfies the known high $\omega$ behavior \disp{moment-expansion-2}, and therefore reproduces the correct plasma frequency \disp{3d-plasmon}. It should also be clear that with obvious changes to the variables, the above formulas \disp{new-rep,new-rep-2} can be useful for other physical situations such as the homogeneous electron gas etc.


\subsection{Approximate formulas for the irreducible susceptibility $ \hatchi_{\rho\rho}(\vq,\nu)$ }

    It is very convenient to calculate the susceptibility starting from formulas \disp{new-rep}. The input variables in \disp{new-rep,new-rep-2} are found from the ECFL theory, using suitable approximations described next. We make extensive use of the {\em bubble approximation}, where in taking the derivative with respect to the external potential in \disp{chis-22}, the $\G$ is assumed to depend on this potential {\em only} through the explicit terms as in \disp{G0-2}, and the implicit dependence via the other factors are thrown out. For $\hatchi_{\rho\rho}$ we find an approximate expression from this {bubble approximation} 
\beq
\hatchi_{\rho \rho}^{\mbox{\tiny Bub}}(\vec{q}, \omega)&=& - q_e^2 \sum_{k \si} \G(k) \G(k+q) \label{Inter-1}, 
\eeq
and evaluating $\hatchi_{W W}$ within the bubble approximation
\beq
\hatchi_{W W}^{\mbox{\tiny Bub}}(\vec{q}, \omega)&=& - q_e^2 \sum_{k\si} \G(k) \G(k+q)(\varepsilon_k-\varepsilon_{k+q})^2. \label{Inter-2}
\eeq
Using the spectral representation \disp{spectral} for $\G$ the latter reduces to
\beq
\hatchi_{WW}^{\mbox{\tiny Bub}}(\vec{q}, \omega)&=& 2 q_e^2 \sum_{k } (\varepsilon_k-\varepsilon_{k+q})^2  \nn \\ && \times \int_{\nu_1 \nu_2} \frac{f(\nu_1)-f(\nu_2)}{\nu_2-\nu_1-\omega - i 0^+} A(k, \nu_1) A(k+q, \nu_2), \;\;\nn \\ \label{Inter-1.1} 
\eeq
where $\int_{\nu}= \int_{-\infty}^\infty d\nu$. The density response $\hatchi_{\rho \rho}^{\mbox{\tiny Bub}}(\vec{q}, \omega)$ is found by dropping the factor $(\varepsilon_k-\varepsilon_{k+q})^2$ in this formula. The spectral functions in our model (see \figdisp{spectrum}) consist of a quasiparticle part with a much reduced weight $Z\ll1$, and an extended background part. The indicated integrations can be performed numerically.

    Our two starting points are susceptibilities found from these bubble estimates and \disp{scr-7} 
\beq
\hatchi_A(\vq,\omega) &=& \hatchi^{\mbox{\tiny Bub}}_{\rho\rho}(\vq,\omega) \label{chiA} \\
\hatchi_B(\vq,\omega) &=&\frac{N_s \kappa(\vq)}{\omega^2} \left( \hatchi^{\mbox{\tiny Bub}}_{WW}(\vq,\omega)/\hatchi^{\mbox{\tiny Bub}}_{WW}(\vq,0)-1 \right). \label{chiB} 
\eeq
 The estimate $\hatchi_A$ provides a reasonable estimate in the static limit for the susceptibility. The magnitude of the compressibility, found by taking the $\vq\to 0$ limit, is much smaller than the band value, as seen in \figdisp{fig-compressibility}. It is comparable for most densities to that found from thermodynamical evaluation of $\frac{d n}{d \mu}$ (see \figdisp{fig-compressibility}). At finite $\vq$ its shape is compared to that of the band susceptibility apart from some interchanges of magnitudes between different directions (see \figdisp{static}). The imaginary part of $\hatchi_A$ shows a quasiparticle contribution of the type $\chi''\propto \frac{\pi \omega}{|\vq| v_F}$ for very small $\omega < |\vq| v_F Z$. For larger $\omega$, it has a broad  contribution from the background spectral functions, but does not give the first moment of frequency, and is therefore not satisfactory.

For $\hatchi_B$ we verify that $\hatchi_{W W}^{\mbox{\tiny Bub}}(\vec{q}, 0)$  agrees closely with $N_s \kappa(\vq)$, calculated independently using a single Green's function $\G$ from \disp{kappa}, at all $\vq$ (see \figdisp{chiBubWWstatic}). The estimate $\hatchi_B$ is expected to be satisfactory at finite (high) frequencies since it is constructed to satisfy the first moment of frequency in the high $\omega$ limit. However  at low $\omega$ it is does not capture the quasiparticle contribution discussed above. Further the static limit --- found from the ${\cal O}(\omega^2)$ limiting behavior of $ \hatchi^{\mbox{\tiny Bub}}_{WW}(\vq,\omega)$ --- does not display the behavior expected for an incompressible system discussed above. Thus the two estimates are successful in almost non-overlapping regimes of frequency.

    Before proceeding we note that the two expressions \disp{chiA,chiB} lead to two different self energies
\beq
\Psi_A(\vq,\omega) + \frac{\omega^2}{\widetilde{\omega}_A^{(1)}(\vq)}=\frac{N_s q_e^2}{\hatchi_A(\vq,0)}- \frac{N_s q_e^2}{\hatchi_A(\vq,\omega)} \label{PsiA} \\
\Psi_B(\vq,\omega) + \frac{\omega^2}{\widetilde{\omega}_B^{(1)}(\vq)}=\frac{N_s q_e^2}{\hatchi_B(\vq,0)}- \frac{N_s q_e^2}{\hatchi_B(\vq,\omega)} \label{PsiB}.
\eeq  
The first frequency moment $\widetilde{\omega}_B^{(1)}(\vq)$ in the second equation \disp{PsiB} is in fact exact, i.e. $\widetilde{\omega}_B^{(1)}(\vq)=\widetilde{\omega}^{(1)}(\vq)$, as explained above. The corresponding frequency $\widetilde{\omega}_A^{(1)}(\vq)$ is not correct, and we show that it is possible to avoid using it altogether.

\sr{ We next construct two approximations to the irreducible susceptibility
\beq
\hatchi_{\rho\rho}^{(I)}(\vq,\omega) \; \mbox{ and } \hatchi_{\rho\rho}^{(II)}(\vq,\omega). \label{begin-hatchi}
\eeq
 When the context is clear we drop the subscript and use the simplified notation
\beq
\hatchi_{\rho\rho}^{(I,II)}(\vq,\omega) \leftrightarrow  \hatchi^{(I,II)}(\vq,\omega). \label{briefly-hatchi}
\eeq
}
\sr{    Consider the approximate susceptibility $\hatchi^{(I)}$ combining the two susceptibilities $\hatchi_A,\hatchi_B$ in the form
\beq
\hatchi^{(I)}(\vq,\omega)&=& \left\{ \frac{1}{\hatchi_A(\vq,0)}- \frac{1}{\hatchi_B(\vq,0)}+ \frac{1}{\hatchi_B(\vq,\omega)} \right\}^{-1}. \label{appx0} \\
\eeq 
We can rewrite this using \disp{PsiB} in the form
\beq
\hatchi^{(I)}(\vq,\omega)&=& N_s q_e^2\left\{ \frac{N_s q_e^2}{\hatchi_A(\vq,0)} - \frac{\omega^2}{\widetilde{\omega}^{(1)}(\vq)} - \Psi_B(\vq,\omega) \right\}^{-1}. \label{appxI}
\eeq
Since $\Psi_B(\vec{q},0)=0$, we see that $\hatchi^{(I)}(\vq,\omega)$ has the correct static limit, and since $\Psi_B(\vec{q},\omega)$ vanishes at high $\omega$,  the approximate $\hatchi^{(I)}(\vq,\omega)$ also has the correct plasma frequency, while respecting the strong local correlations. It therefore serves as a reasonable first approximation over the entire frequency domain. }

    A feature that is missing from $\hatchi^{(I)}$ in \disp{appxI}, is the quasi-particle contribution. This was present in \disp{chiA}, but was left out in \disp{appxI} since we threw out all the frequency dependence of $\hatchi_A$. We can incorporate this contribution, again approximately, by making a correction to $\Psi_B$ taken from $\Psi_A$. Inspection shows that for small $\vq,\omega$ the quasiparticle feature in $\hatchi_A$ arises from a contribution  $\Im \, \Psi_A \propto \frac{\omega}{|\vq| v_f} $. It is analogous to the familiar correction that arises in the Lindhard function from quasiparticles \cite{Nozieres,AGD,Fetter-Walecka}. This quasiparticle contribution  leads to $|\Im \, \Psi_A(\vq,\omega)| > |\Im \, \Psi_B(\vq,\omega)|$ for small enough $\omega$ at a fixed $\vq$, while for larger $|\omega|$ we
find
$ |\Im \, \Psi_B(\vq,\omega)| \gg |\Im \, \Psi_B(\vq,\omega)|$. To further refine the approximation, we keep this observation in mind and add the incremental $\delta \Psi_{QP}(\vq,\omega)$ containing the quasiparticle damping to $\Psi_B$,  
\beq
\hatchi^{(II)}(\vq,\omega)
&=& N_s q_e^2\left\{ \frac{N_s q_e^2}{\hatchi_A(\vq,0)} - \frac{\omega^2}{\widetilde{\omega}^{(1)}(\vq)}  -\Psi_B(\vq,\omega) - \delta \Psi_{QP}(\vq,\omega) \right\}^{-1} .\nn \\\label{appxII}
\eeq
\sr{ 
In order to determine the appropriate correction term  $ \delta \Psi_{QP}(\vq,\omega)$ in the above expression, we argue as follows.}
Since $\Im \, \delta \Psi_{QP}$ should add the damping due to quasiparticles, with $\omega>0$ we choose 
\beq
\Im \,\delta \Psi_{QP}(\vq,\omega)  + \Im \, \Psi_{B}(\vq,\omega) = \mbox{Max} \{\Im \, \Psi_{A}(\vq,\omega),\Im \, \Psi_{B}(\vq,\omega) \}. \label{imdeltaPsi} 
\eeq
\sr{This construct  isolates the excess damping present in $\Psi_B(\vq,\omega)$ over and above that in $\Psi_A(\vq,\omega)$, due to quasiparticles at low $\omega$. In slightly more technical terms } 
 $\Im \, \delta \Psi_{QP}(\vq,\omega)$ vanishes outside  the region  $|\Im \, \Psi_A(\vq,\omega)| > |\Im \, \Psi_B(\vq,\omega)|$. For $\omega<0$ a similar argument can be used keeping in mind the odd-ness of $\Im \, \Psi's$ in $\omega$, we use $\mbox{Min}$ instead of $\mbox{Max}$ in \disp{imdeltaPsi}.  The  real part of $\delta \Psi_{QP}$ can be calculated using the Kramers-Kronig relation, i.e.   by taking the real part in \disp{Psi-chi-inverse}
\beq
\Re\,\delta \Psi_{QP}(\vq,\omega) =- {\cal P} \frac{1}{\pi}  \int_{-\infty}^\infty \, d\nu \, \frac{\Im\,\delta \Psi_{QP}(\vq,\nu) }{\omega- \nu }- \frac{1}{\pi}   \int_{-\infty}^\infty \, d\nu \, \frac{\Im\,\delta \Psi_{QP}(\vq,\nu)}{ \nu }, \label{redeltaPsi}
\eeq 
whereby we guarantee that $\delta \Psi_{QP}(\vq,0)=0$.

    On further separating the complex self-energies, these two approximate susceptibilities \disp{appxI,appxII} lead to expressions analogous to \disp{new-rep-2}, with the same static susceptibility \disp{gamma} but slightly different characteristic frequencies $\Omega$ in \disp{Omega}. 

    With these approximations $\hatchi^{(I)}(\vq,\omega),\hatchi^{(II)}(\vq,\omega)$, the 2-d dielectric function can be written in the form
\beq
\varepsilon^{(I,II)}(\vq,\omega)& =& 1 + \frac{2 \pi q_e^2}{|\vq|a^2_0 N_s q_e^2 \varepsilon_\infty} {\hatchi}^{(I,II)}(\vec{q}, \omega) \nn \\
&=&1+ \frac{g_c}{|\vq|a_0} \left( \frac{t}{q_e^2 N_s} {\hatchi}^{(I,II)}(\vec{q}, \omega)\right), \label{app-dielectric}
\eeq
where the dimensionless Coulomb constant is defined by
\beq
g_c= \frac{2 \pi q_e^2}{\varepsilon_\infty \, a_0 t }.  \label{gc}
\eeq
With the 2-d lattice constant  $a_0=3.81\AA$, $t=0.45 eV$ and $\varepsilon_\infty=1.76$, we get $g_c\sim 30.0$. For the material BSCCO used in \cite{MEELS-1,MEELS-2} the authors estimate dielectric constant $\varepsilon_\infty\sim 4.5$, giving $g_c \sim 11.5$, with the same $t$.
Since the basic parameter $t$  can vary somewhat depending on the theory, we present results for typical values $g_c=10,50,100$ in the following.



\subsection{Related  variables irreducible susceptibility
 $\Im \, \hatchi_{\rho\rho}$,  optical conductivity  $\Re \, \bar{\sigma}$,  current susceptibility  $\Im \,\hatchi_{JJ}$ }

 We next record a useful relation between a triad of variables defined below, that follows from  conservation of charge. These variables are
    the \sr{dimensionless conductivity variable} $\Re \,\bar{\sigma}(\vq,\omega)$ is related to the physical (i.e. dimensional) conductivity through  (see \disp{cond-chiJJ-2d}) \beq \Re \, {\sigma}(\vq,\omega)= \frac{h}{ q_e^2 c_0} \Re \, \bar{\sigma}(\vq,\omega) \label{sigmabar}, \eeq
     where $c_0$ is the separation between two copper oxygen planes in the cuprates.
Detailed results from the ECFL theory on  the  resistivity,  optical conductivity and inelastic Raman cross sections   have been recently published by us in \cite{ECFL-2d,ECFL-2d-Mai-a,ECFL-2d-Mai-b, ECFL-Raman}, over a wide  set of parameters, but corresponding to the $\vq=0$ limit only. These are extended to finite $\vq$ here.     
Let us first  note the relationships between the three sets of variables $\Re \, \bar{\sigma}(\vq,\omega)$, $\Im \, \hatchi_{JJ}(\vq,\omega)$ and $\Im \, \hatchi_{\rho \rho }(\vq,\omega)$. Combining \disp{WtoJ,scr-6} we find
\beq
\mbox{For $|\vq| a_0 \ll 1 $,    } \Im \, \hatchi_{JJ}(\vq,\omega)= \frac{\omega^2}{|\vq|^2} \Im \,\hatchi_{\rho \rho}(\vq,\omega), \label{rel-J-rho}
\eeq
which is a form of the charge conservation law. Combining further with \disp{cond-chiJJ-2d})  we get the important relation valid in the regime $|\vq| a_0 \ll 1 $: 
\beq
  \Re \, \bar{\sigma}(\vq,\omega) = \frac{1}{\omega} \left( \frac{\Im \, \hatchi_{JJ}(\vq,\omega)}{q_e^2 N_s} \right)=  \frac{\omega}{|\vq|^2}  \left( \frac{\Im \, \hatchi_{\rho \rho}(\vq,\omega)}{q_e^2 N_s} \right). \label{rel-all}
\eeq
As mentioned above the electron diffraction experiments  reported in \cite{MEELS-1,MEELS-3,MEELS-3} measure $\Im \, \hatchi_{\rho \rho}(\vq,\omega)$ at essentially arbitrary  $\vq$. We 
 point out below that the other two variables in \disp{rel-all} are are also  measurable, at least if we make suitable assumptions regarding the approximate correlation between Raman scattering intensities and the current susceptibility     $\Im \, \hatchi_{JJ}(\vq,\omega)$, at sufficiently low $\vq$.
After accounting for   explicit $\vq$ dependent terms  arising from the  conservation laws, if the remaining  $\vq$ dependence is  assumed to be  mild, then \disp{rel-all} acts as a     constrain $\Im \, \hatchi_{\rho \rho}(\vq,\omega)$  for small non-zero $\vq$ as well. 
We  discuss this relation extensively below in Sec.(\ref{Sec6.7})  with regard to the theoretical calculations, and comment about  the $\vq$ dependent peaks in $\omega$  of this triad of variables.

\subsection{Characteristic frequency scale $\Omega_p(\vq)$ revisited \label{sec-4.2.3}} 
    This turn-around occurs at the peak frequency $\Omega_p(\vq)$ \sr{defined} in \disp{Omega-p}. The magnitude of the turn-around frequency $\Omega_p(\vq)$, typically a small fraction of $t$ can, depending upon the choice of the hopping parameter $t$, be very small. We can estimate this further as follows. Using \disp{Omega,Omega-p} together with the expression for the first moment
$\widetilde{\omega}^{(1)}(\vq)$ in \disp{moment-2,mu-dispersion,low-q-w1} we express $\Omega_p(\vq)$ explicitly as a function of $\vq$. At small $\vq$ this simplifies further to
\beq
\lim_{\vq \to 0}\Omega_p(\vq) = |\vq| \sqrt{\frac{{\cal T}}{q_e^2} \frac{d\mu}{d n}}, \label{emergentOmega}
\eeq 
where the velocity $\sqrt{\frac{{\cal T}}{q_e^2} \frac{d\mu}{d n}}$ is determined by the ratio of ${\cal T}$ \disp{Big-Tau} that shrinks as the density $n\to 1$, and the compressibility \figdisp{fig-compressibility}. We comment further on this turn-around in Sec.~(\ref{Sec6.7}). 

    Given the interesting role played by this energy scale $\Omega_p(\vq)$, a natural question is whether it has a more direct origin and interpretation. For this purpose we construct a positive definite spectral-shape function $\varphi(\vq,\omega)$ from the complex susceptibility $\hatchi_{\rho \rho}(\vq,\omega)$ as
\beq
\varphi(\vq,\omega) = \frac{1}{ \hatchi_{\rho \rho}(\vq,0)} \left[ \frac{ \Im \, \hatchi_{\rho \rho}(\vq,\omega)}{\pi \omega} \right] \label{emergentPhi}.
\eeq
Using a dispersion relation for $\hatchi_{\rho \rho}(\vq,\omega)$ \disp{dispersion-relation-11}, we verify the normalization condition
\beq
\int_{-\infty}^{\infty} \, d\omega \, \varphi(\vq,\omega) =1,
\eeq
and also the even-ness $\varphi(\vq,-\omega)= \varphi(\vq,\omega)$. The second frequency moment of this spectral-shape function is given by
\beq
\int_{-\infty}^{\infty} \, d\omega \,  \omega^2\,  \varphi(\vq,\omega) =\Omega^2_p(\vq), \label{emergentOmega-2}
\eeq
where we used \disp{moment-expansion-2,dispersion-relation-11,equality-1,gamma} to relate the result of the integration to the expression in \disp{Omega-p}. Thus $\Omega_p(\vq)$ provides a characterization of the dynamics of $\hatchi_{\rho\rho}(\vq,\omega)$. As noted above, our theory identifies this energy as the peak frequency, or equivalently the turn-around scale for $\Im \, \hatchi_{\rho\rho}(\vq,\omega)$ (see Sec.~(\ref{Sec6.7})).



    In experiments a reasonable estimate of $\Omega_p(\vq)$ might be obtained by
an integration over a {\em finite} frequency window in \disp{emergentOmega-2}, if $\varphi(\vq,\omega)$ falls off rapidly with $\omega$ \cite{FN3}. From \disp{emergentOmega,emergentOmega-2,emergentOmega-3}, we see that this energy scale results from a ratio of two diminishing scales, the bandwidth reduction and the compressibility reduction, both due of the Gutzwiller-Hubbard correlations.



\begin{figure*}[h]
\centering
    \subfigure[\;\;]  {\includegraphics[width=0.45\columnwidth]{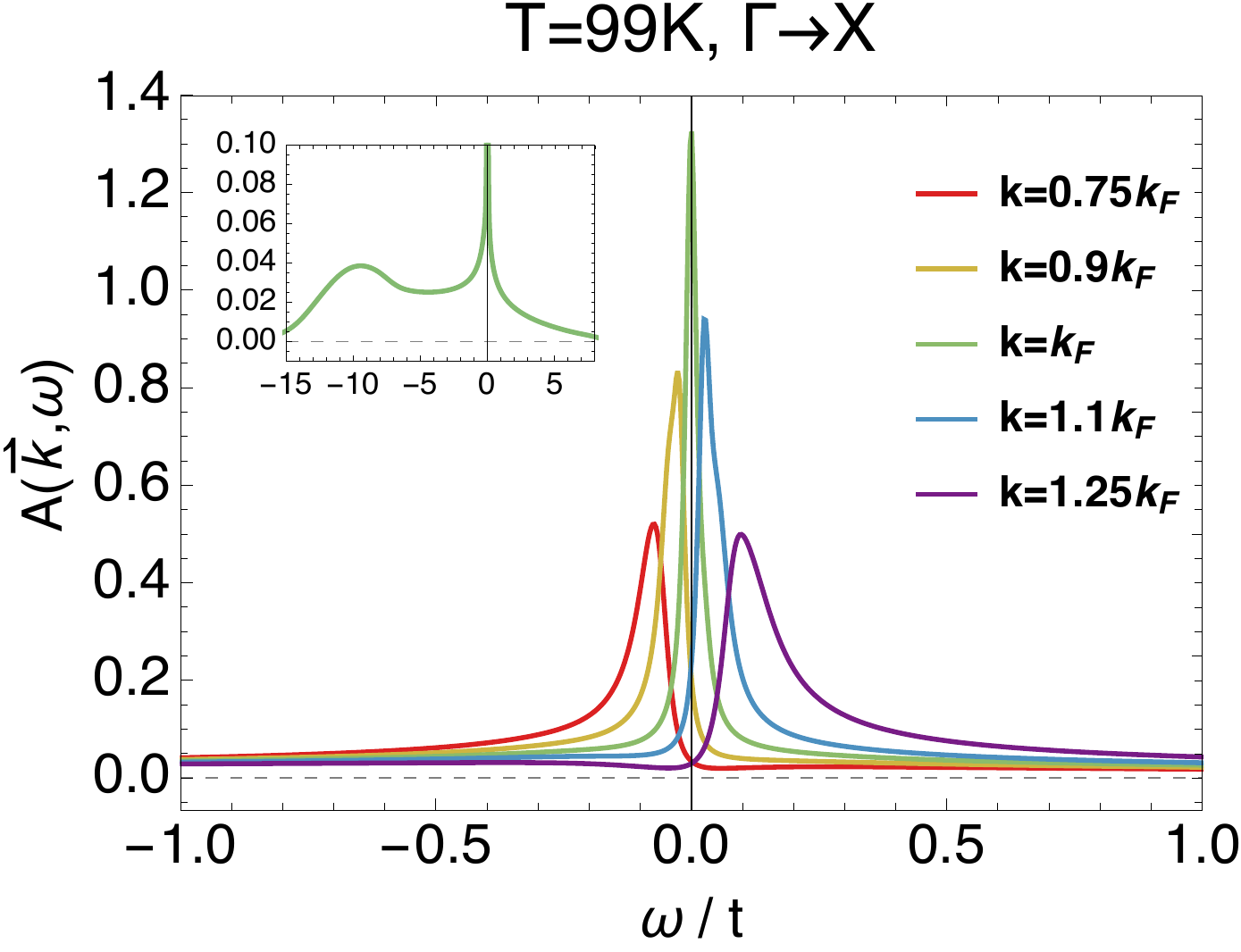}}
    \subfigure[\;\;]  {\includegraphics[width=0.45\columnwidth]{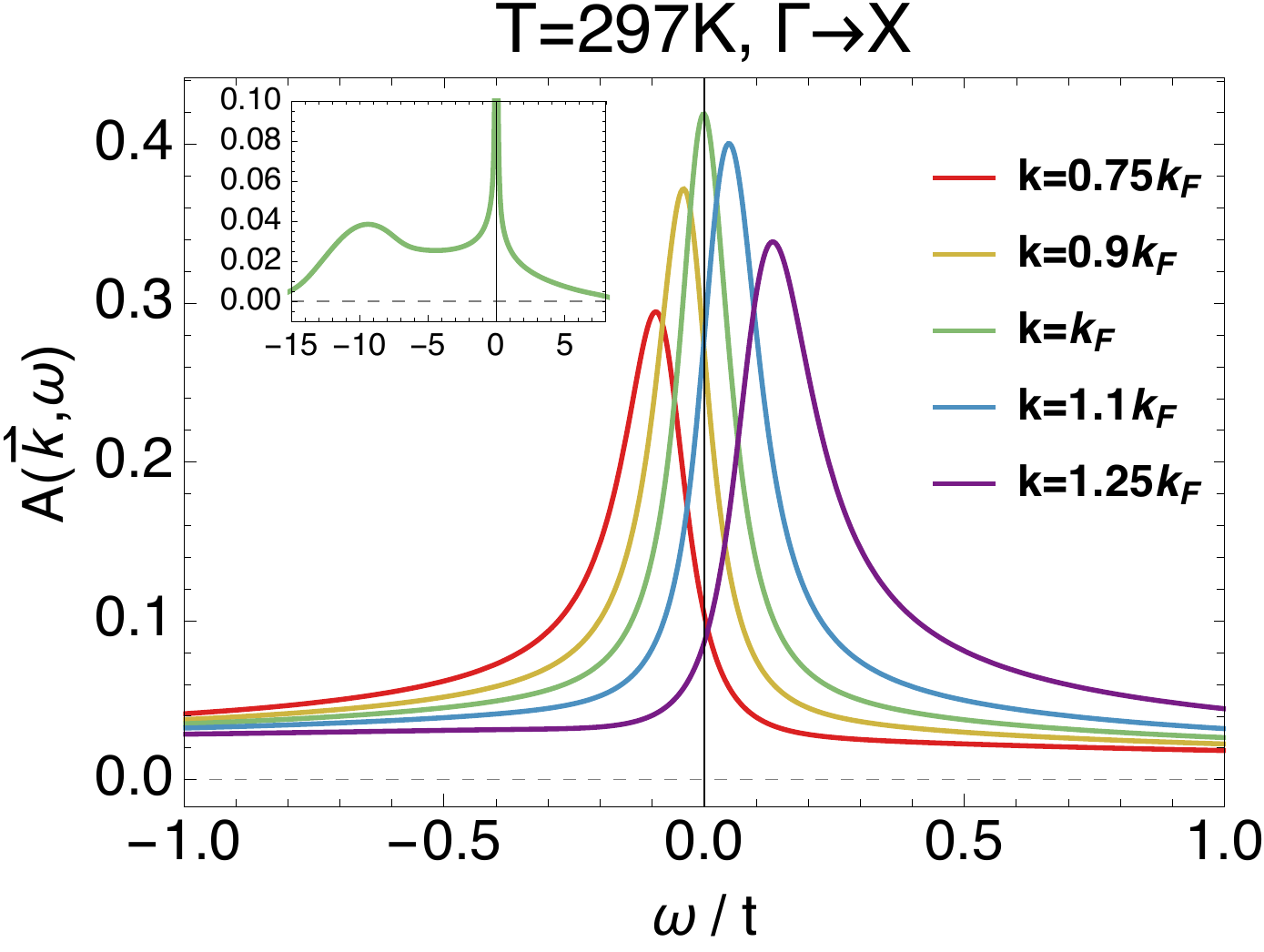}}
    \caption{\footnotesize The (single) electronic spectral functions for the ECFL Green's function  at two temperatures: (a) $T=99$K and (b) $T=297$K at $n=0.85$,   computed from  system sizes  $N_\omega=2^{14}, \; L_x\times L_y=64\times 64$. The insets  show the spectral function at $k_F$ against $\omega/t$,  over a  wide  energy scale. The Fermi wave vector is $k_F a_0=1.36$, and the quasiparticle weight at the Fermi wave vector $Z_{k_F}$ (abbreviated as $Z$) is very small compared to unity: $Z = 0.06, 0.09$ for $T=99$K and $T=297$K respectively. The reduced quasiparticle weight is also reflected in a small (Migdal) jump in the momentum distribution function \figdisp{occ-num}. 
The insets show that    the small area  under the quasiparticle peak at $\omega\sim0$, (due to a tiny  $Z$),  is  compensated by broad features at very high excitation energies $\sim10$t.   In evaluating the spectral functions, an implicit energy smearing of ${\cal O}(t/L_x)$ is implicit. Analogous figures for the spectral function at other densities and temperatures over an wider energy window for this theory can be found in \cite{ECFL-2d-Mai-a} (Figs.~(1,2))
   }
     \label{spectrum}
\end{figure*}

\begin{figure*}[h]
\begin{center}
    \subfigure[\;\;] {\includegraphics[width=0.5\columnwidth]{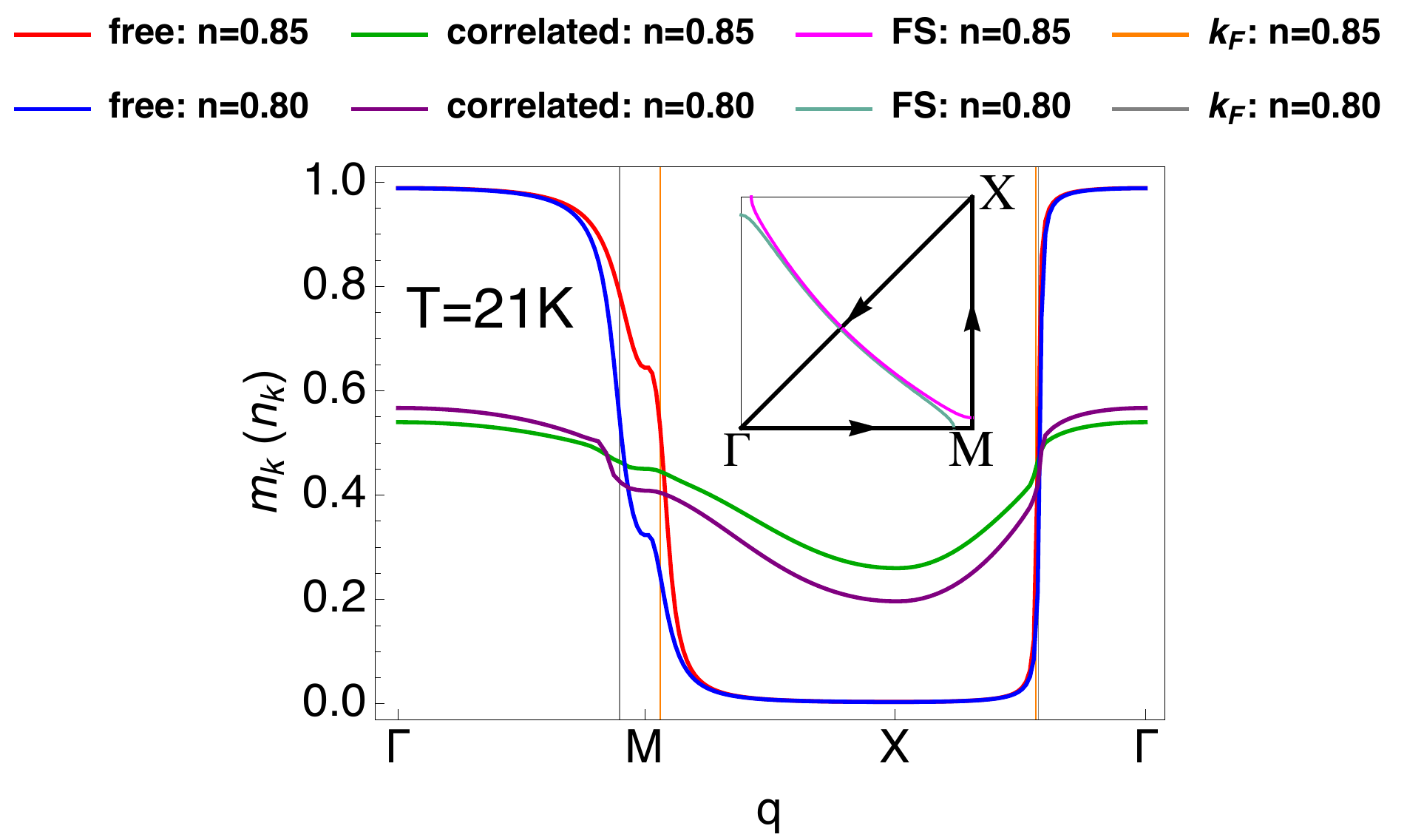}}
    \caption{\footnotesize The momentum distribution function $m_k$ for correlated electrons $m_k = \langle \wt{C}^{\dagger}_{k \uparrow} \wt{C}_{k \uparrow}\rangle$ found from \disp{mk} in purple ($n=0.80$) and  green ($n=0.85$) over the Brillouin zone. For comparison the  
  analogous function $n_k$ for the uncorrelated Fermi gas in blue ($n=0.80$) and red ($n=0.85$). The Fermi momentum is indicated by the vertical dashed lines in red ($n=0.85$) and grey ($n=0.80$). The inset shows the location of the noninteracting Fermi surface for the two densities. The system size used in the computation is $N_\omega=2^{14}, \; L_x\times L_y=64\times 64$. Here  we used $t=0.45,  J=0.17$ eV, $t'=-0.2 t$ and $T=21$K.   The theory satisfies the Luttinger-Ward theorem and hence the Fermi surface (FS) is unshifted by interactions. The wave vector $q$ traverses the octant of the Brillouin Zone, with corners $\Gamma = (0, 0)$, M=$(\pi,0)$, and $X = (\pi,\pi)$ and the green lines locate the non-interacting Fermi surface.  We note that the Fermi surface crossing of the interacting theory is missing in the $\Gamma \to M$ direction, it is roughly visible in the $M\to X$ direction and most clearly seen in the $X\to \Gamma$ direction.
       A sharp reduction of the quasiparticle weight $Z_{k_F}$, which equals the disconinuity in $m_{k_F}$ at $T$$=0$ by Migdal's theorem, is evident from the flattening of the correlated distribution $m_k$ in this figure.
   } \label{occ-num}
   \end{center}
\end{figure*}


   
\begin{figure}[h]
\begin{center}
   \includegraphics[width=0.36\columnwidth]{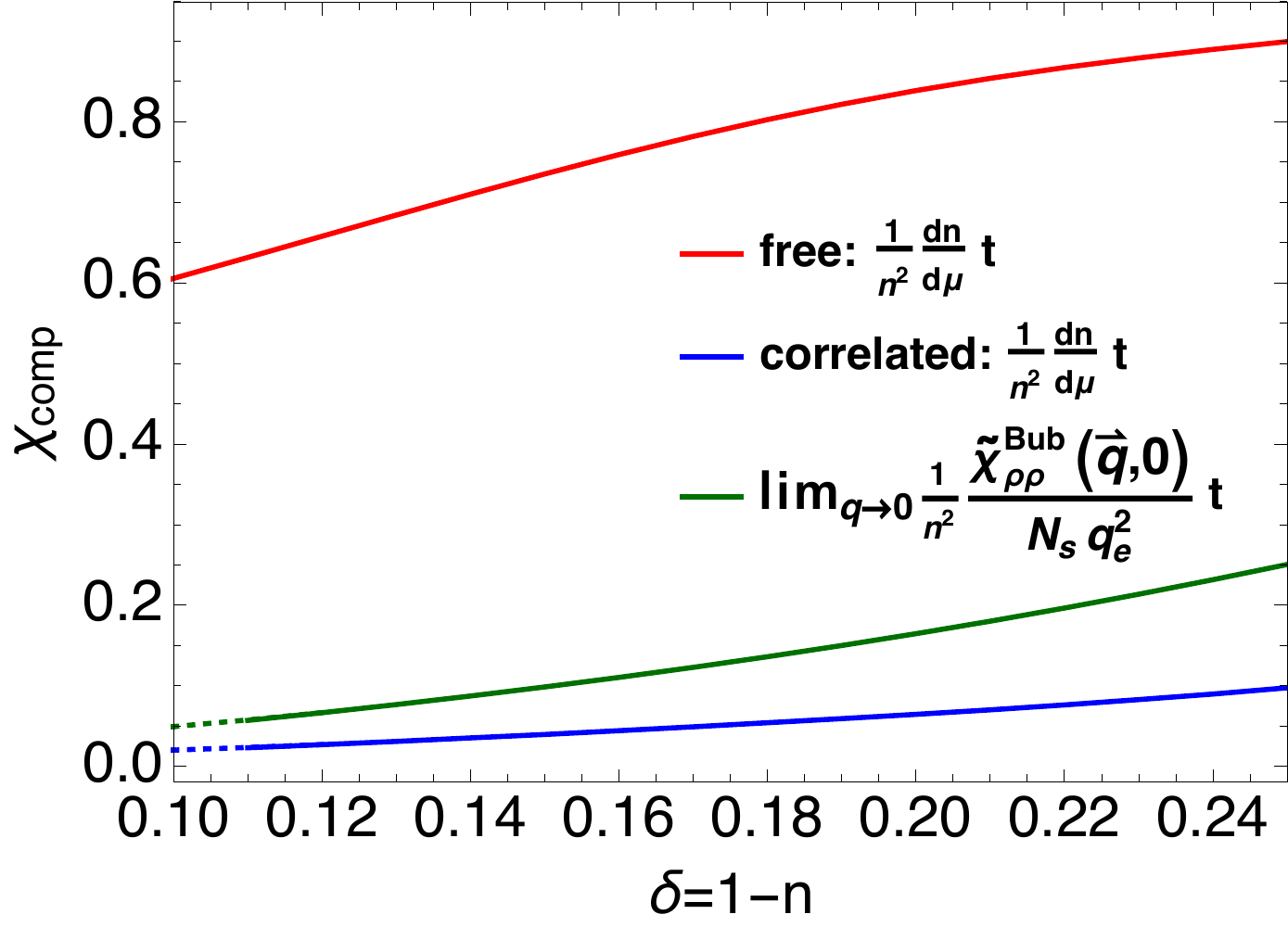}
    \caption{ \footnotesize The compressibility \disp{compressibility-2} at  $T=297$K versus doping $\delta = 1 - n$, where blue curve is the correlated case and red curve is the uncorrelated case.  In the correlated case $\frac{dn}{d \mu}$ is found numerically from the computed $\mu(n)$ for $\delta \geq 0.15$.  Correlations are seen to suppress the compressibility as $\delta$ decreases towards the insulating limit. The green curve is calculated numerically from the static uniform limit of the susceptibility $\frac{1}{q_e^2 N_s}\lim_{\vec{q}\to0}\hatchi^{\mbox{\tiny Bub}}_{\rho\rho}(\vec{q},0)$ (\disp{Inter-1}). If an exact calculation, going beyond the bubble approximation was possible, the corresponding green and blue curves would coincide.    } 
    \label{fig-compressibility}
    \end{center}
\end{figure}

\begin{figure*}[h]
\centering
    \subfigure[\;\;]  {\includegraphics[width=0.36\columnwidth]{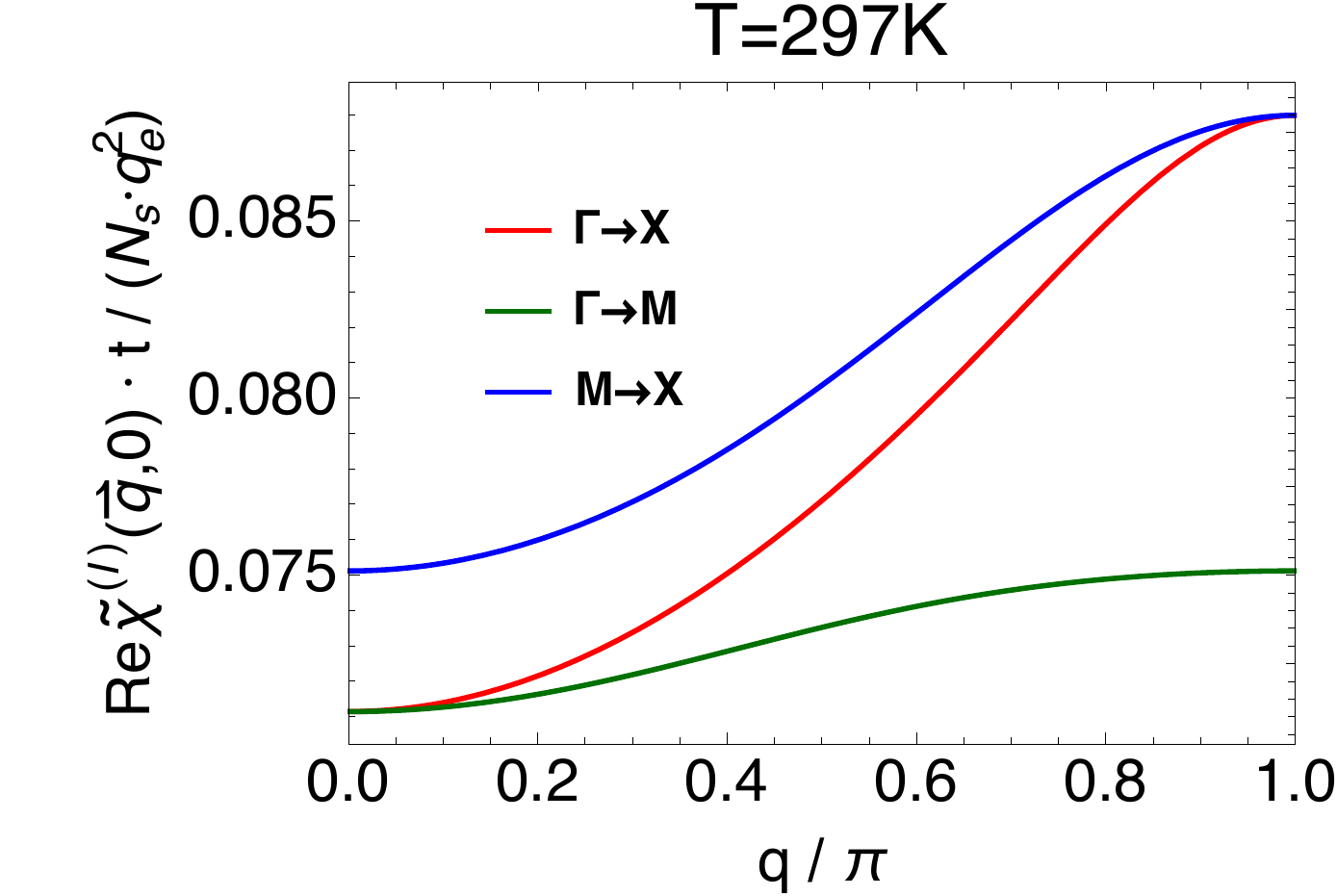}}
    \subfigure[\;\;]  {\includegraphics[width=0.36\columnwidth]{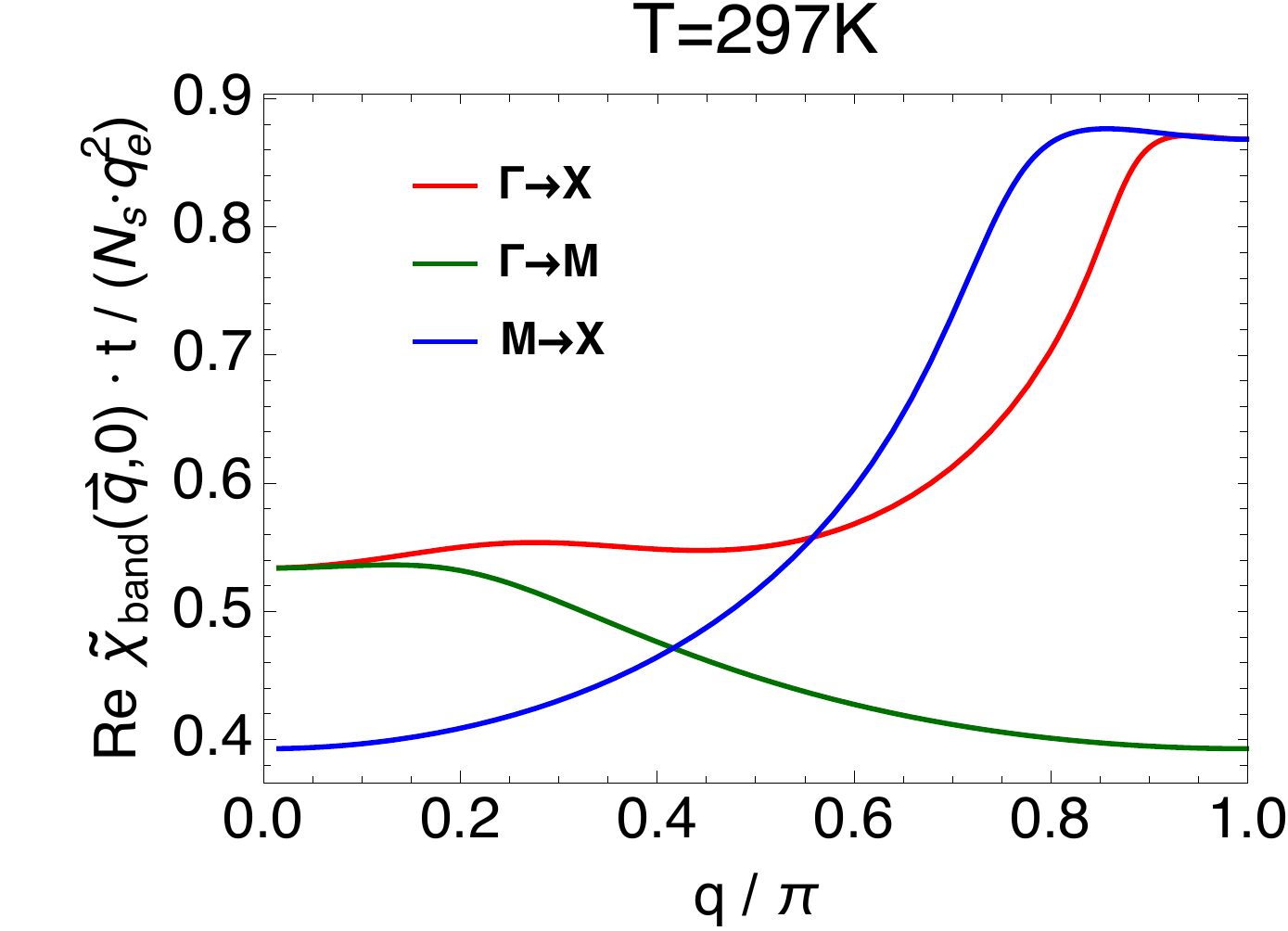}}
    \caption{ \footnotesize (a) The wave vector dependence of (a) the ECFL static susceptibility $\hatchi^{(I)}(\vec{q},0)$ (\disp{appxI}), for different paths in the BZ and (b) the (noninteracting) band structure case (\disp{RPA-1}). The density $n=0.85$ and $q$ is the relevant component of $\vq$ connecting the (high symmetry) points $\Gamma=(0,0), X= (\pi,\pi),M=(\pi,0)$ in the 2-d square lattice BZ. Correlations are seen to suppress the magnitudes of the susceptibilities. The relative locations of the three curves for the correlated system undergoes a surprising reshuffle relative to the band susceptibilities. } 
    \label{static}
\end{figure*}

    \begin{table*}[t]
\begin{center}
\begin{tabular}{|p{0.5in}| p {1. in} |p{ 1. in} |p{1.  in} |}\hline 
\;\;\;\;\;n& &Uncorrelated& Correlated\\ \hline
\;\;\;0.80&$\langle \cos k_x \rangle_{ave}$& 0.188847 & 0.056881 \\
&$\langle \cos k_x \cos k_y \rangle_{ave}$&0.032757 &0.00661296\\ \hline
\;\;\;0.85&$\langle \cos k_x \rangle_{ave}$& 0.190954 &0.0400778 \\
&$\langle \cos k_x \cos k_y \rangle_{ave}$&0.018181&-0.0079378\\ \hline
\end{tabular} 
\caption{The averages used in \disp{mu-dispersion} to calculate $\kappa(\vq)$ in  \figdisp{sumrule-dispersion}. The flattened distribution function $m_k$ in \figdisp{occ-num} leads the much smaller values of these angular averages for the correlated metal. \label{Table1}}
\end{center}
\end{table*}


\begin{figure*}[h]
\includegraphics[width=0.36\columnwidth]{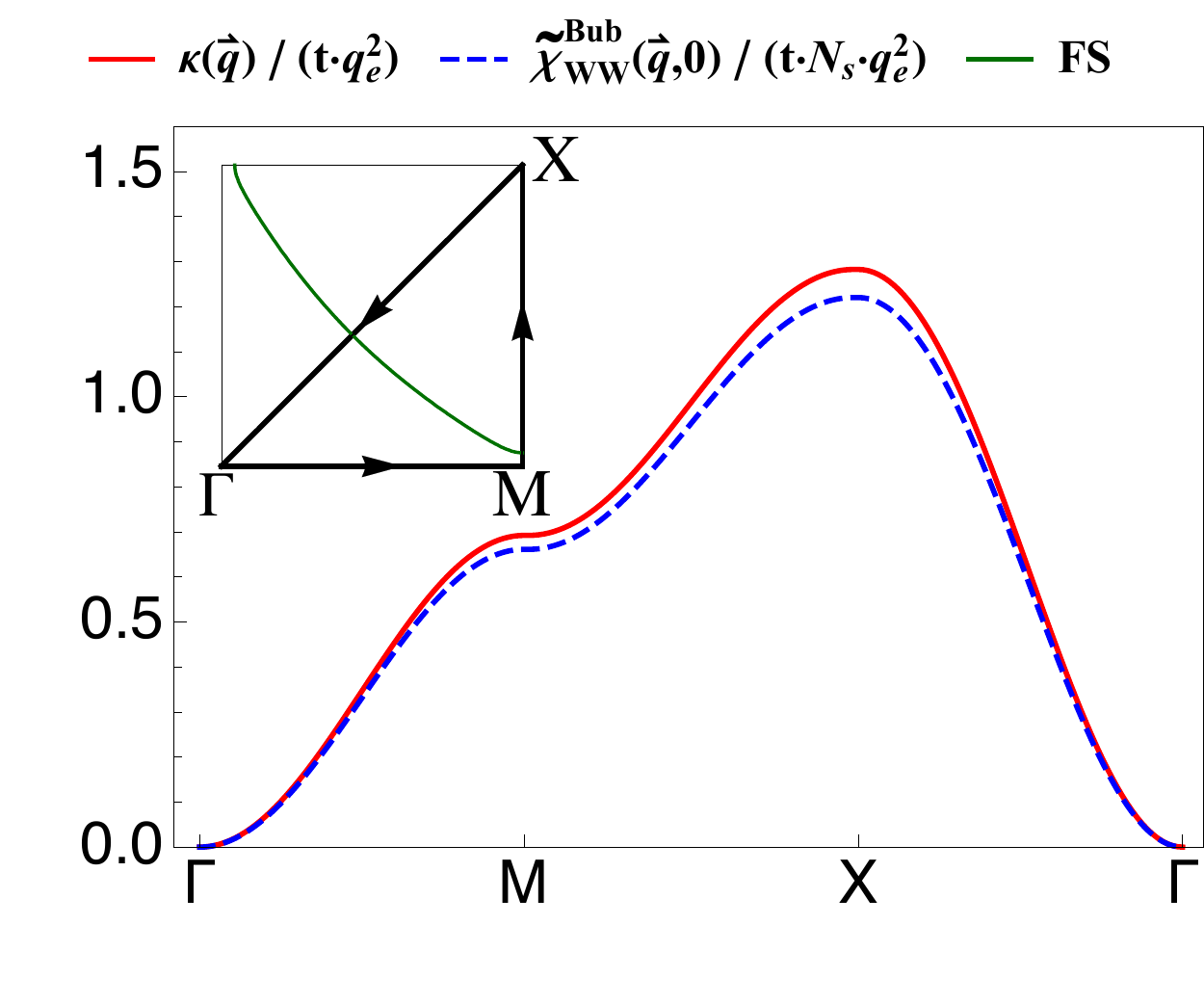}
\centering
\caption{\footnotesize The dimensionless functions $\frac{1}{t q_e^2}\kappa(\vec{q})$ from \disp{kappa} and $\frac{1}{t q_e^2 N_s} \hatchi^{\mbox{\tiny Bub}}_{WW}(\vec{q},0)$ from \disp{Inter-2} plotted over the Brillouin zone are approximately identical for a system at $n = 0.85$ and $T = 297$K. The curves are  coincident near $|\vq|\sim0$, but separate out  at higher $|\vq|$. In an ideal exact calculation (going beyond the bubble approximation), these two curves are expected to coincide identically  at all $|\vq|$. The mismatch is a measure of the error made in the bubble approximation employed  (using the correlated Greens functions).   }
 \label{chiBubWWstatic}
\end{figure*}

\begin{figure*}[h]
\centering
 \subfigure[\;\;] {\includegraphics[width=0.36\columnwidth]{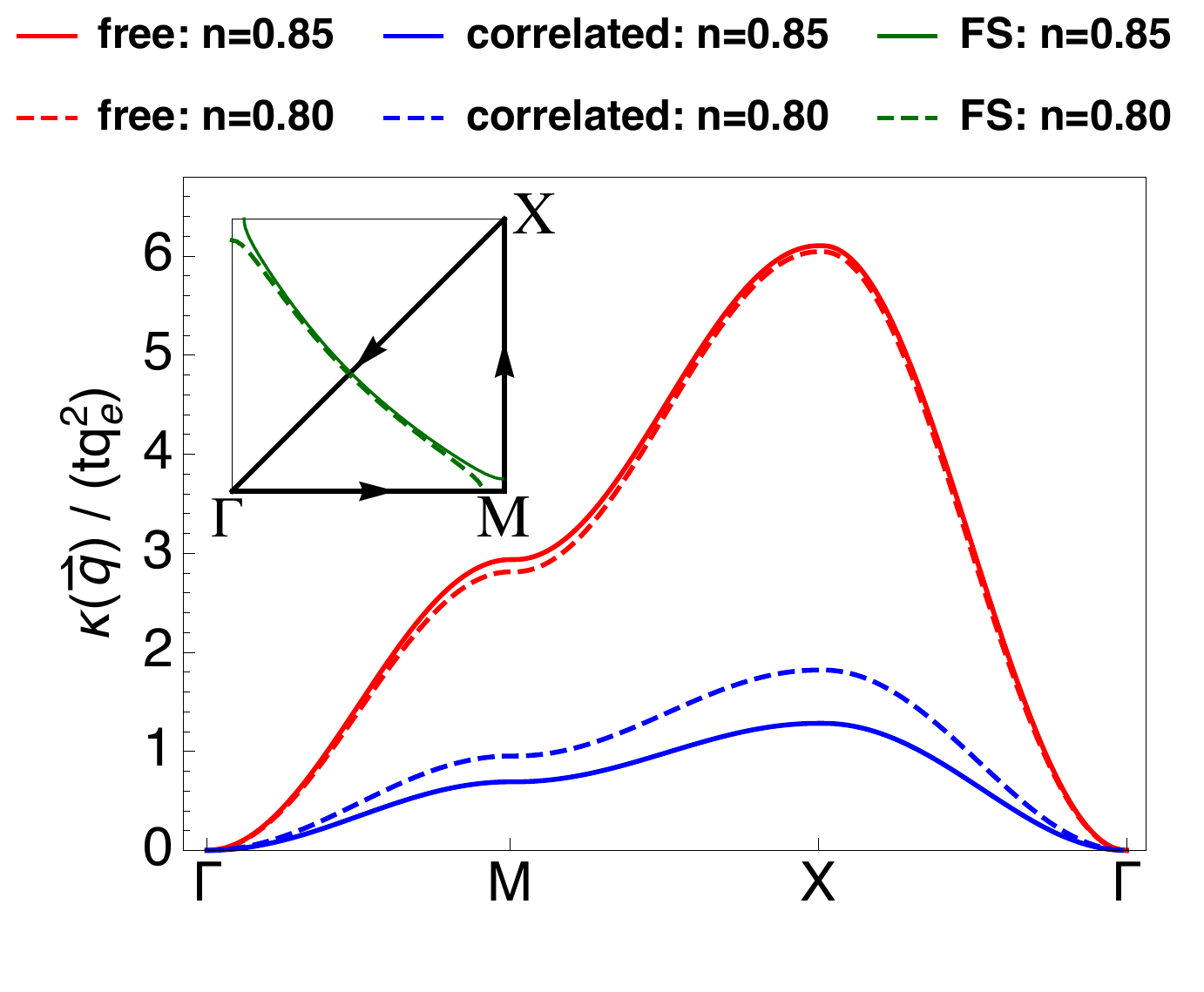}}
 \subfigure[\;\;] {\includegraphics[width=0.36\columnwidth]{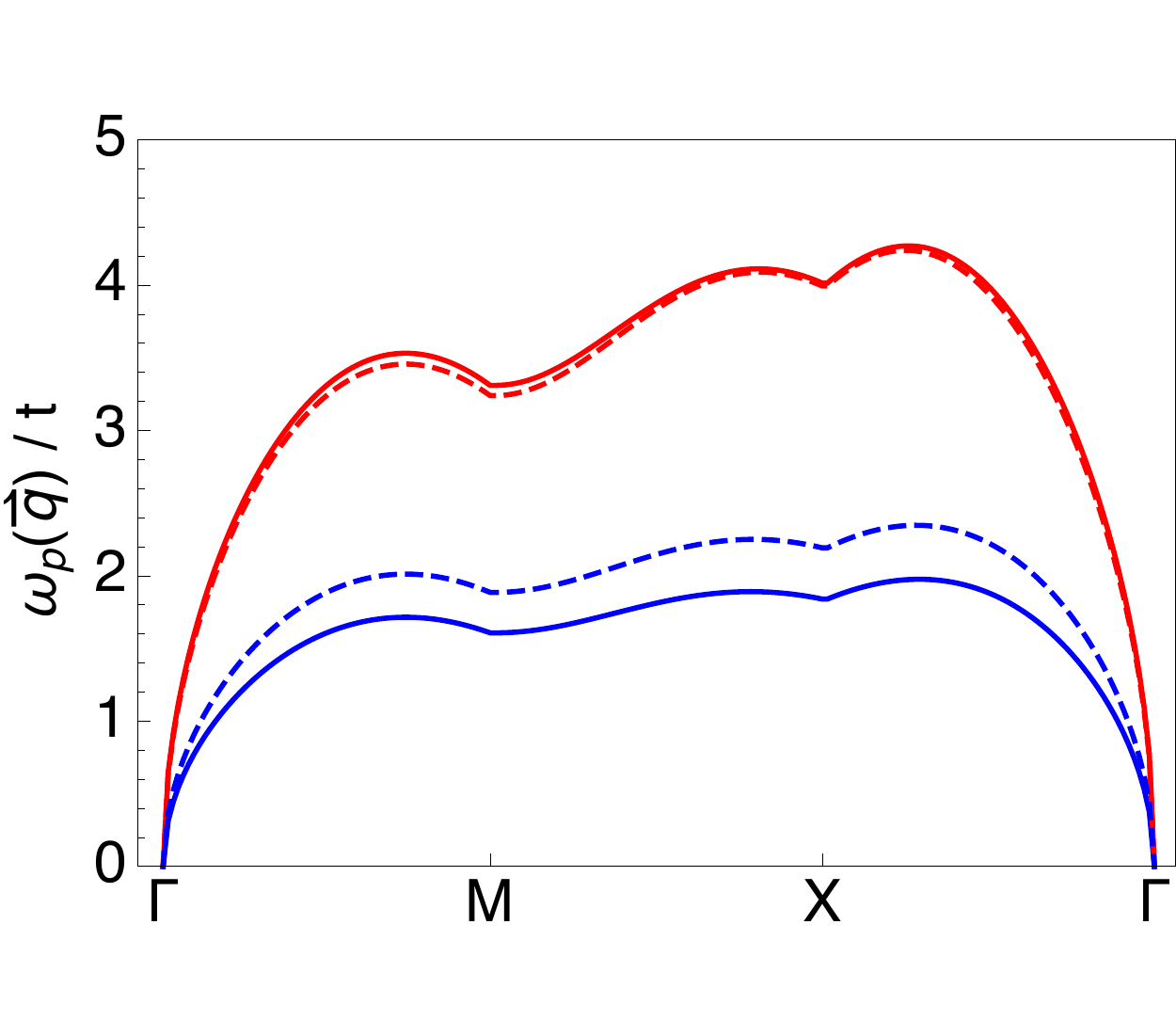}}
    \caption{\footnotesize ({\bf a}) The function $\frac{1}{t q_e^2}\kappa(\vq)$ (\disp{kappa}), or upon using \disp{moment-2},  the first moment $\widetilde{\omega}^{(1)}(\vq)/t$  over the BZ (indicated in the inset)  at $T=297$K. We used \disp{mu-dispersion}, at two densities $n=0.85$ (solid curves) and $n=0.8$ (dashed curves) for the uncorrelated (red) and correlated (blue) systems.     Recall from \disp{equality-1,moment-2}, that $\widetilde{\omega}^{(1)}(\vq)$ can in  be inferred in principle from experiments by e.g. using \disp{dispersion-relation-11,moment-expansion-2}. ({\bf b}) The plasmon dispersion $\omega_p(\vq)$ in 2-d from \disp{plasmons-4,3d-plasmon,Coulomb-2d} for the same parameters, and $\varepsilon_\infty=4.5$ (i.e. $g_c$$\sim$$11.5$), for the uncorrelated (red) and correlated (blue) systems.  In the latter the characteristic $\omega_p\propto |\vq|^\half$ behavior of 2-dimensional plasmons, is followed by a broad continuum at an energy  scale $\omega_p\sim 1.50$t, which is considerably lower than the energy scale without interactions.  }
     \label{sumrule-dispersion}
\end{figure*}

\begin{figure*}
\centering
\subfigure[\;\;]{\includegraphics[width=0.3\columnwidth]{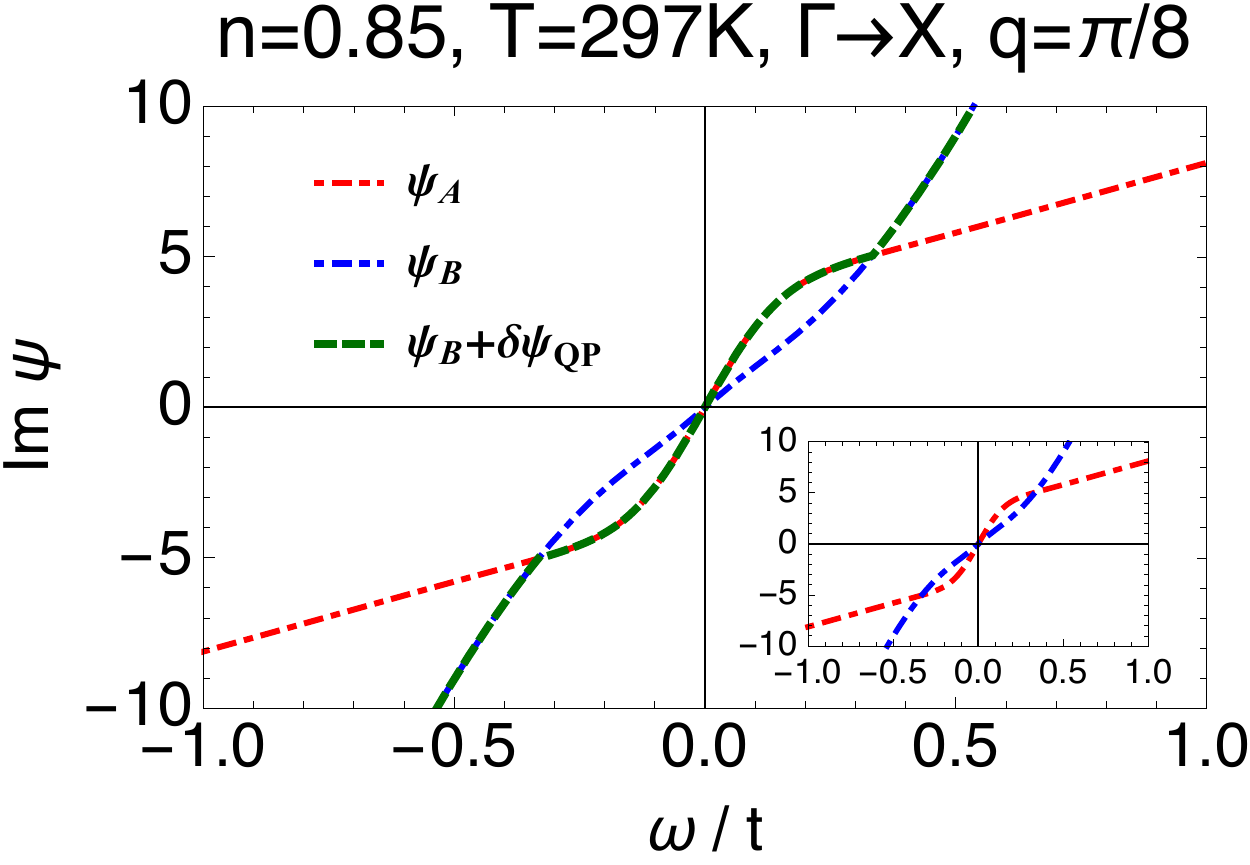}}
\subfigure[\;\;]{\includegraphics[width=0.3\columnwidth]{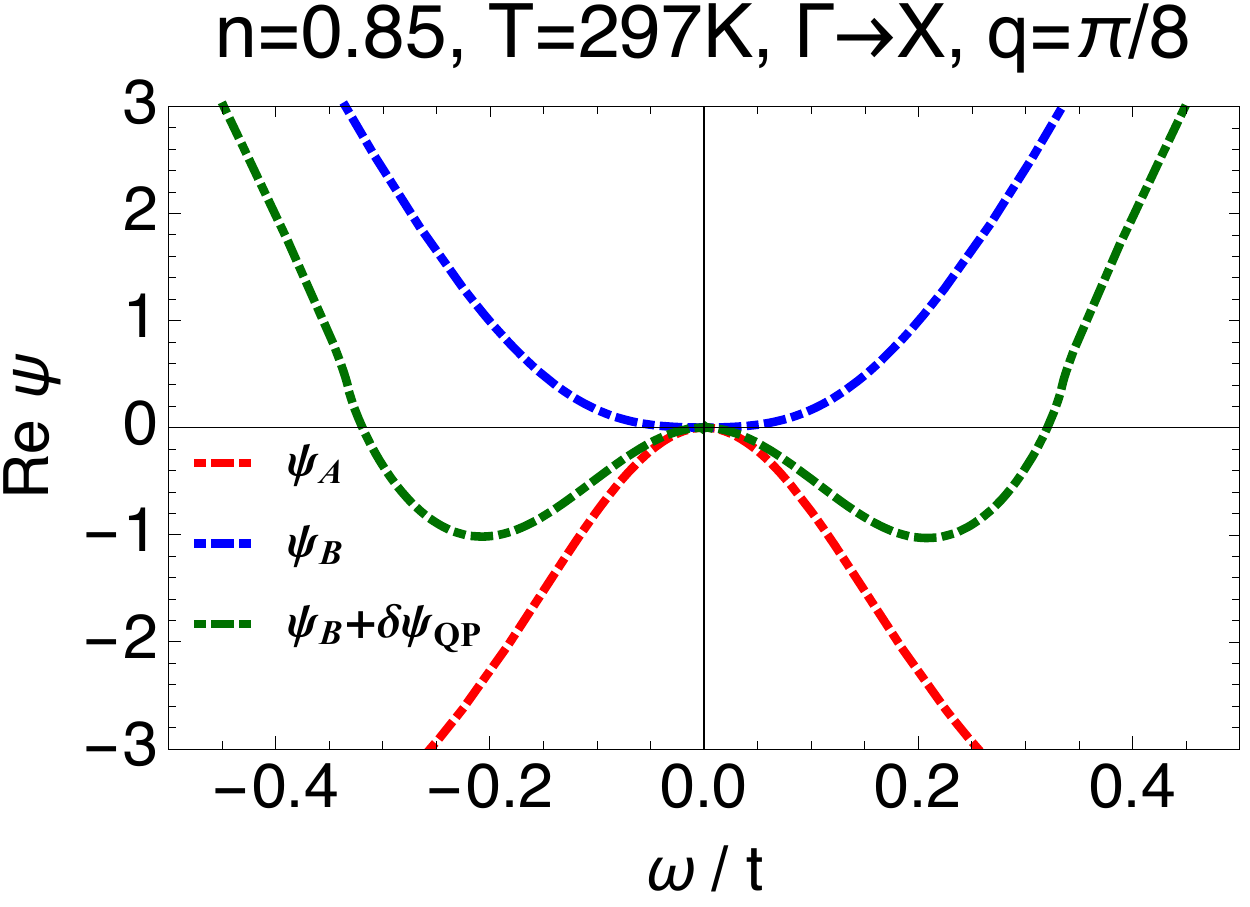}}
\subfigure[\;\;]{\includegraphics[width=0.3\columnwidth]{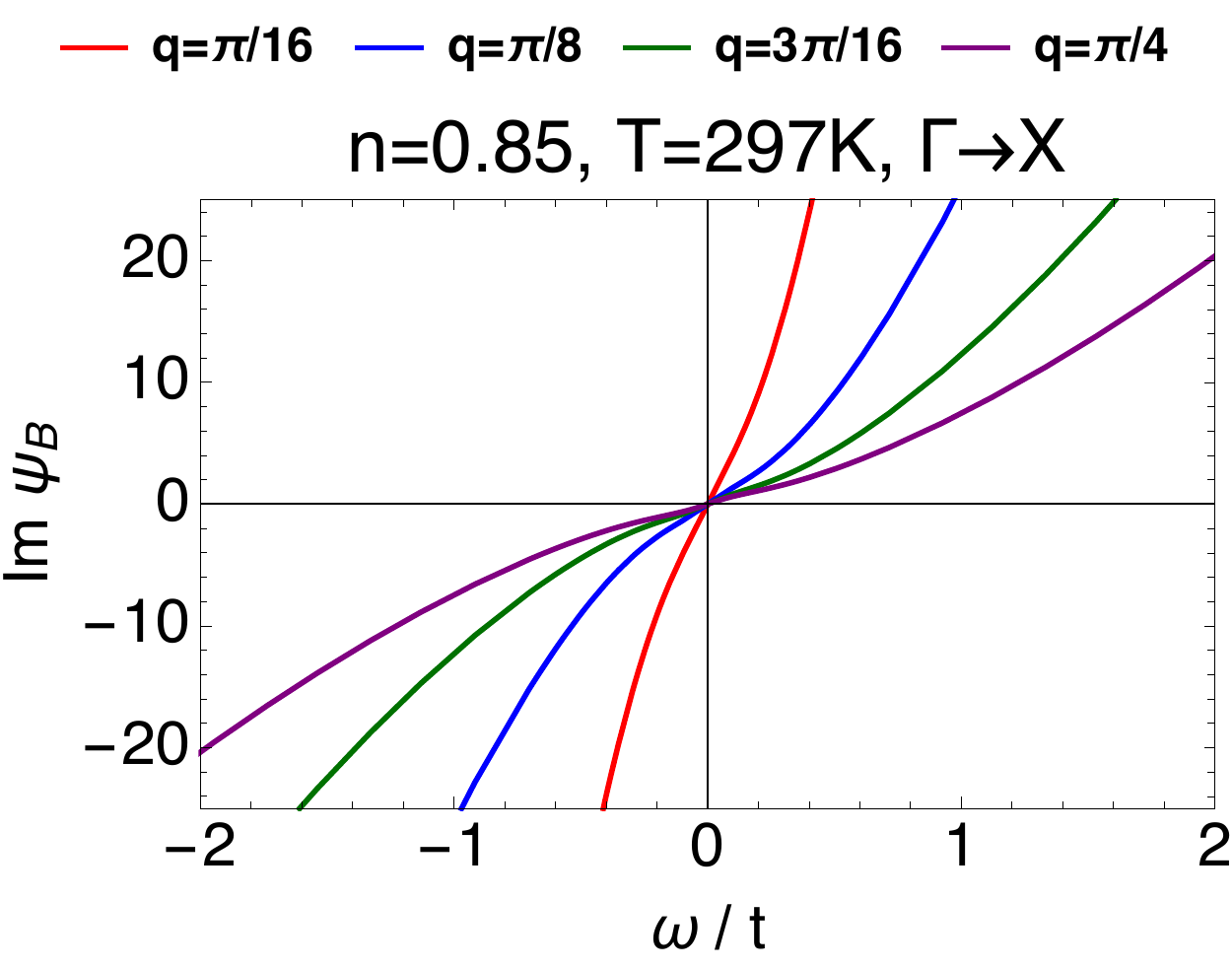}}
\subfigure[\;\;]{\includegraphics[width=0.3\columnwidth]{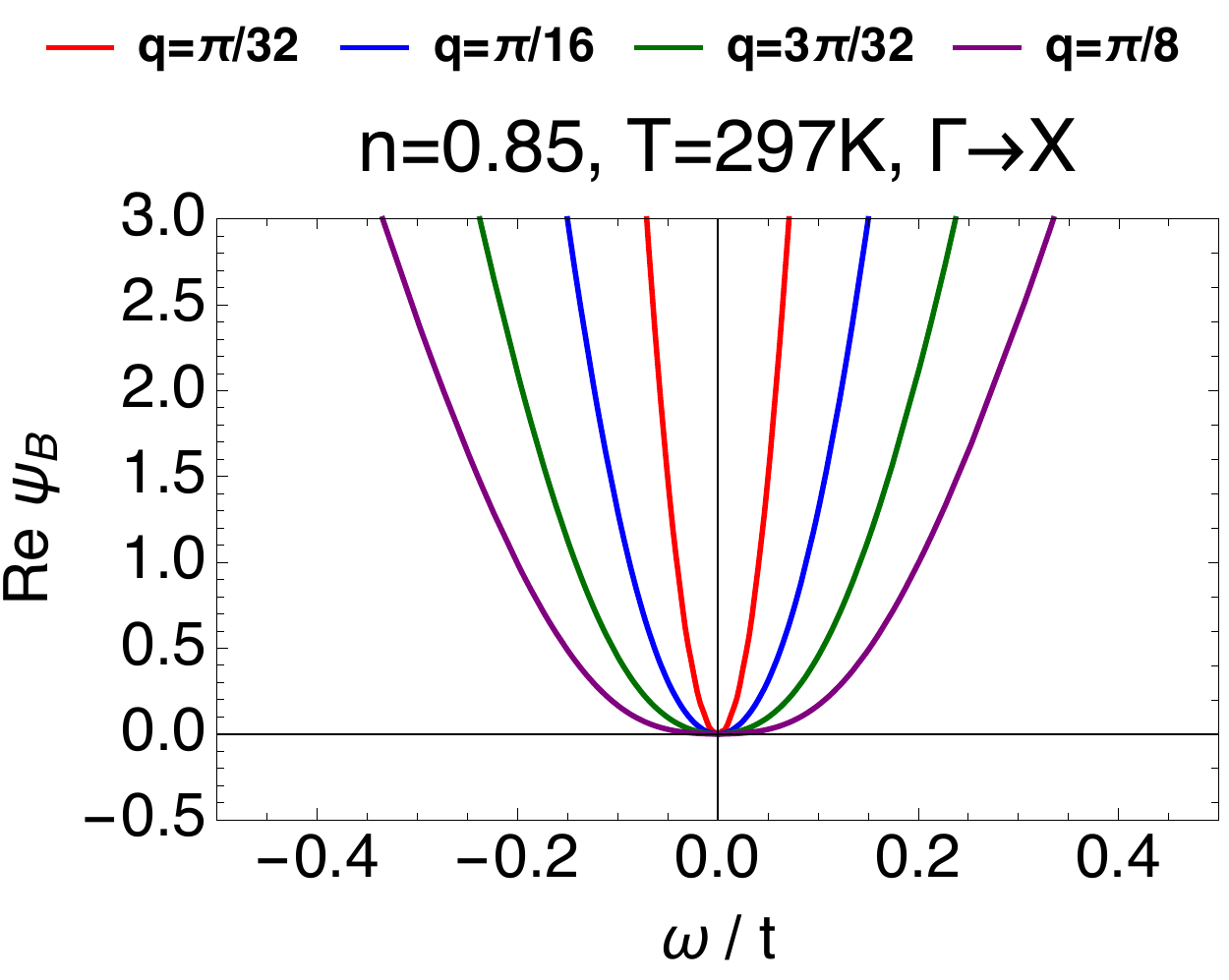}}
\subfigure[\;\;]{\includegraphics[width=0.3\columnwidth]{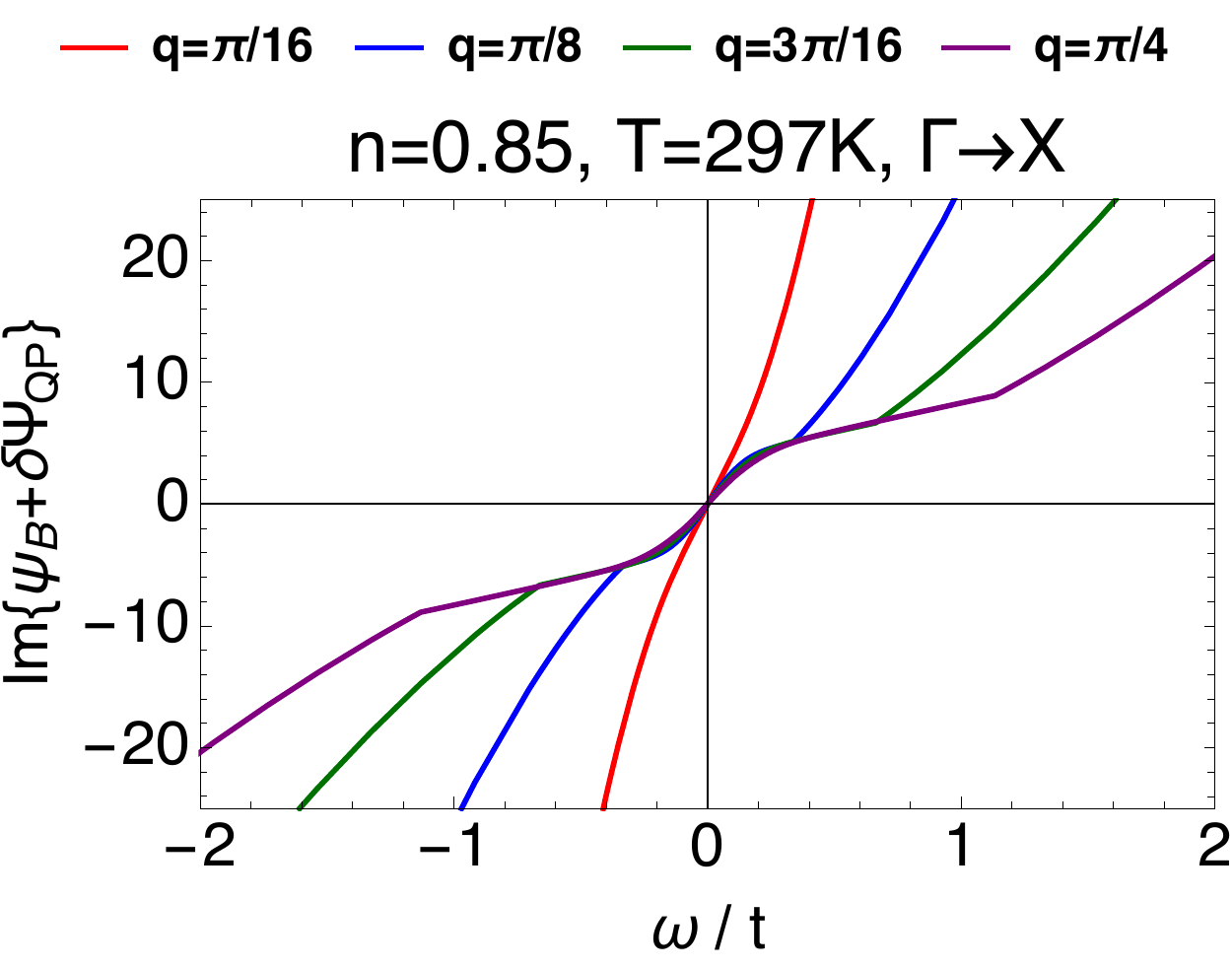}}
\subfigure[\;\;]{\includegraphics[width=0.3\columnwidth]{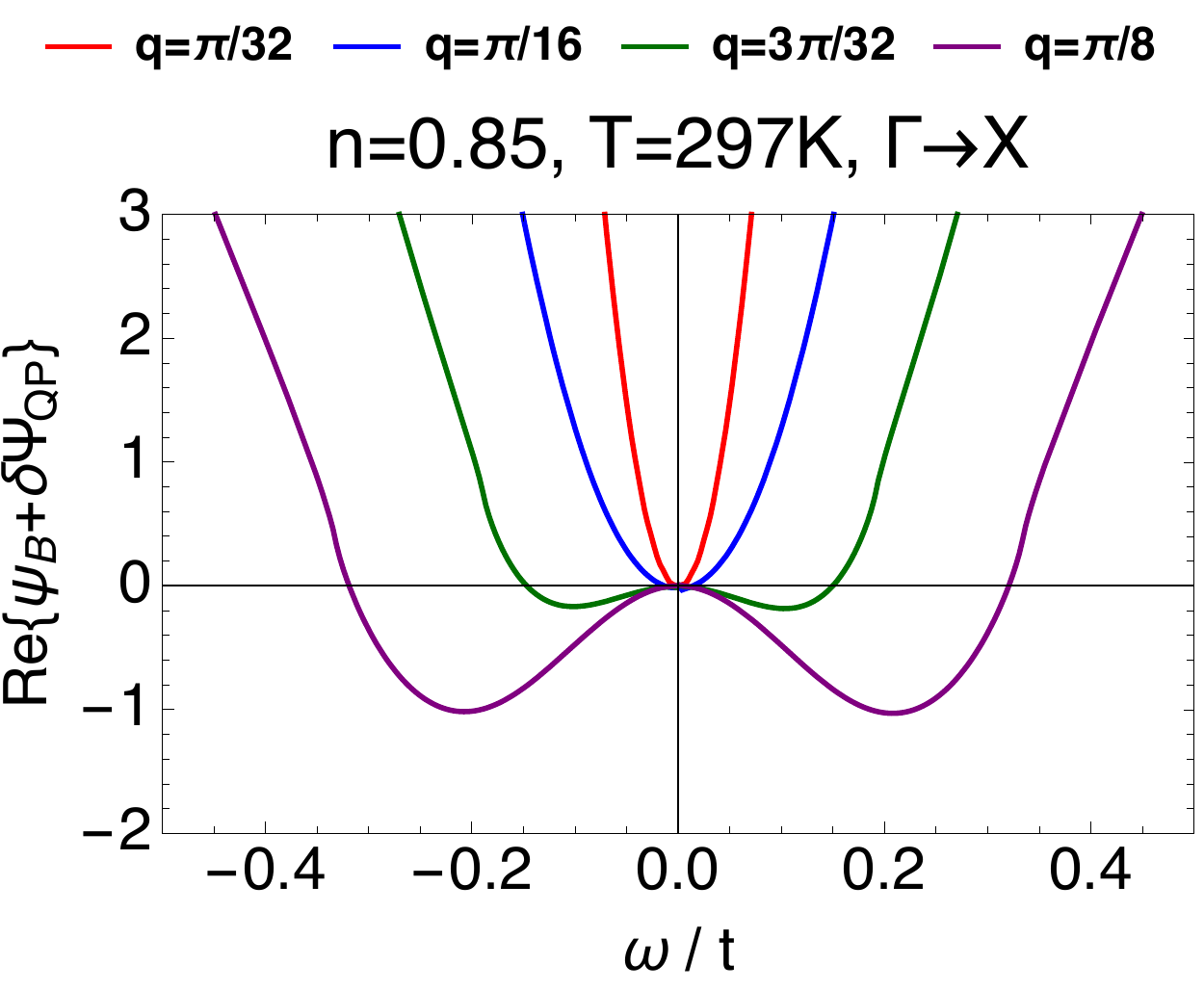}}
\caption{\footnotesize  The different panels illustrate the complex self energies $\Psi(\vq,\omega)$, relevant  for the two successive approximations to   $\hatchi^{(I,II)}$, at a typical density $n=0.85$ and  temperature $T=297$K. The  susceptibilities $\hatchi^{(I)}$ in \disp{appxI}    and $\hatchi^{(II)}$ in \disp{appxII} are constructed using  the self energies  $\Psi_B(\vq,\omega)$
 and      $\Psi_B(\vq,\omega)+ \delta \Psi_{QP}(\vq,\omega) $ respectively. 
 In  panel (a)   at $\vq=\{\pi/8,\pi/8\}$  we show $\Psi''_A$ (red-dotted) and $\Psi''_B$ (blue-dotted), as well as  the imaginary part of the third self energy $\Psi_B+ \delta \Psi_{QP}$ (green-dotted). For $\omega>0$  the latter  is obtained by taking the larger of $\Im \Psi_A$ and $\Im \Psi_B$, while for $\omega<0$ we use the oddness of $\Im \Psi$ to flip the curve.  The imaginary part of $\Psi_B(\vq,\omega)+ \delta \Psi_{QP}(\vq,\omega) $ captures the quasiparticle part contained in $\Im \Psi_A(\vq,\omega)$ at low $\omega$, but otherwise is the same as $\Im \Psi_B$. The  real parts are calculated using the  causality relation \disp{redeltaPsi}. The real parts of these three susceptibilities are  shown in panel (b).  Panels (c) and (d)
  show the real and imaginary parts of $\Psi_B$ at a few typical values of $\vec{q}$. Similar plots for $\Psi_B+\delta \Psi_{QP}$ are shown in panels (e) and (f). In comparing panels (c) and (e), we see the linear in $\omega$ regime near the origin due to the quasiparticle contribution, which in turn creates the double minimum in the real part seen in  panels (b) and (f).
}
\label{Psi}
\end{figure*}

\begin{figure*}
\centering
 \subfigure[\;\;]{\includegraphics[width=0.36\columnwidth]{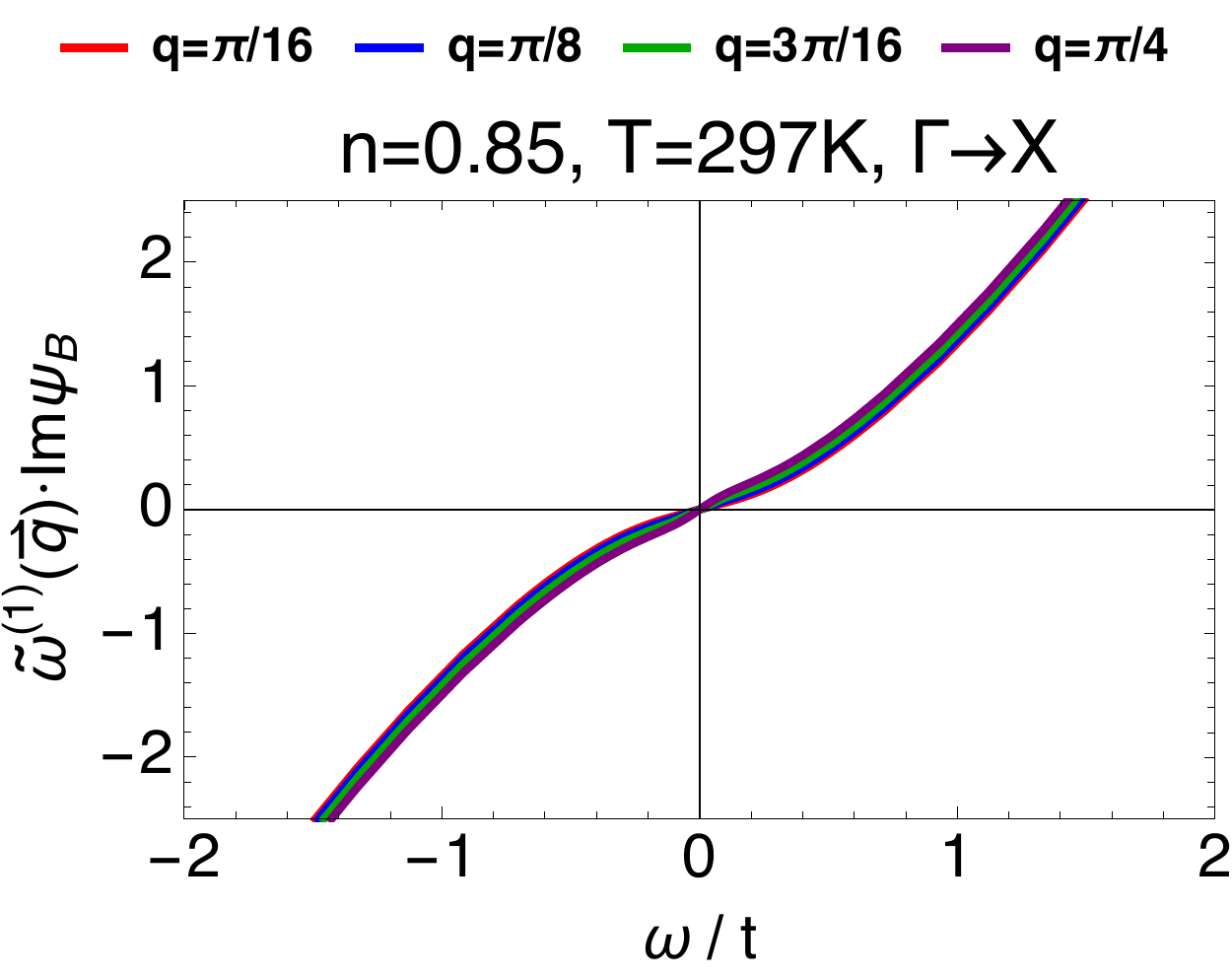}}
 \subfigure[\;\;]{\includegraphics[width=0.36\columnwidth]{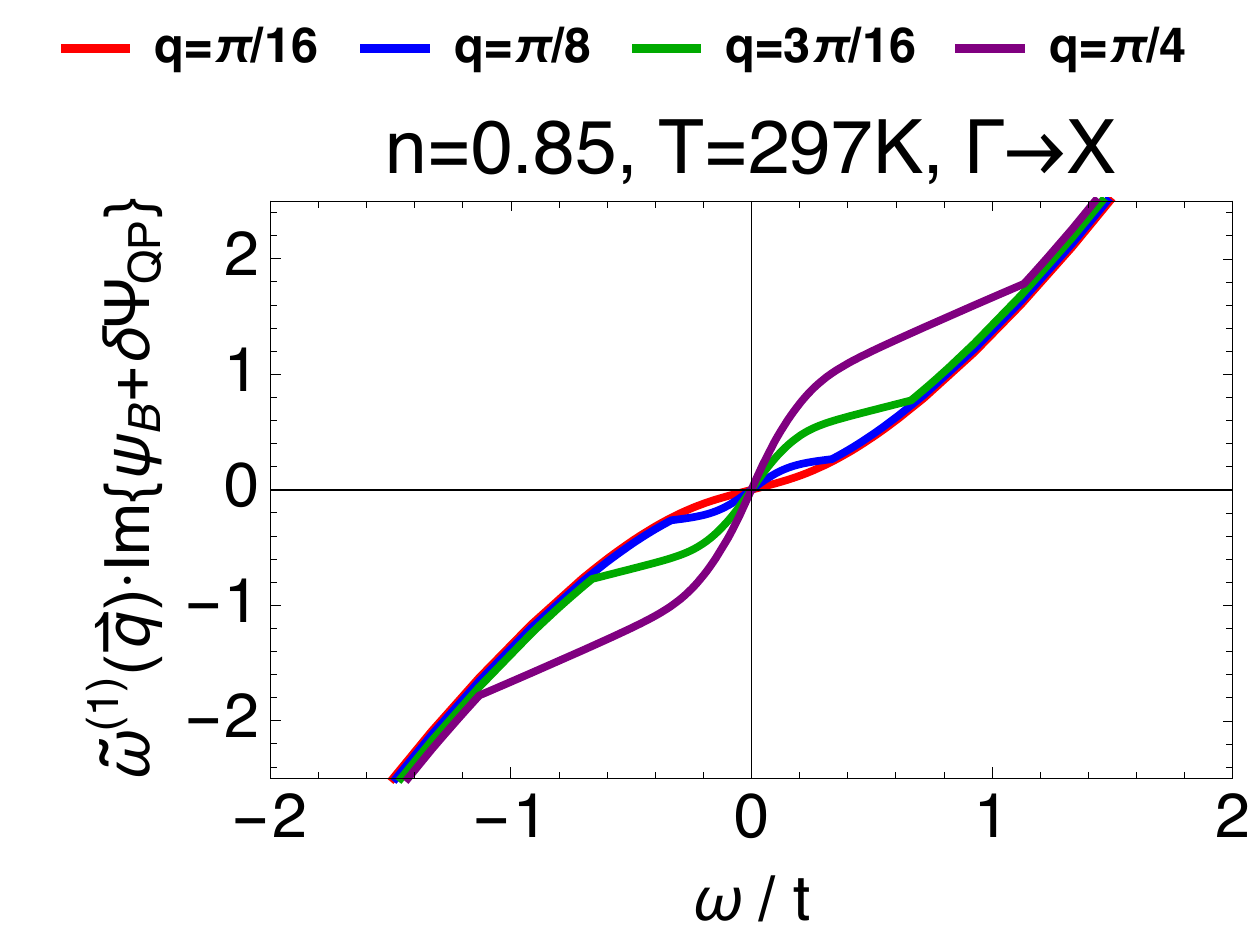}}
 \caption{\footnotesize ({\bf a}) The imaginary part of the self energy $\Psi_B(\vq,\omega)$  (\disp{PsiB} and in \figdisp{Psi}) relevant to $\hatchi^{(I)}$, at different values of $\vec{q}= (q,q)$ (in $\Gamma \to X$ direction) are seen to collapse to a single curve, when scaled by the first moment $\widetilde{\omega}^{(1)}(\vq)$ (\disp{moment-expansion-1} and in Fig.~(\ref{sumrule-dispersion}.(b)). ({\bf b}) For $\Psi_B(\vq,\omega) +\delta\Psi_{QP}(\vq,\omega)$ relevant to $\hatchi^{(II)}$, the imaginary part of this self-energy also  coincide, but only at high frequencies beyond the energy scale of the quasiparticle excitations. 
 }
 \label{Psi-scaling}
\end{figure*}

\begin{figure*}[h]
\centering
 \subfigure[\;\;]{\includegraphics[width=0.36\columnwidth]{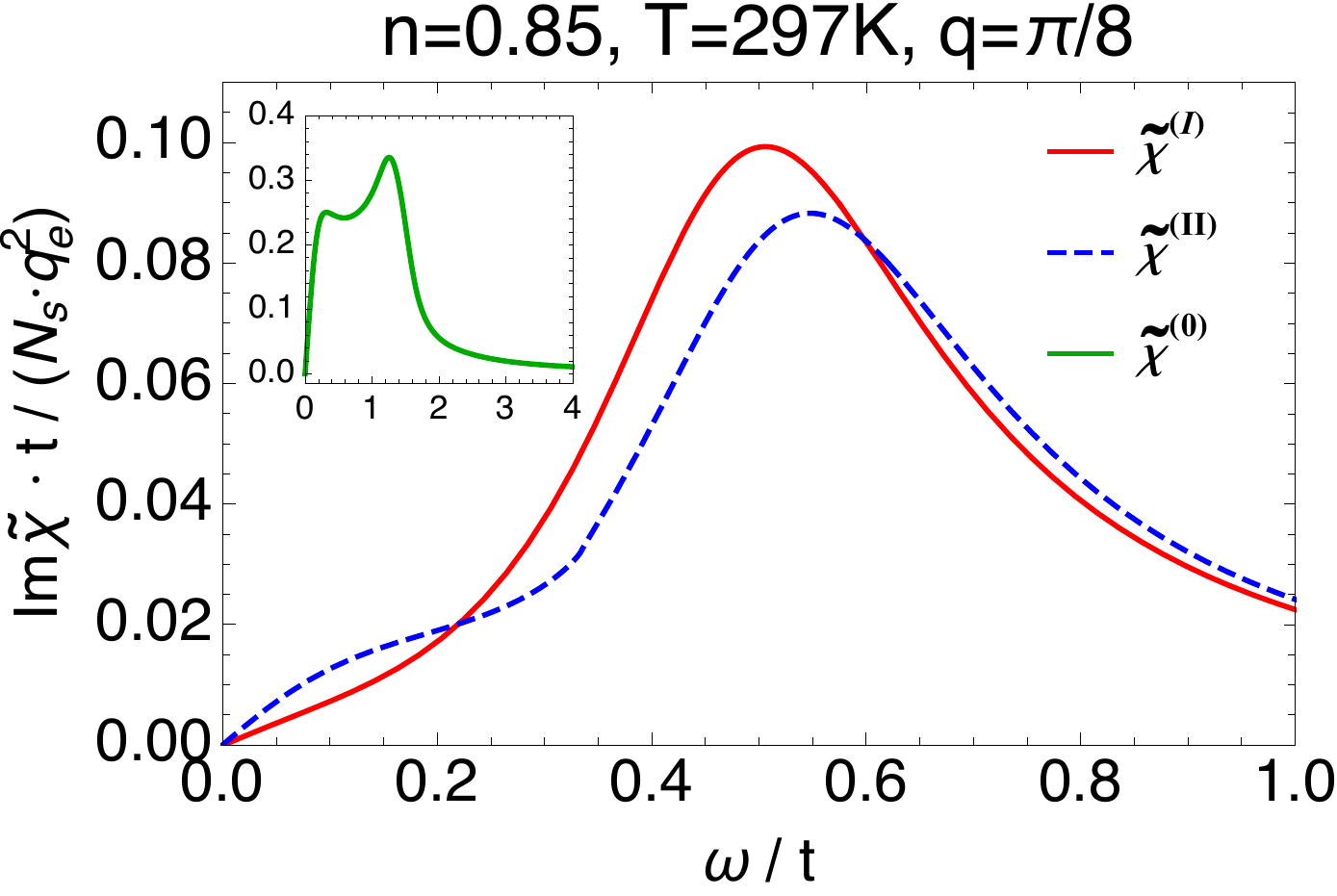}}
 \subfigure[\;\;]{\includegraphics[width=0.36\columnwidth]{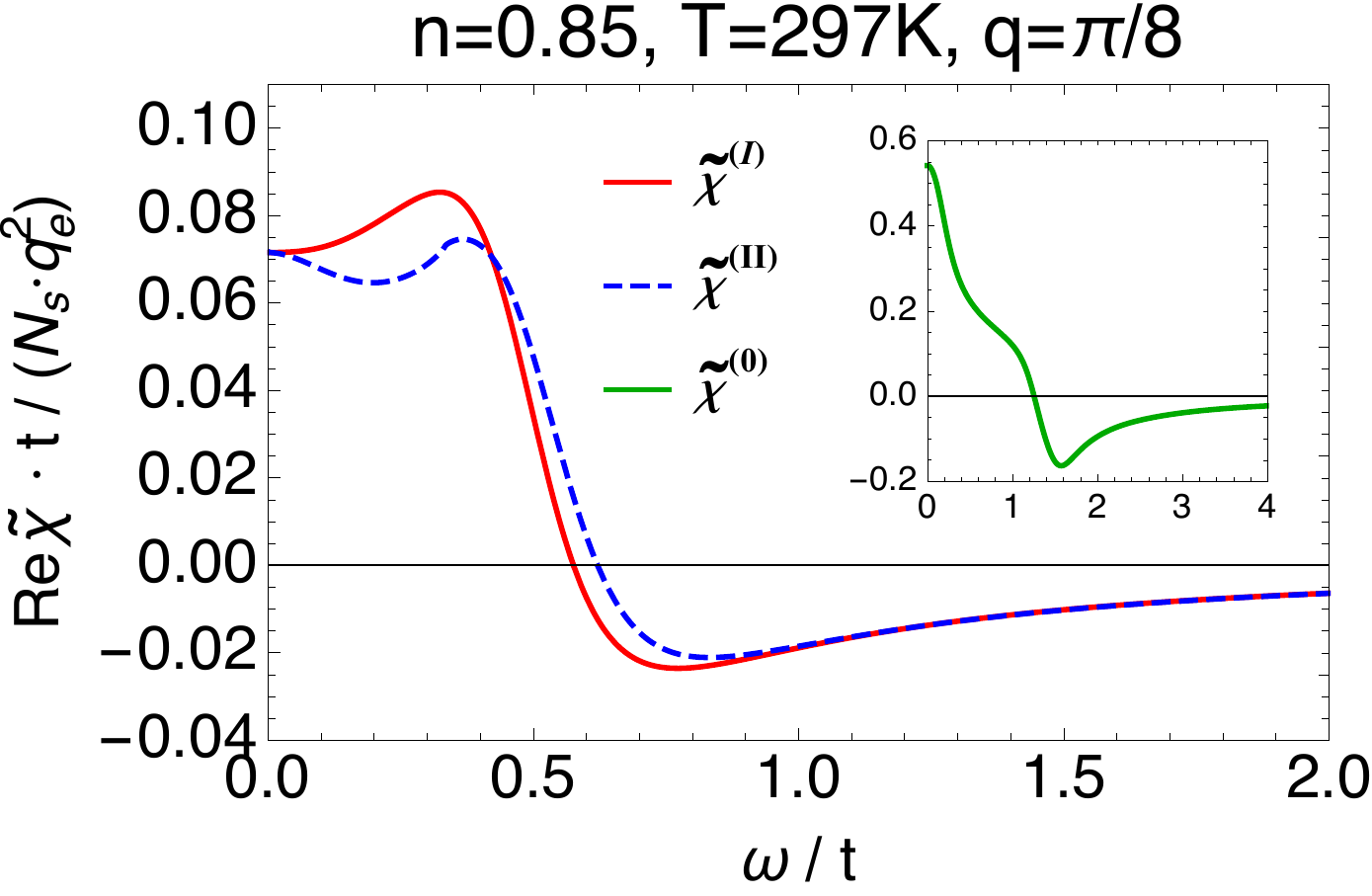}}
 \caption{\footnotesize A comparison between the ({\bf a}) imaginary and ({\bf b}) real parts of the irreducible susceptibilities $\hatchi^{(I)}$ in red (using $\Psi_B(\vq,\omega)$  in  \disp{appxI}),  and $\hatchi^{(II)}$  in dotted-blue (using $\Psi_B+\delta \Psi_{QP}$  in \disp{appxII}).  Note that 
 a quasiparticle (linear in $\omega$)  contribution is visible in ({\bf a})  at low frequencies. If we neglect that regime, 
the two approximations lead to similar results for $\omega\gssim 0.40t$. The inset shows that the  corresponding non-interacting  complex susceptibility given in \disp{RPA-1} for  the same  parameters extend to much higher frequencies $\omega/t$, and have  different vertical scales and shapes.
    }
 \label{chiIvschiII}
\end{figure*}

\begin{figure*}[h]
\centering
 \subfigure[\;\;]  {\includegraphics[width=0.3\columnwidth]{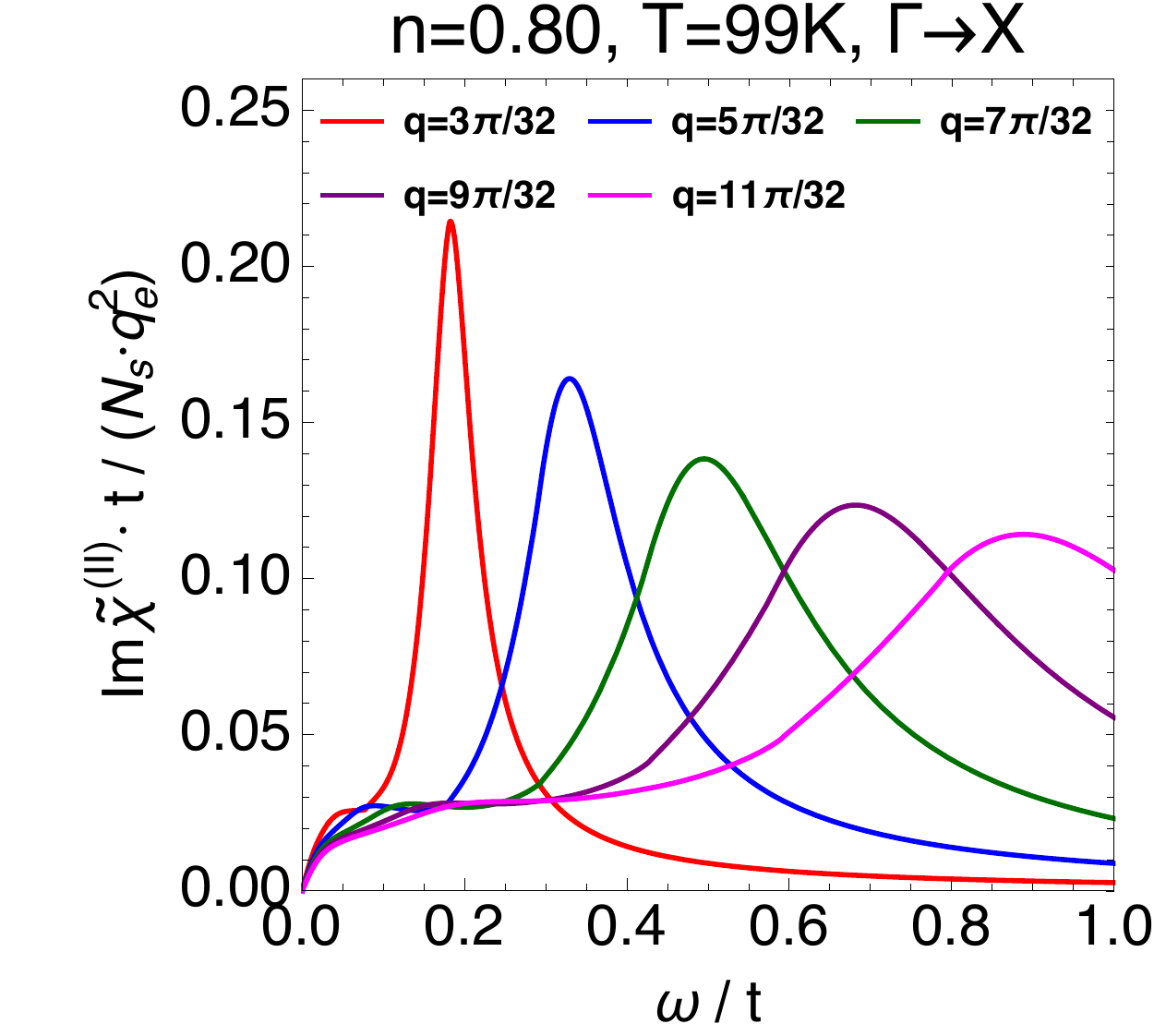}}
 \subfigure[\;\;]  {\includegraphics[width=0.3\columnwidth]{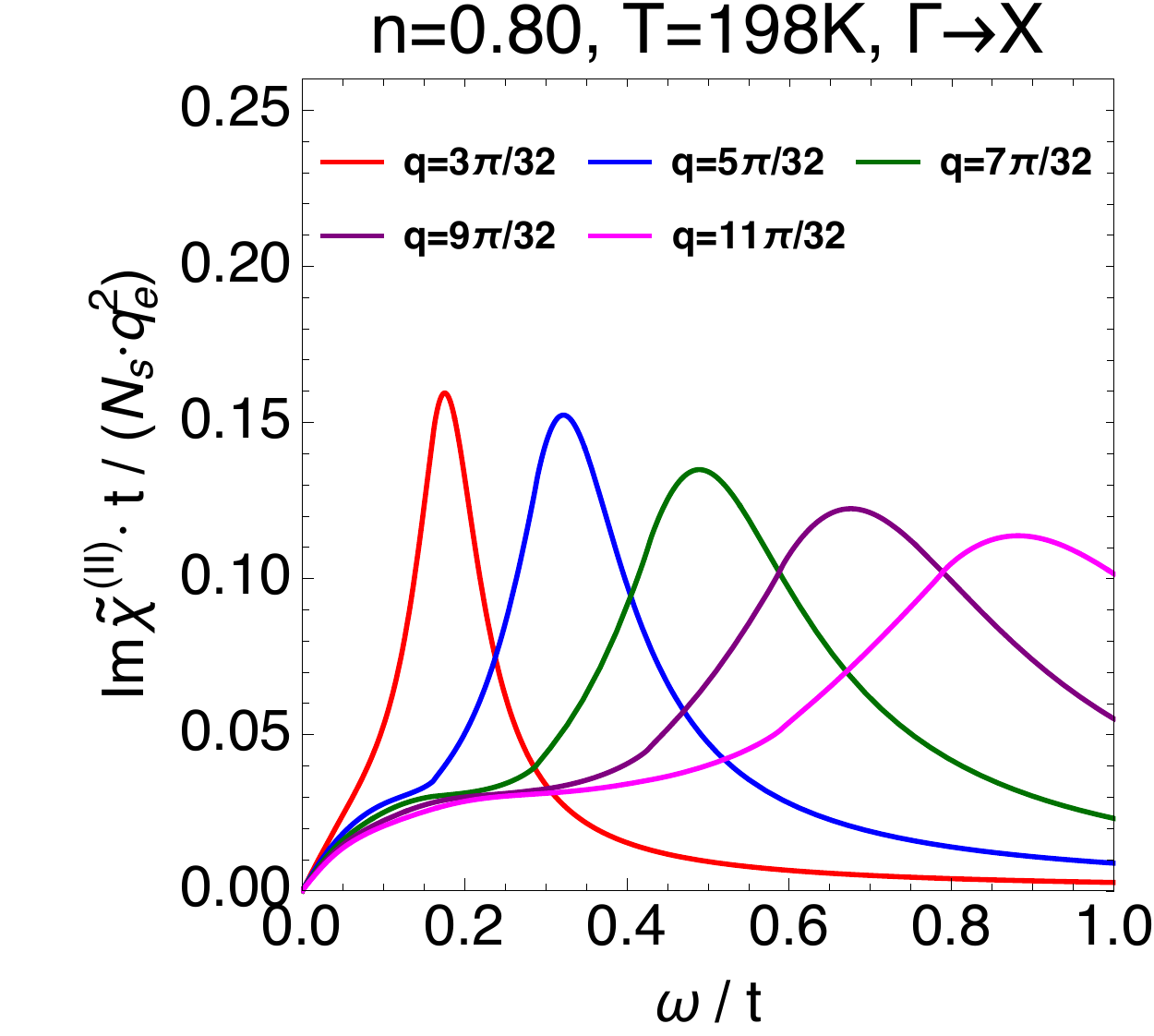}}
 \subfigure[\;\;]  {\includegraphics[width=0.3\columnwidth]{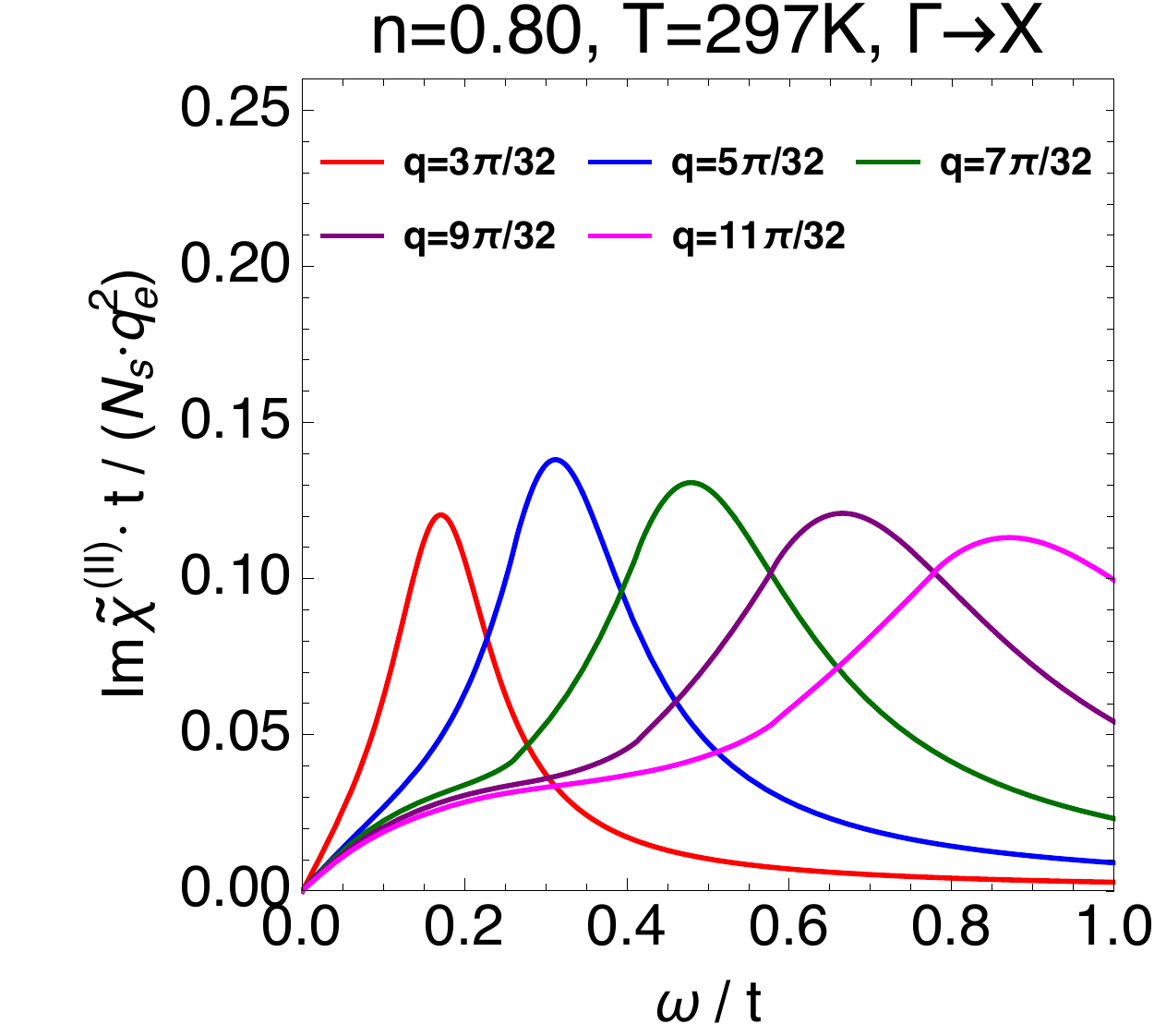}}
 \subfigure[\;\;]  {\includegraphics[width=0.3\columnwidth]{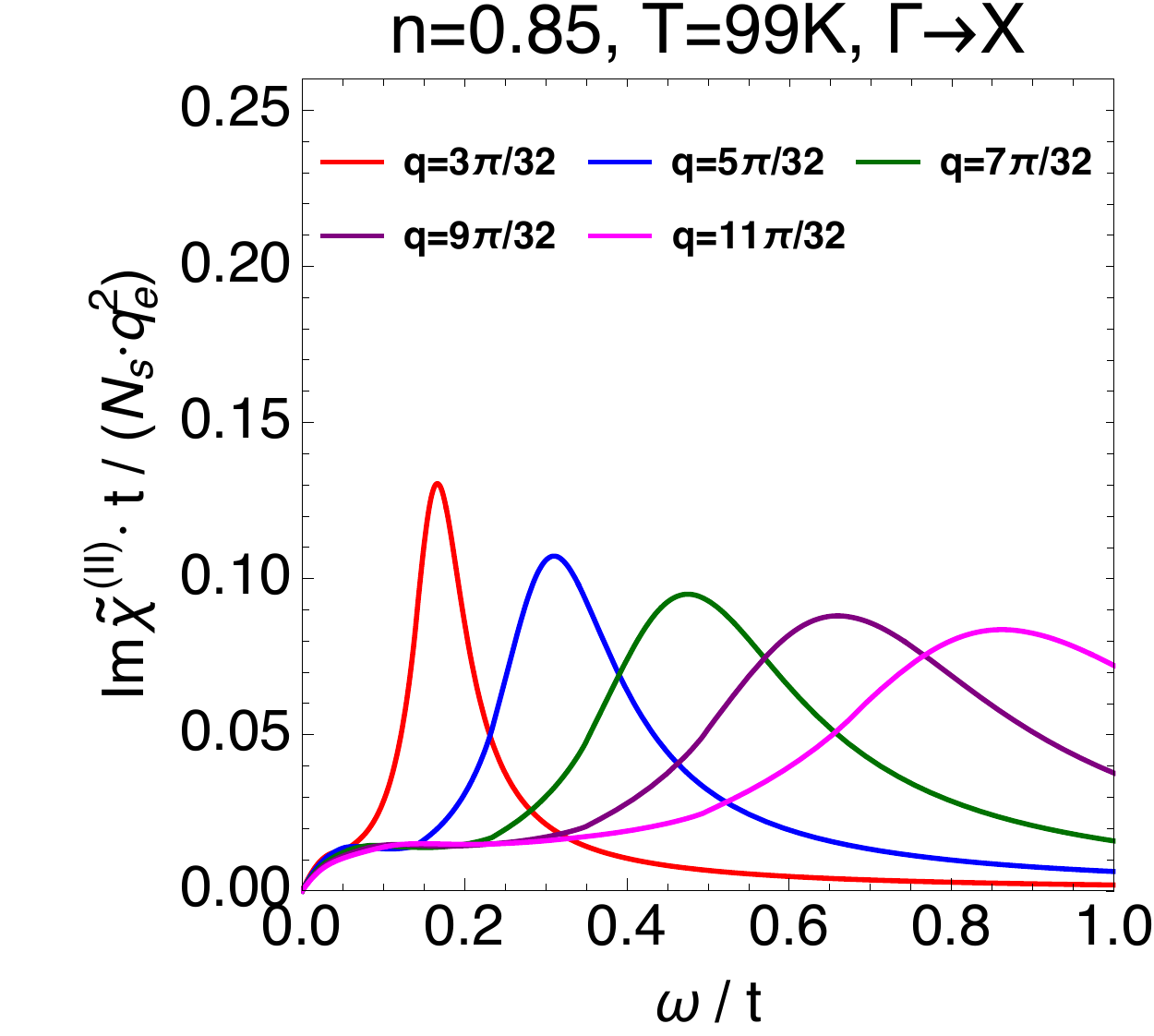}}
 \subfigure[\;\;]  {\includegraphics[width=0.3\columnwidth]{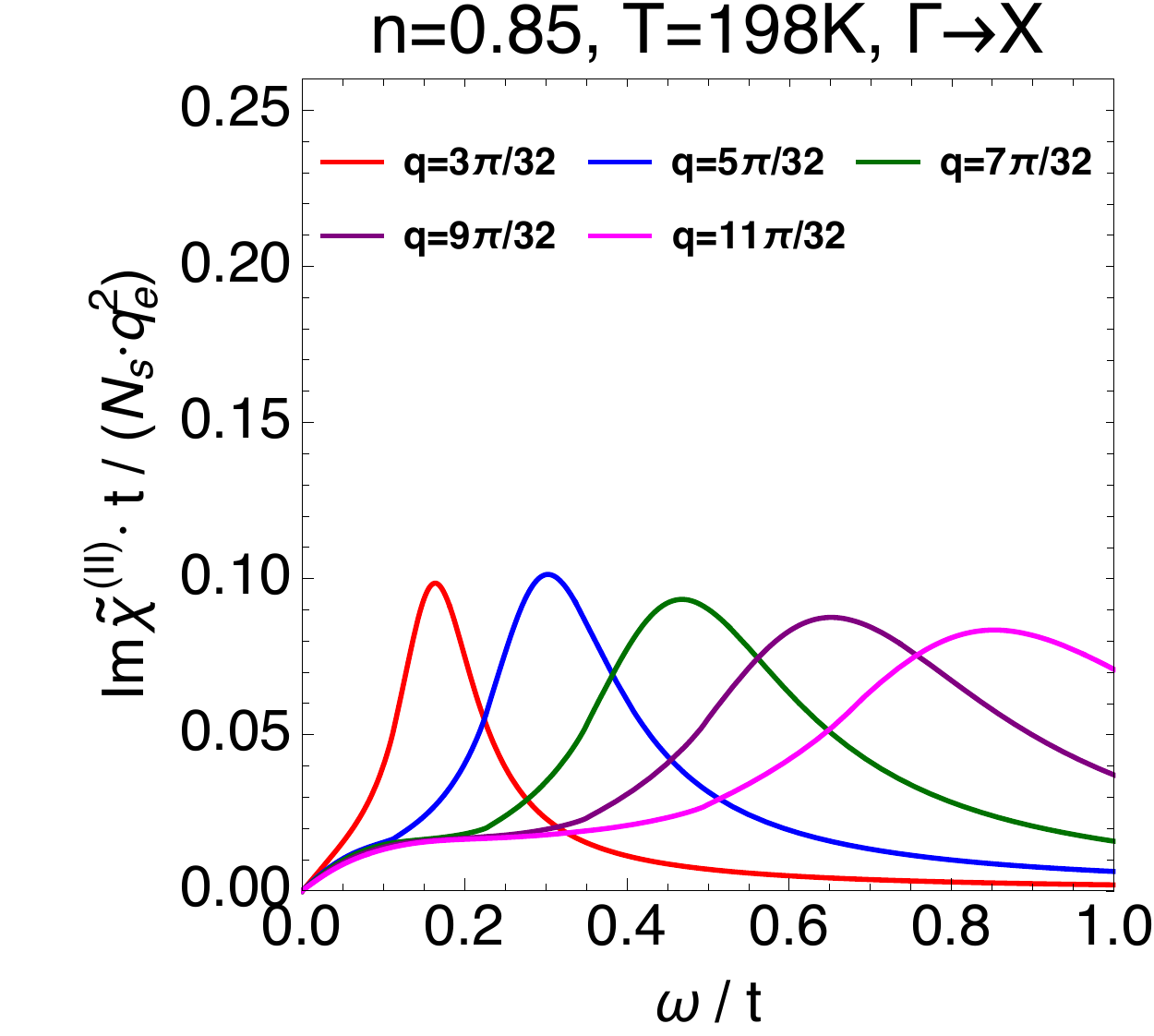}}
 \subfigure[\;\;]  {\includegraphics[width=0.3\columnwidth]{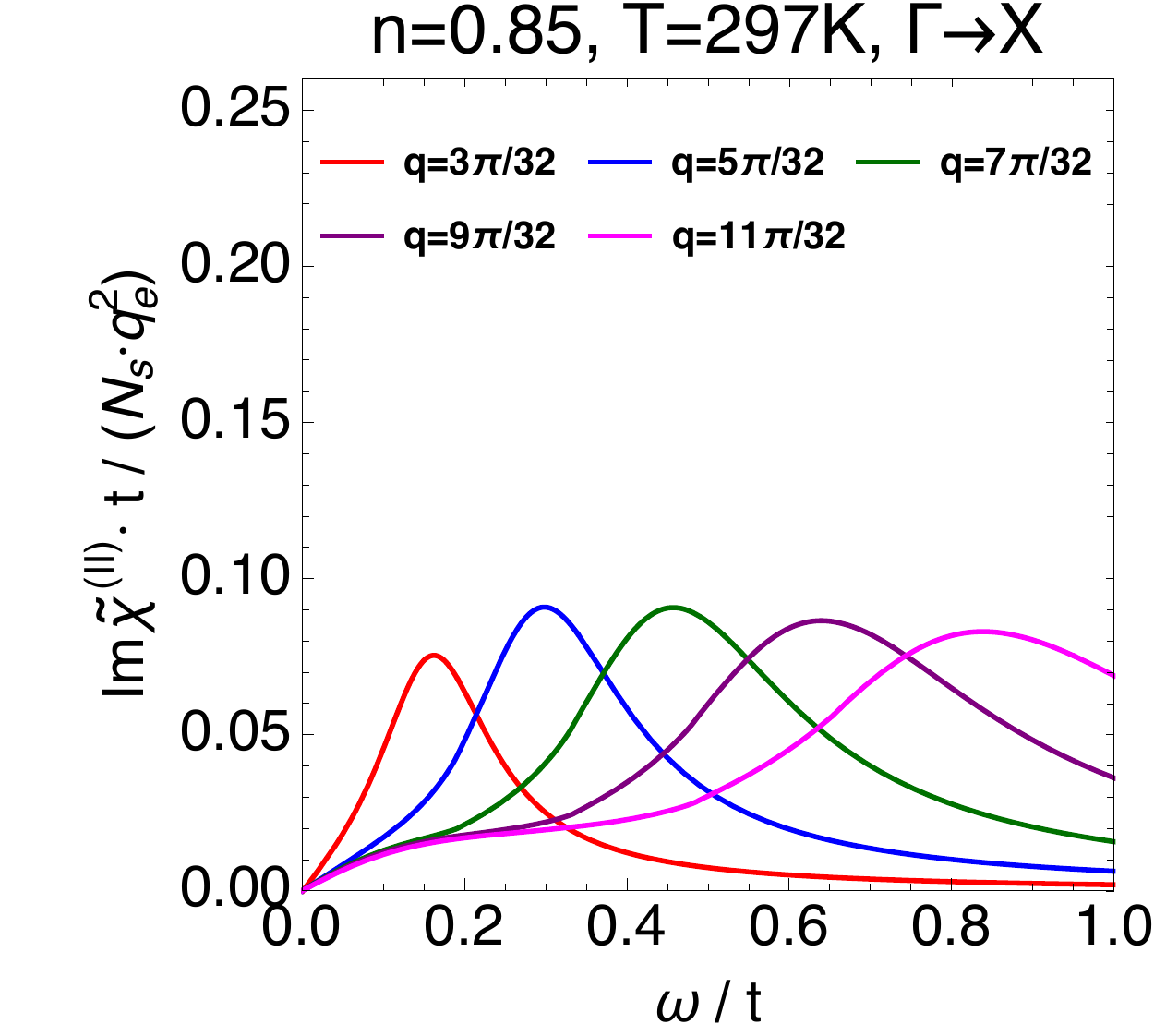}}
 \caption{\footnotesize The $\vq,\omega$ variation of the  imaginary part of the irreducible  susceptibility $\Im \, \hatchi^{(II)}$ \disp{appxII}.  The significant features from the  $\Im \, \hatchi^{(I)}$ (in \disp{appxI}) are qualitatively similar, and hence omitted.
 The figures are  at densities $n=0.8,0.85$ at temperatures $T= \, 99,198,297$K in the $\Gamma \to X$ direction where $\vec{q} = (q,q)$. Other directions in the BZ give similar results for small $|\vq|$, as one might expect. Each  curves exhibit a $\vq$ dependent peak at an energy  $\Omega_p(\vq) \sim \Omega(\vq,0)$ from \disp{Omega-p}. The peak shifts towards lower energies as $q$ is reduced, and for a fixed $q$ the intensity drops rapidly with a modest increase of  $T$. The peak energy is a  (measurable) characteristic energy scale, and discussed further in \figdisp{BigOmega}.
    }
 \label{chiGAMMAtoX}
\end{figure*}


\begin{figure*}[h]
    \centering
    \subfigure[\;\;]{\includegraphics[width=0.36\columnwidth]{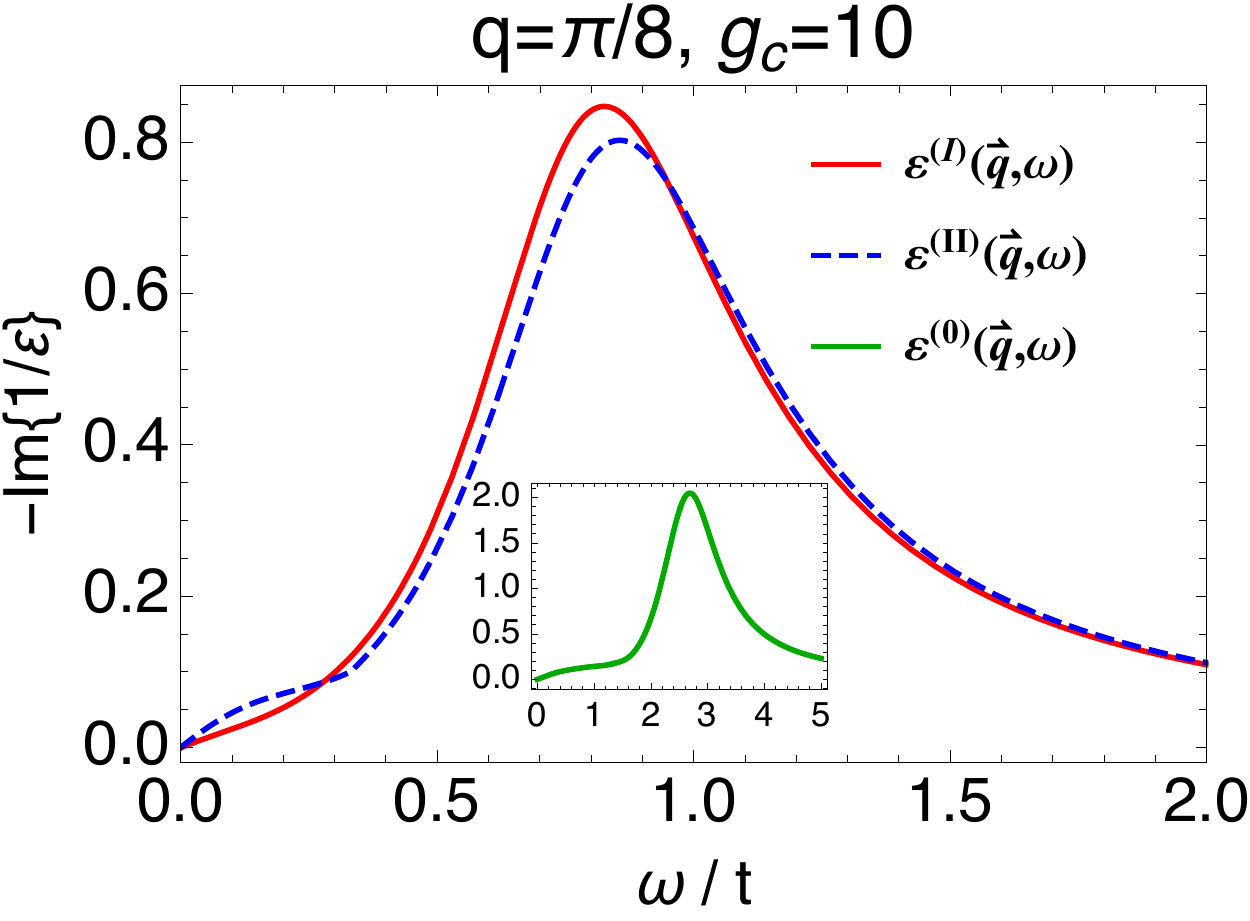}}
    \subfigure[\;\;]{\includegraphics[width=0.36\columnwidth]{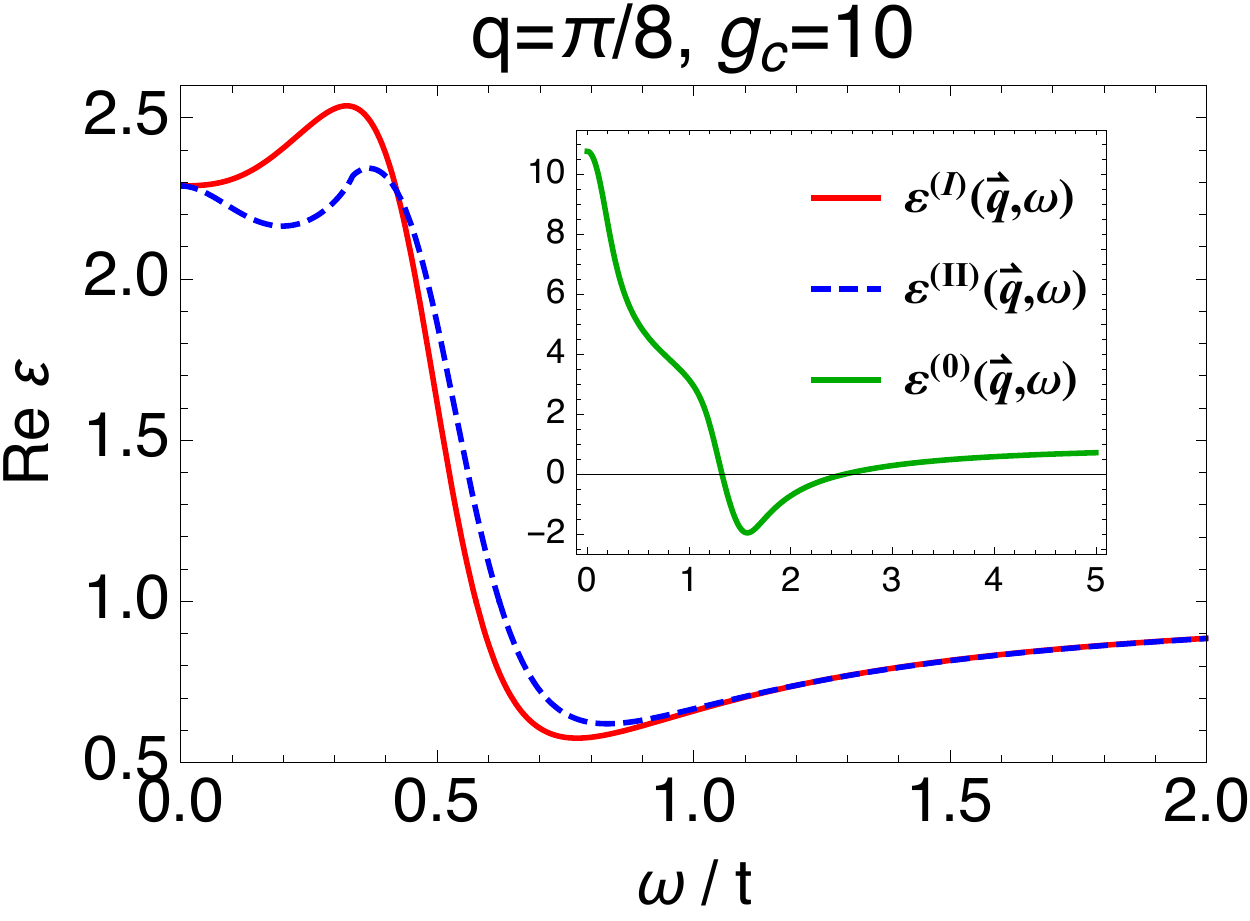}}
    \subfigure[\;\;]{\includegraphics[width=0.36\columnwidth]{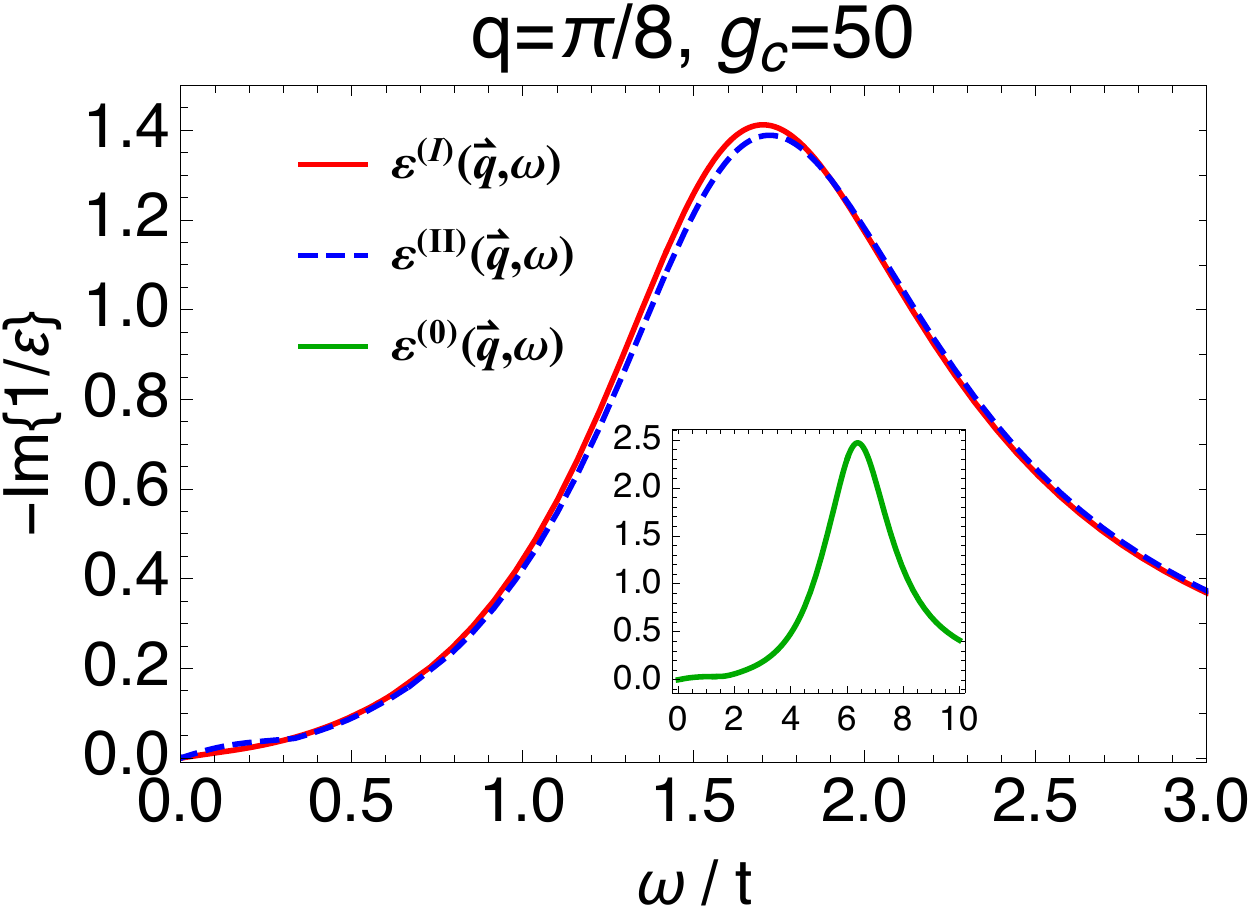}}
    \subfigure[\;\;]{\includegraphics[width=0.36\columnwidth]{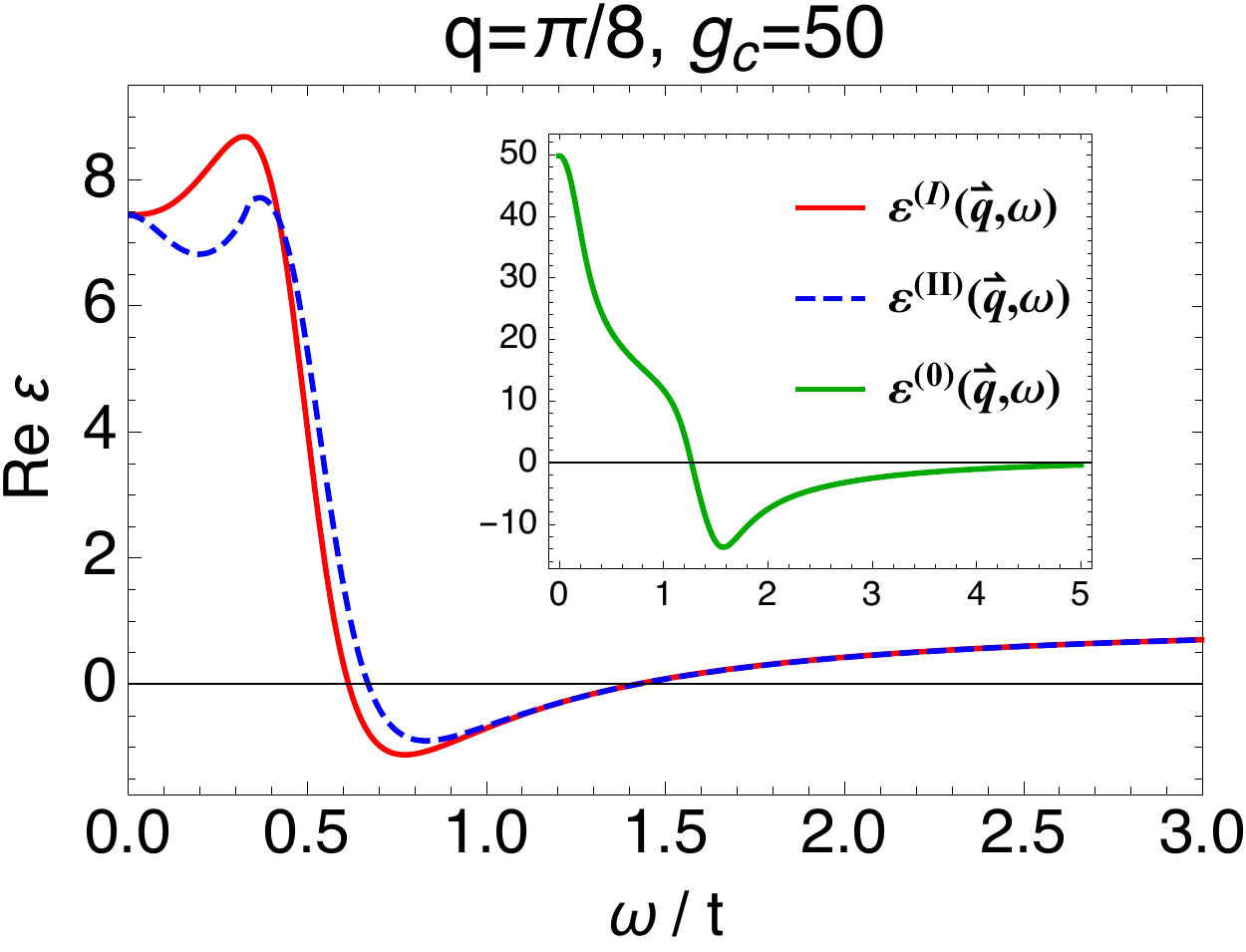}}
    \caption{\footnotesize The dielectric functions $\varepsilon^{(I,II)}(\vq,\omega)$ and their inverse  from \disp{app-dielectric} for a system at $n=0.85$ and $T=297$K, with $\vec{q}=(q,q)$ along $\Gamma \to X$. The insets show the corresponding curves for the  RPA approximation (obtained by using  \disp{RPA-1} for  $\hatchi$ in \disp{app-dielectric}) with the same hopping parameters. 
    In obtaining $\varepsilon^{(I,II)}(\vq,\omega) $ from  $\hatchi$ in \disp{app-dielectric}, we  require the effective Coulomb coupling constant $g_c$  \disp{gc} involving a combination of material parameters $t,a_0,\varepsilon_\infty$. In the BSCCO material used in the experiments of \cite{MEELS-1,MEELS-2}, using  $t\sim 0.45$ eV, $a_0$$\sim3.81$$\AA$ and   $\varepsilon_\infty$$\sim$$4.5$, we find $g_c$$\sim 11.5$, while  using $t\sim 0.16$eV  gives $g_c\sim$$32.0$.  We provide a results for a few typical values of this parameter, since the basic parameters  vary for different  materials.
     Here panel ({\bf a}), ({\bf c}) is the imaginary part while panel ({\bf b}), ({\bf d}) is the real part for $g_c$$=10, 50$ respectively. The curves $\Re \, {\varepsilon^{(I,II)}}$ do not vanish in this range at $g_c$$=10$ (panel({\bf b})), while they do so when $g_c=50$ (panel({\bf d})). This is unlike plasmon in weakly interacting electron gas for both $g_c$ as seen in the insets. In the latter, as discussed in textbooks \refdisp{Fetter-Walecka}, a zero crossing  of $\Re\,\varepsilon(\vq,\omega)$  determines the plasmon frequency, which is  also visible as a peak in $\Im\,\frac{1}{\varepsilon}$.}
     \label{epsIepsII}
\end{figure*}

\begin{figure*}[h]
    \centering
    \subfigure[\;\;]{\includegraphics[width=0.36\columnwidth]{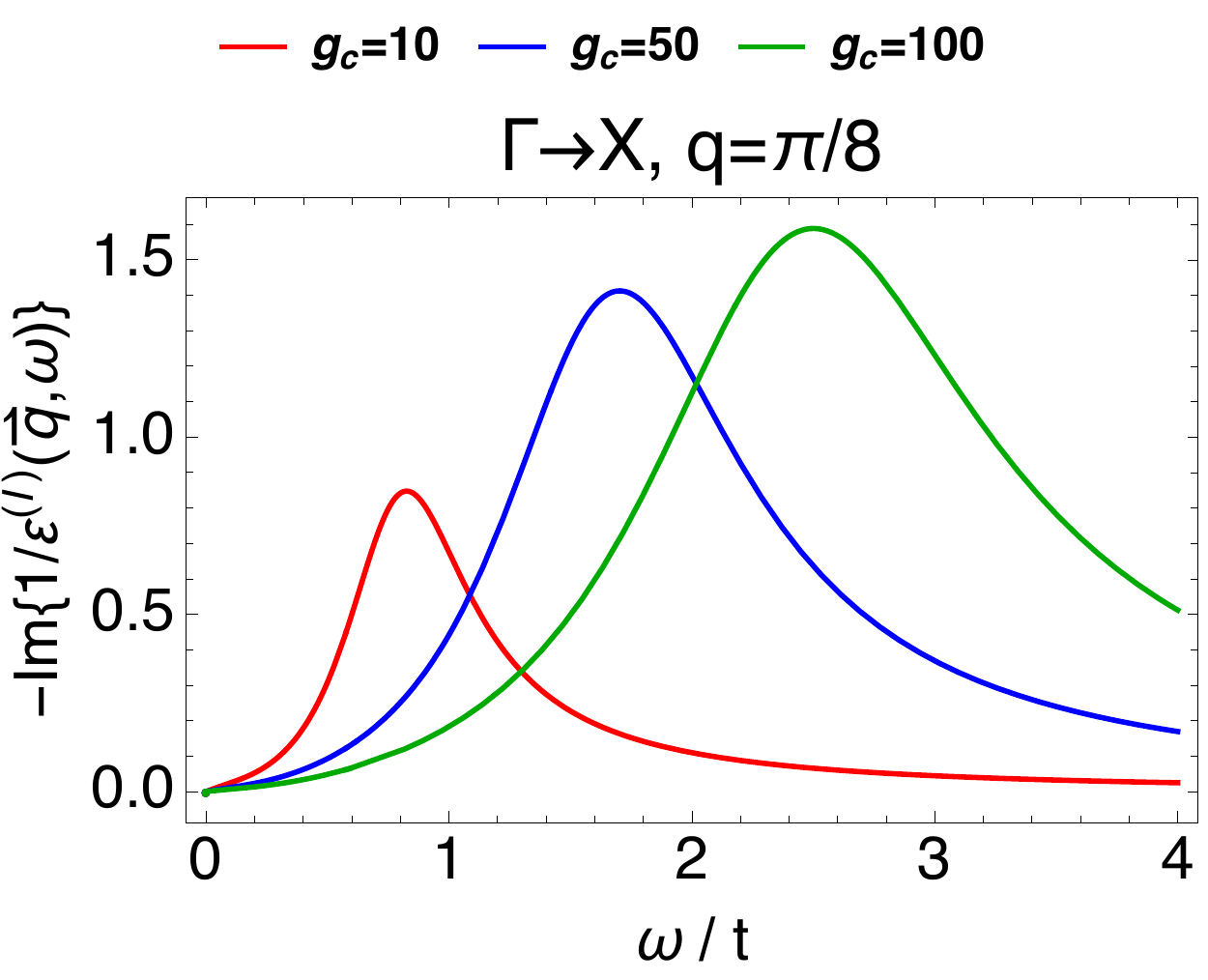}}
    \caption{\footnotesize Imaginary part of inverse dielectric function (\disp{app-dielectric}) at $n=0.85$ and $T=297$K  with $\vq=\{q,q\}$ at representative values of the Coulomb coupling $g_c$ \disp{gc}. The peaks in the $\Gamma \to M$ direction are similar at low $\vq$. The variation with q at given $g_c$ is shown in \figdisp{epsilongc}.
}
    \label{epsilon}
\end{figure*}

\begin{figure*}[h]
    \centering
    \subfigure[\;\;]{\includegraphics[width=0.3\columnwidth]{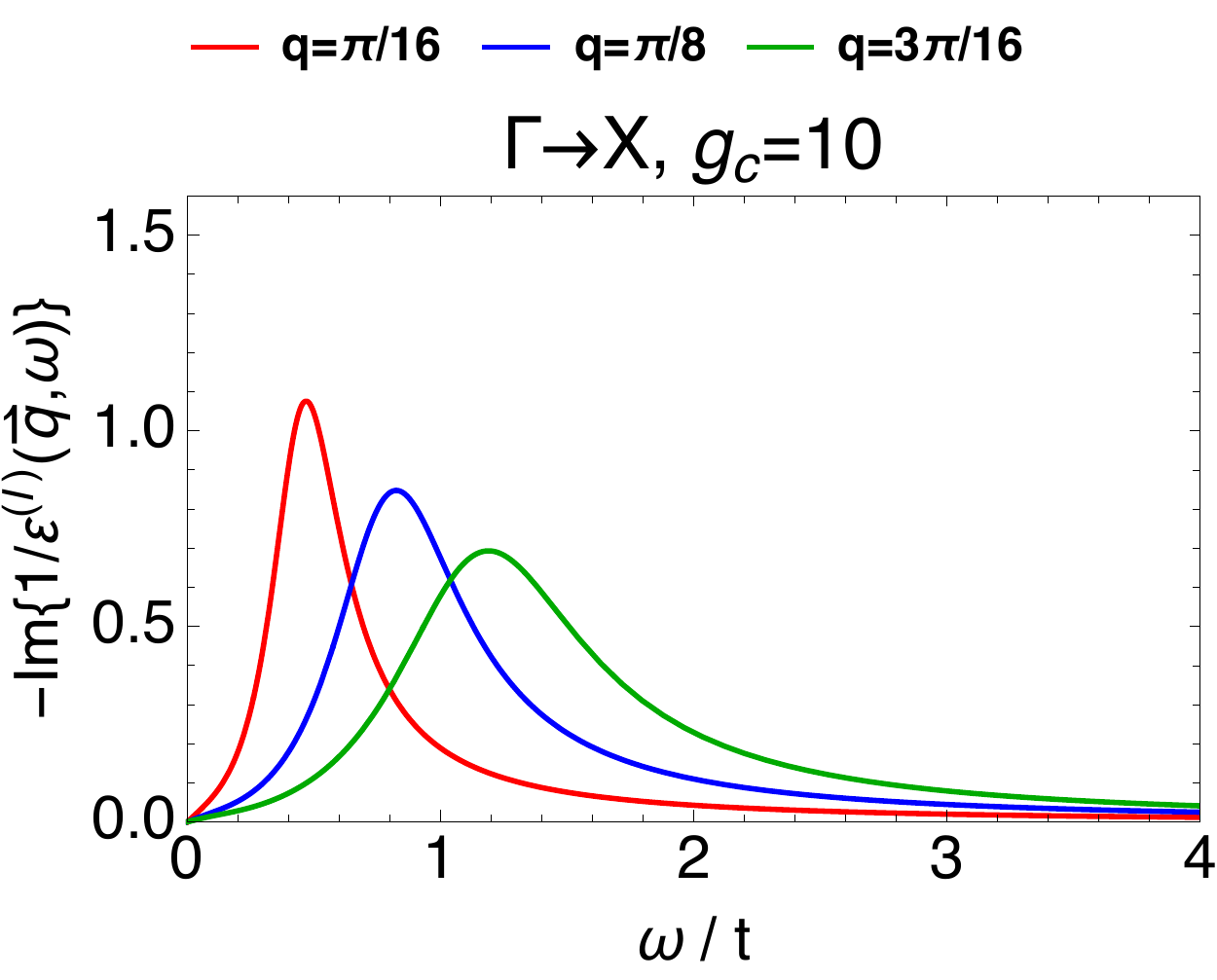}}
    \subfigure[\;\;]{\includegraphics[width=0.3\columnwidth]{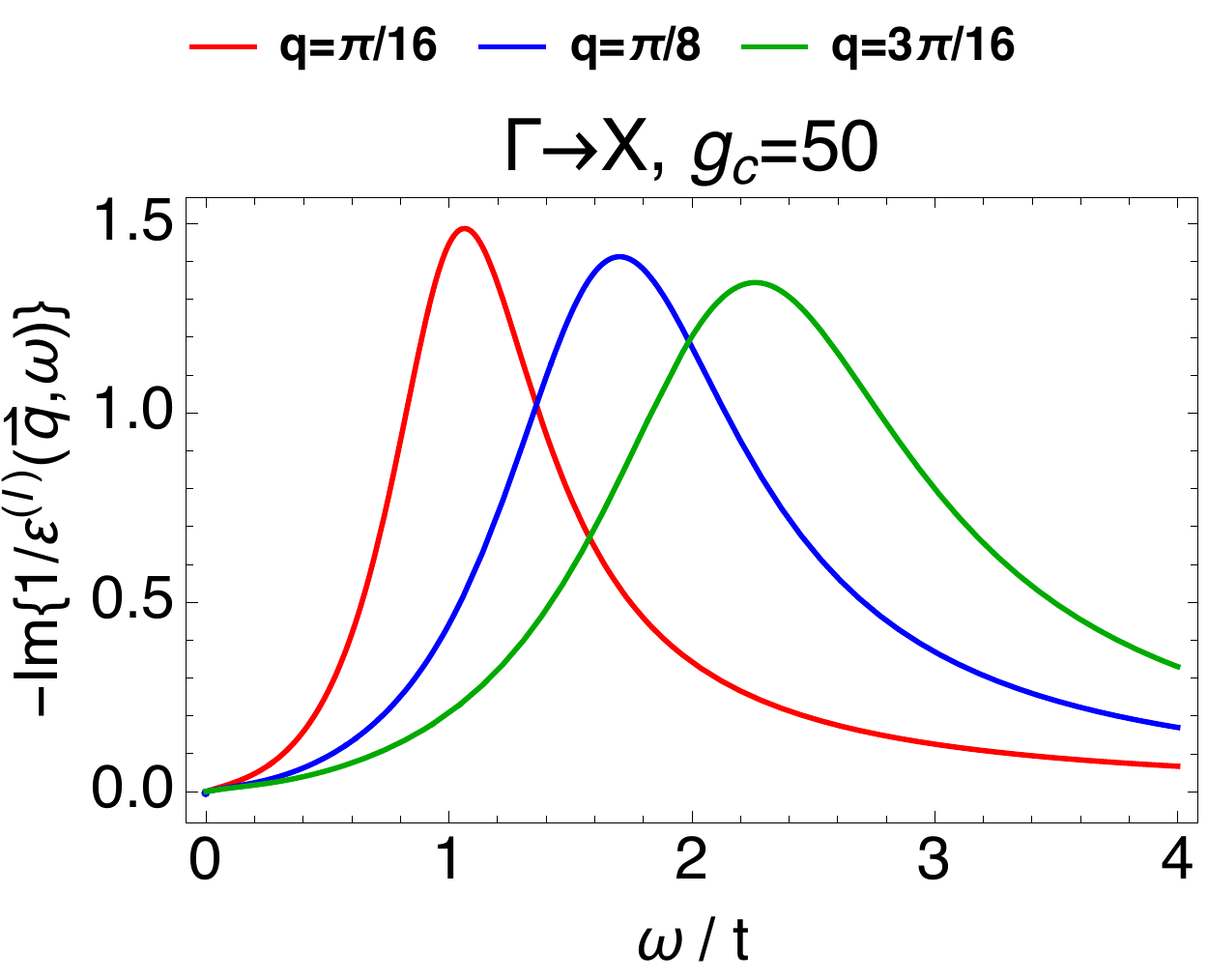}}
    \subfigure[\;\;]{\includegraphics[width=0.3\columnwidth]{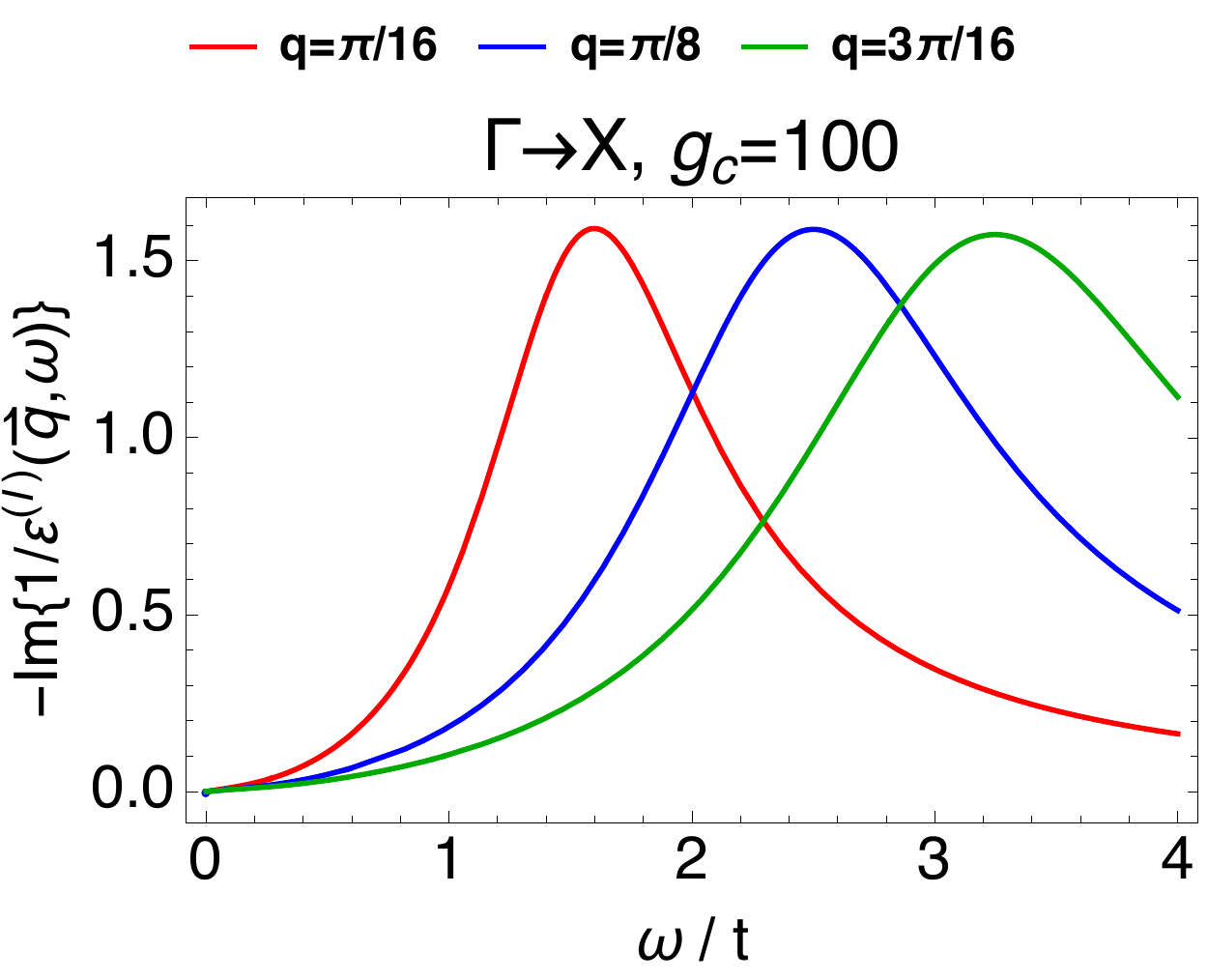}}
    \caption{\footnotesize The $\vq=\{q,q\}$ variation of imaginary part of inverse dielectric function (\disp{app-dielectric}) at $n=0.85$ and $T=297$K at representative values of the Coulomb coupling $g_c$ \disp{gc}.  As $g_c$ increases, we note a shift of peaks to higher frequencies as well as a  broadening.
 }
    \label{epsilongc}
\end{figure*}

\begin{figure*}[h]
    \centering
    \subfigure[\;\;]{\includegraphics[width=0.27\columnwidth]{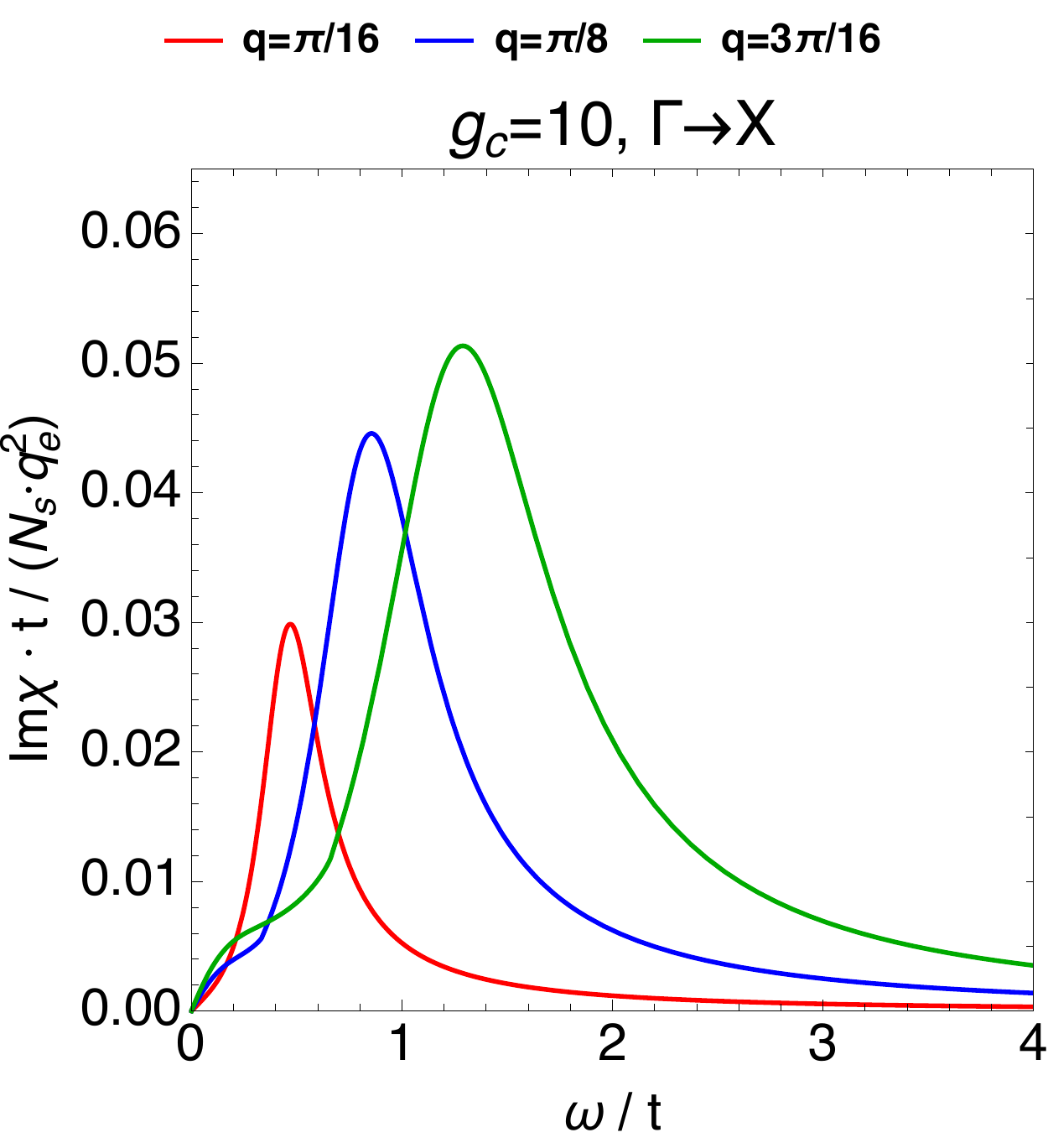}}
    \subfigure[\;\;]{\includegraphics[width=0.27\columnwidth]{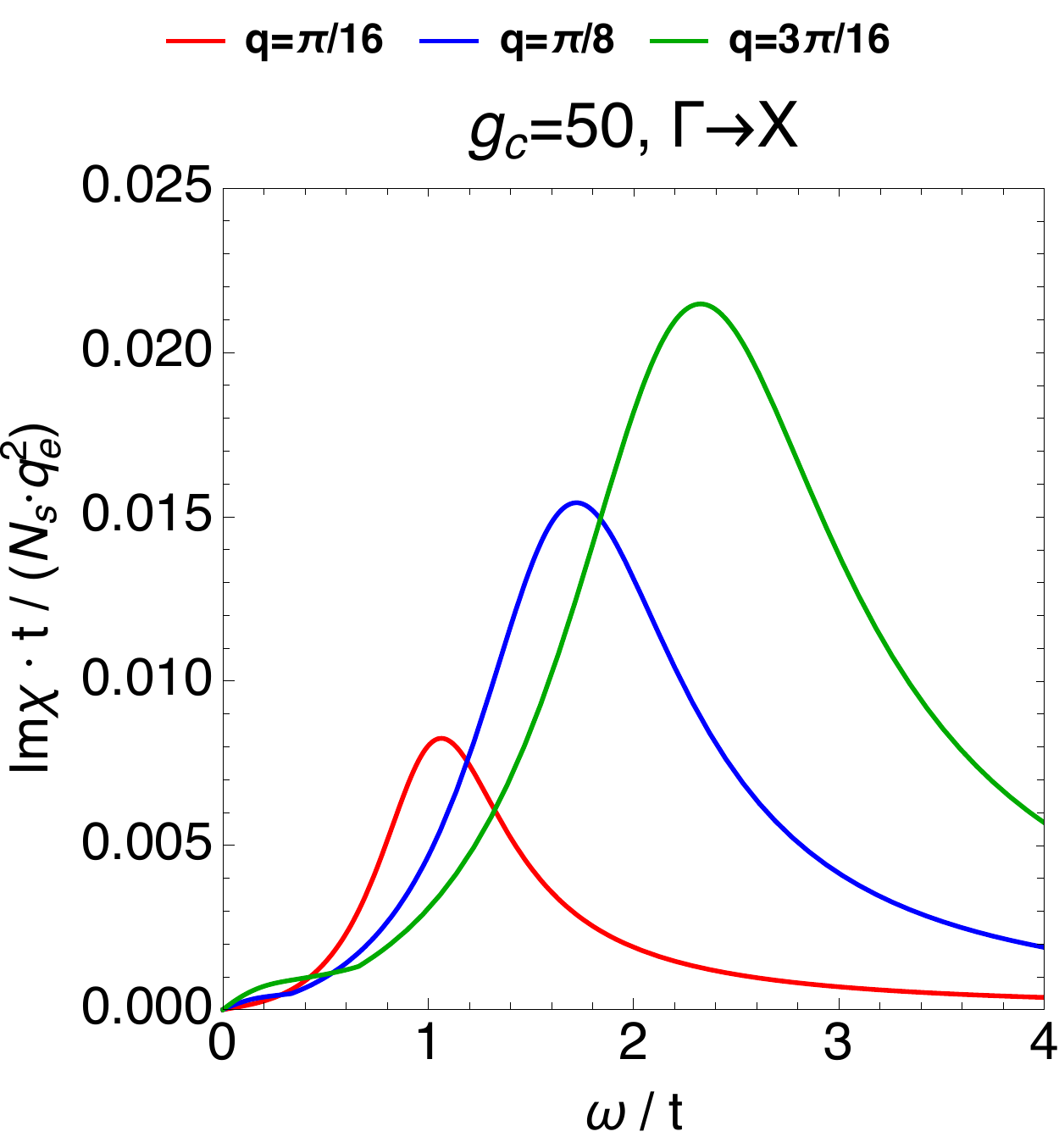}}
    \subfigure[\;\;]{\includegraphics[width=0.27\columnwidth]{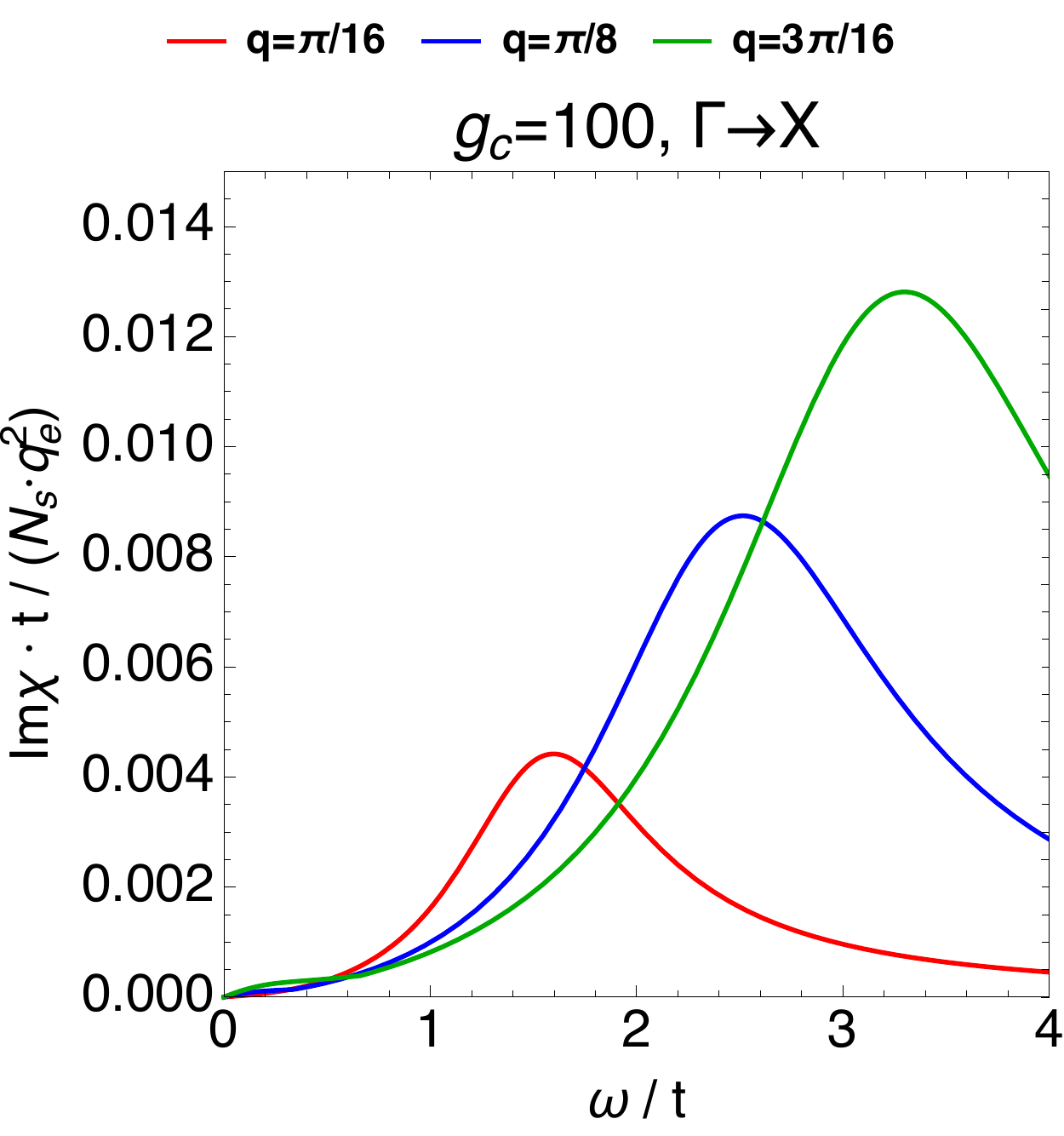}}

    \subfigure[\;\;]{\includegraphics[width=0.27\columnwidth]{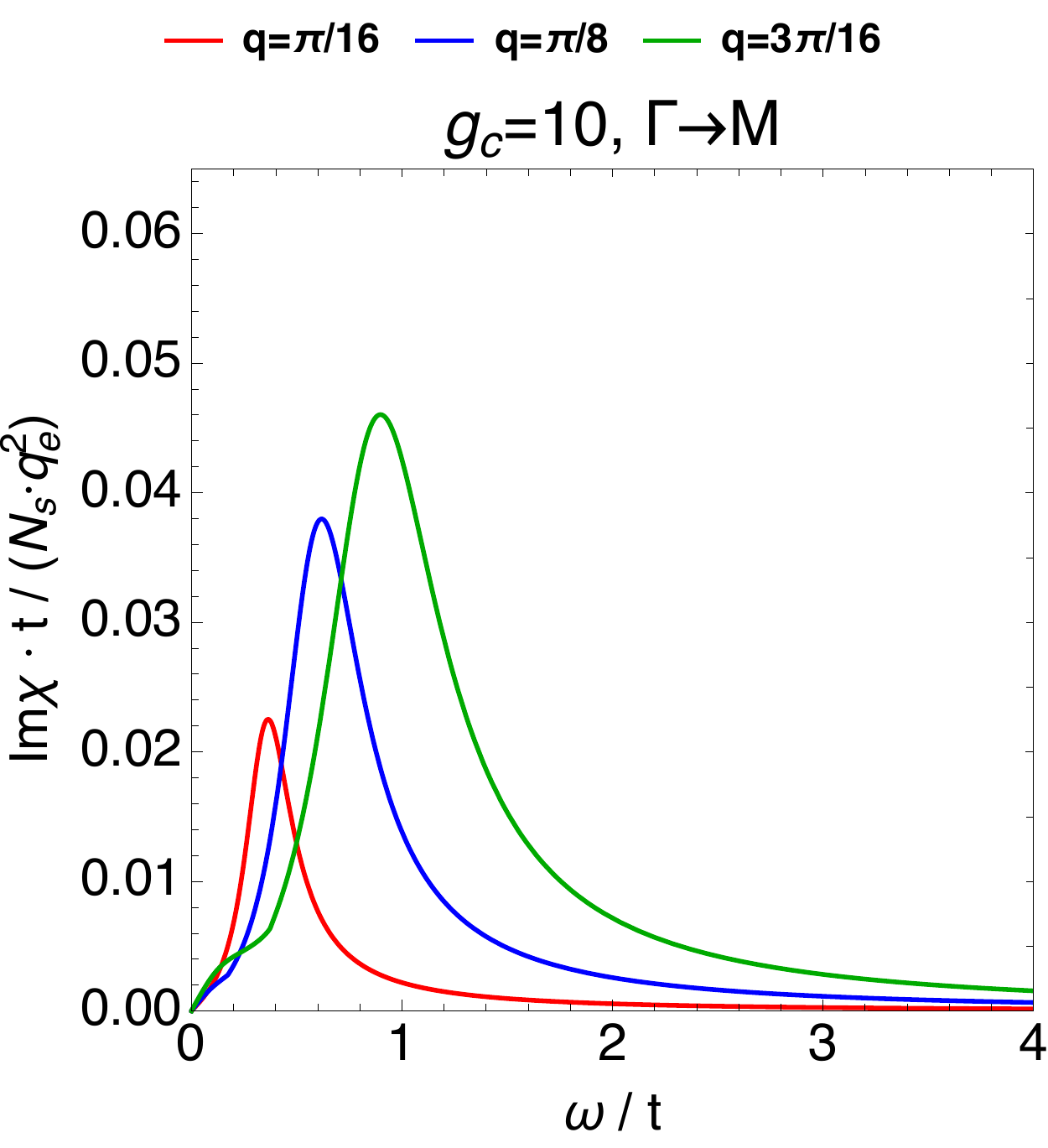}}
    \subfigure[\;\;]{\includegraphics[width=0.27\columnwidth]{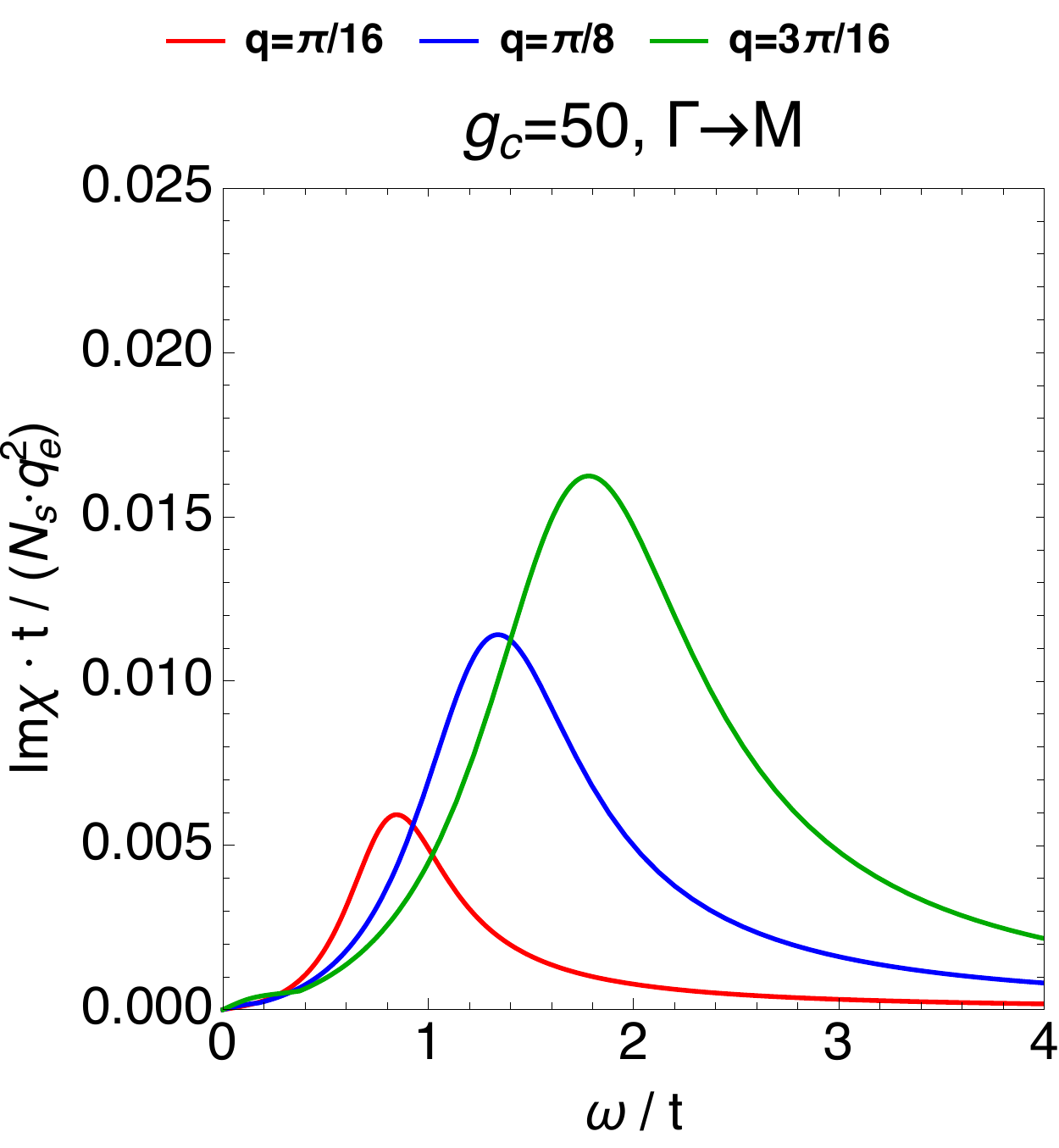}}
    \subfigure[\;\;]{\includegraphics[width=0.27\columnwidth]{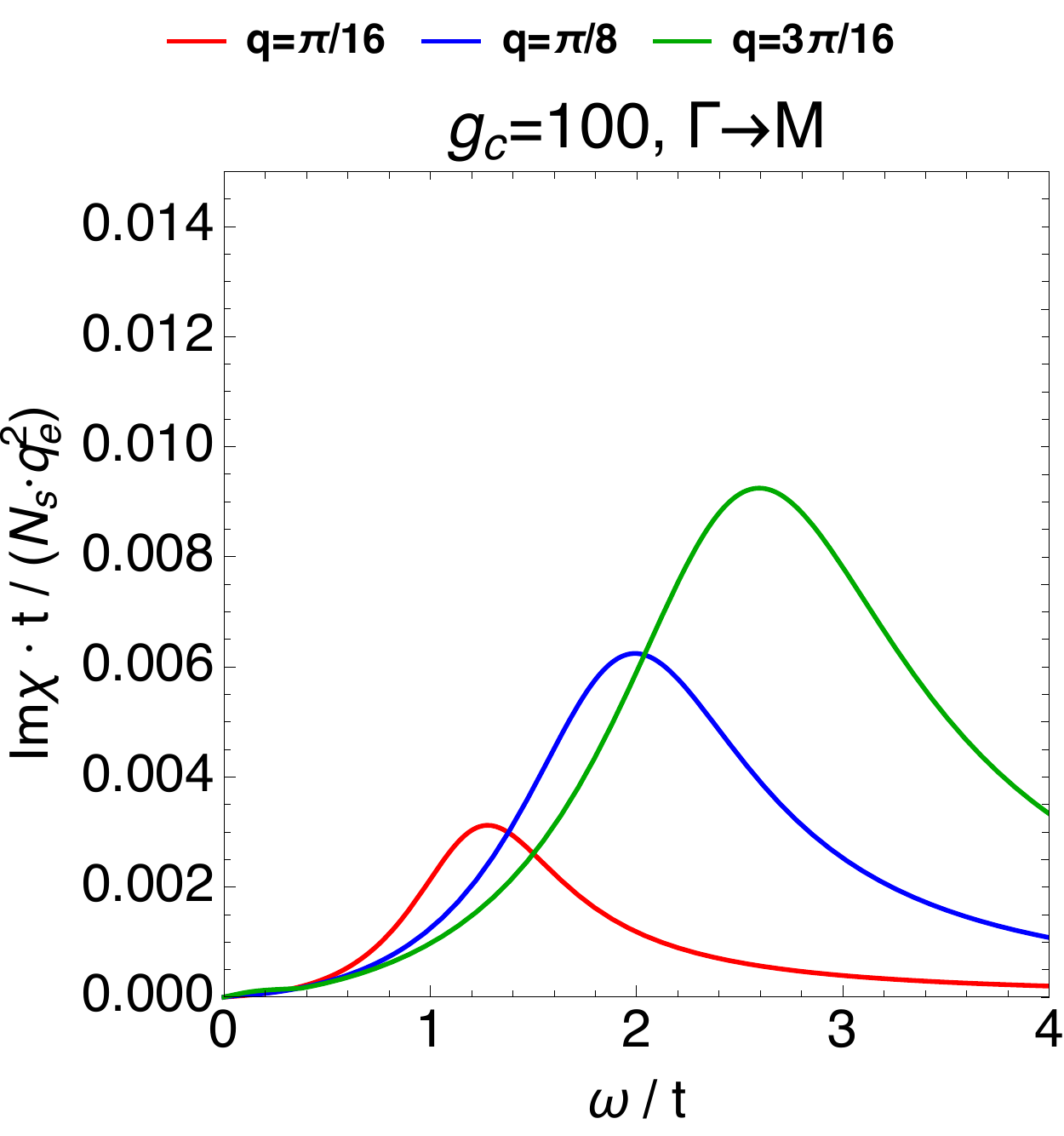}}
    \caption{\footnotesize The reducible (i.e.  unscreened) susceptibility $\Im \, \bigchi_{\rho\rho}$ (\disp{relations-0}) at $n=0.85$ and $T=297$K. We note from \disp{structure-direct}  that this   is the most directly accessible  object in experiments.
        Panels (a,b,c) show wavevectors $\vq=\{q,q\}$ and panels (d,e,f) show wavevectors $\vq=\{q,0\}$ at three values of q, using representative values of the Coulomb coupling $g_c$ \disp{gc}.     Results using  $\hatchi^{(I)}$ are similar apart from the region of smallest $\omega$, and omitted for brevity. 
In   all  panels   the peak  magnitudes     decrease  as $q\to0$, as a  consequence of the conservation of charge.  We observe that as $g_c$ increases, the peaks in $ \Im \, \bigchi_{\rho\rho}$ are broadened  and pushed  to higher energies,  as also seen in \figdisp{epsilongc}. }
    \label{reducible-chi}
\end{figure*}

\begin{figure*}[h]
\centering
 \subfigure[\;\;]{\includegraphics[width=0.27\columnwidth]{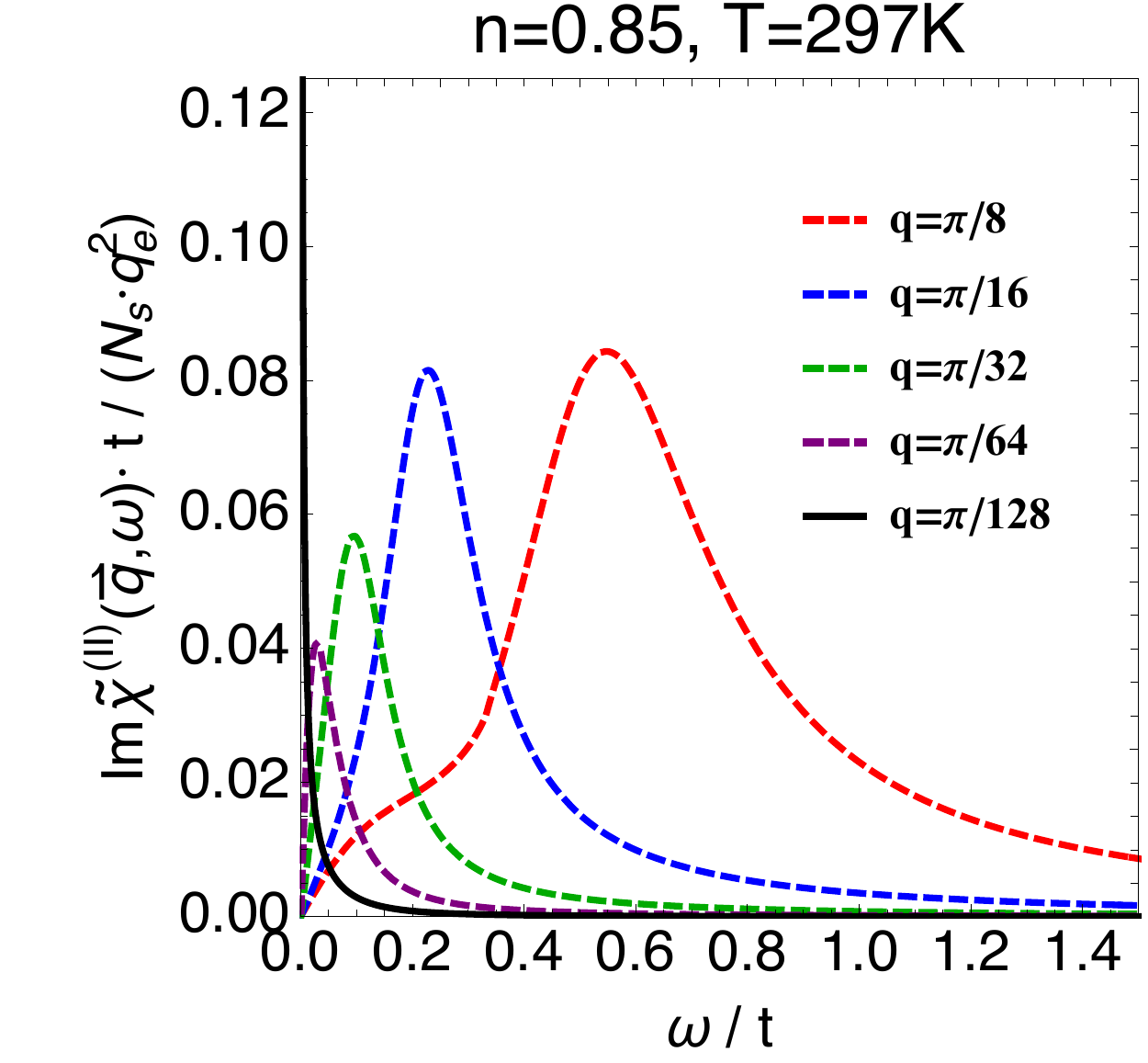}}
 \subfigure[\;\;]{\includegraphics[width=0.27\columnwidth]{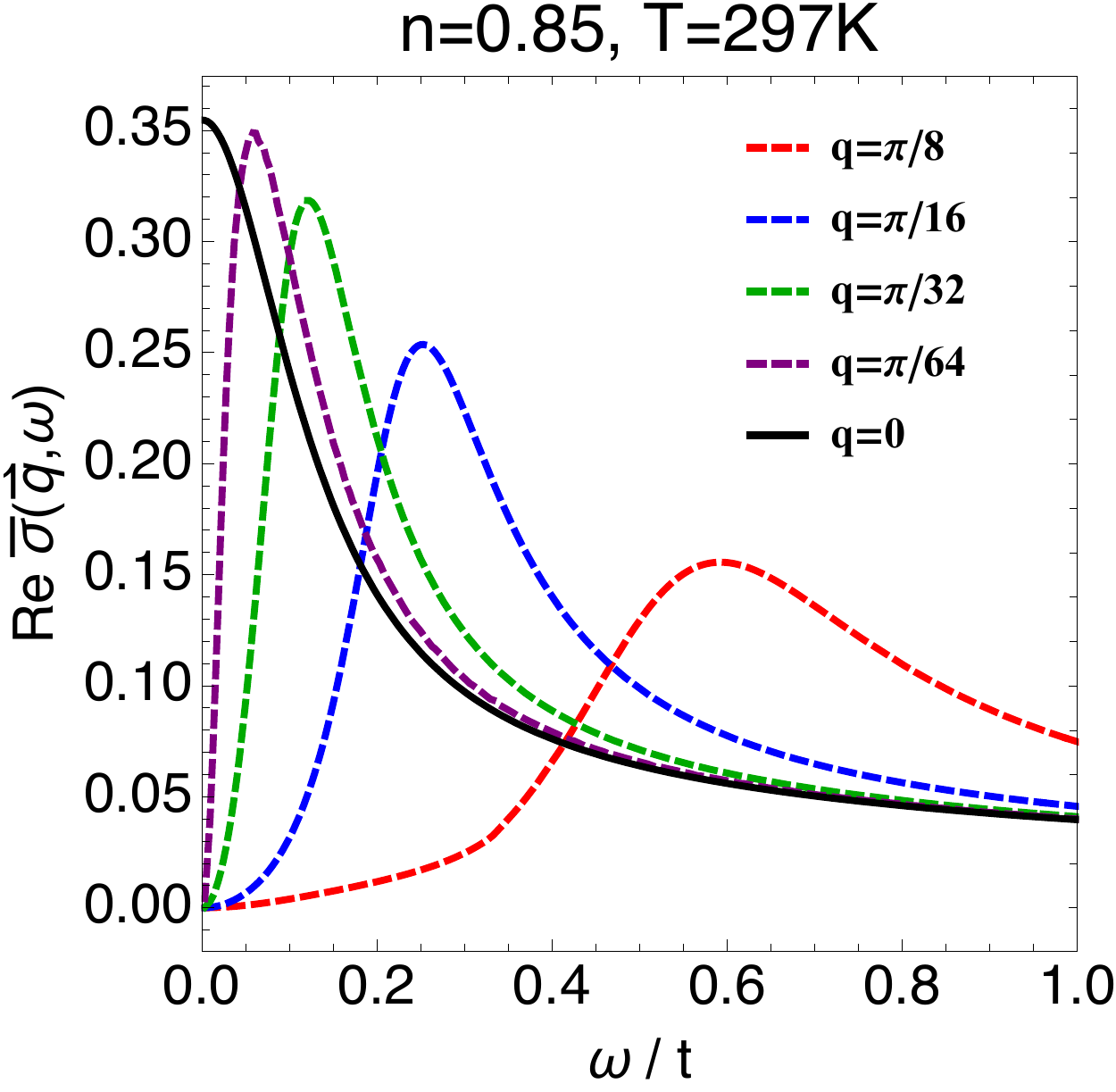}}
 \subfigure[\;\;]{\includegraphics[width=0.27\columnwidth]{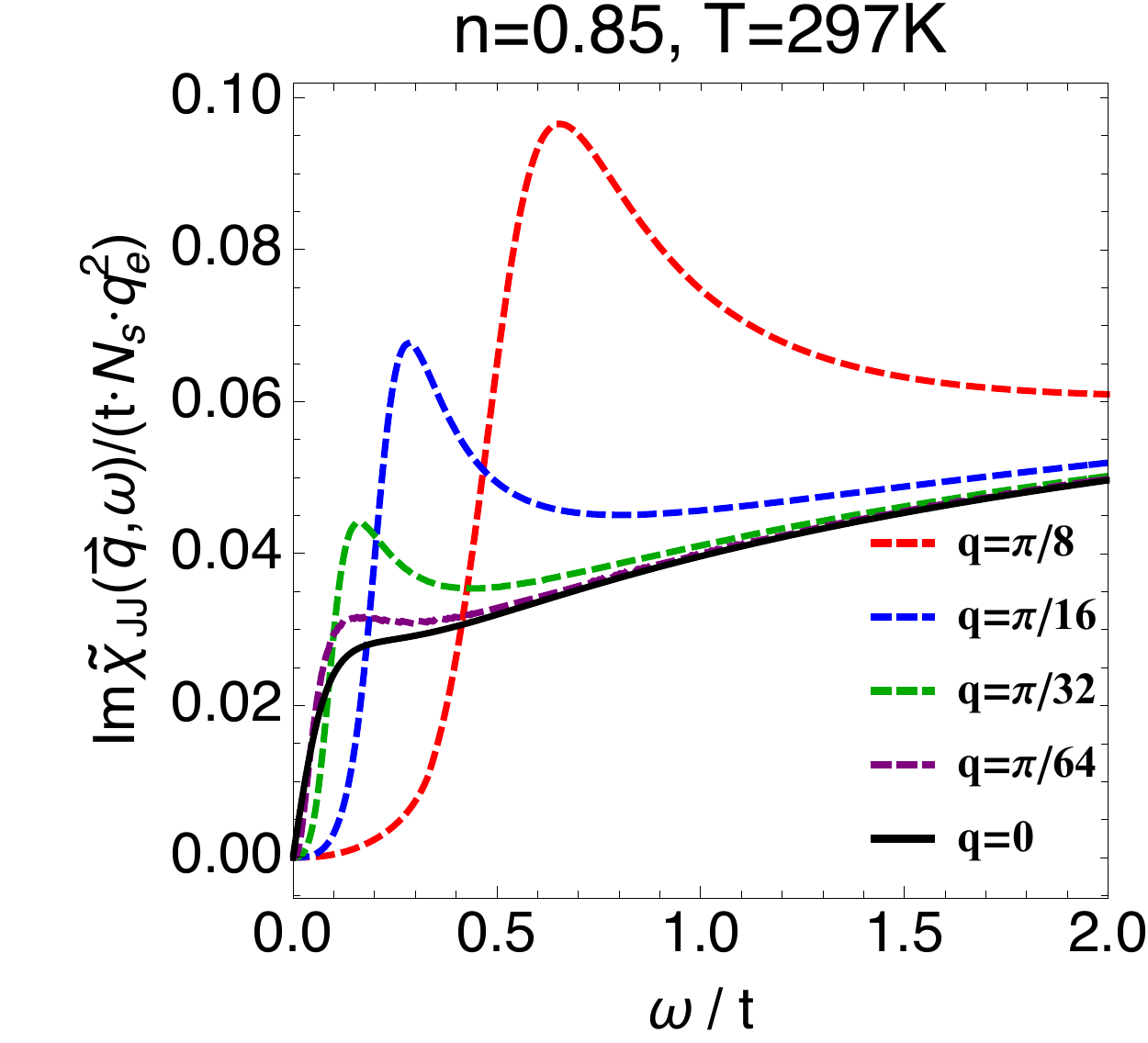}}
 \subfigure[\;\;]{\includegraphics[width=0.27\columnwidth]{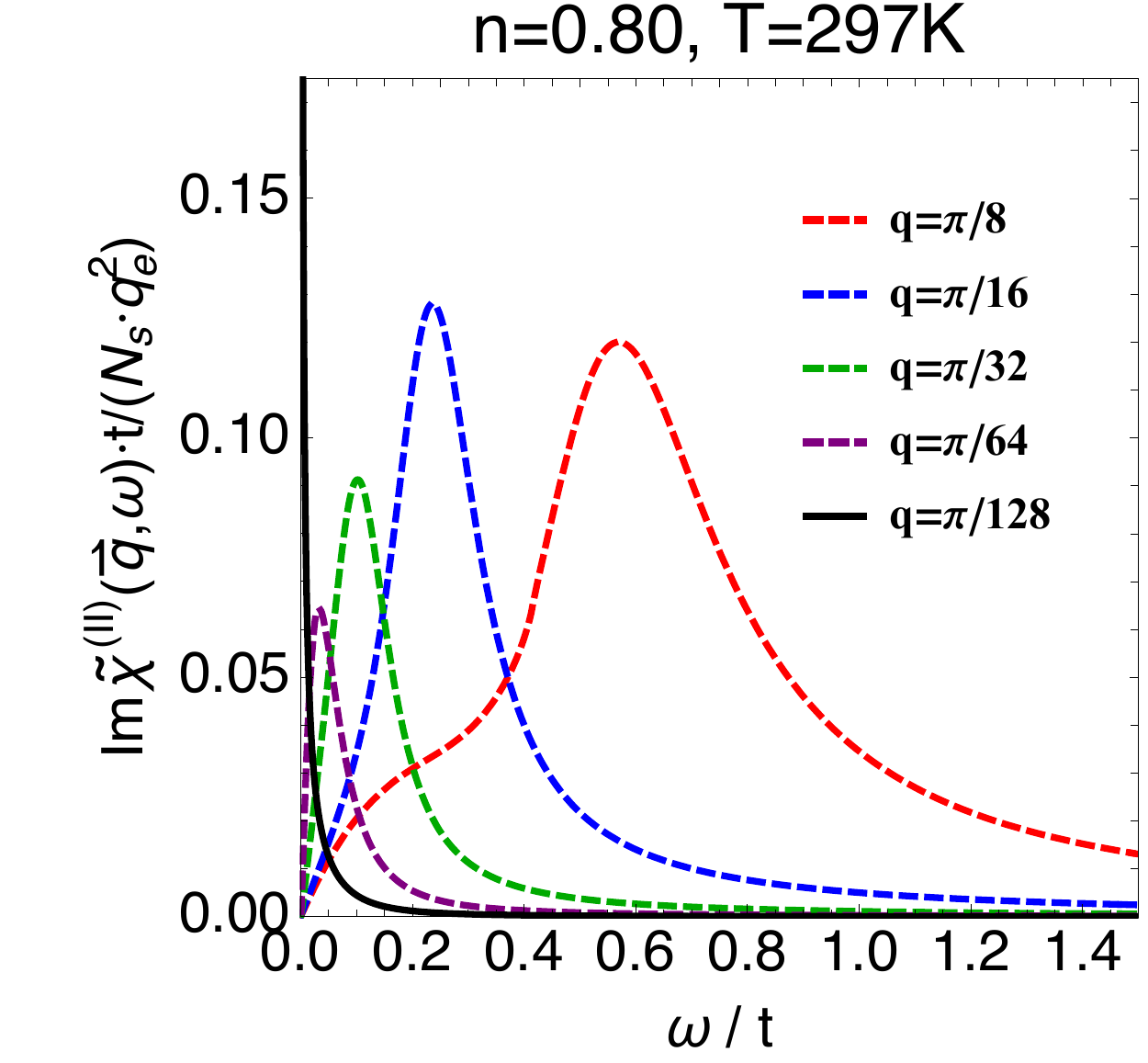}}
 \subfigure[\;\;]{\includegraphics[width=0.27\columnwidth]{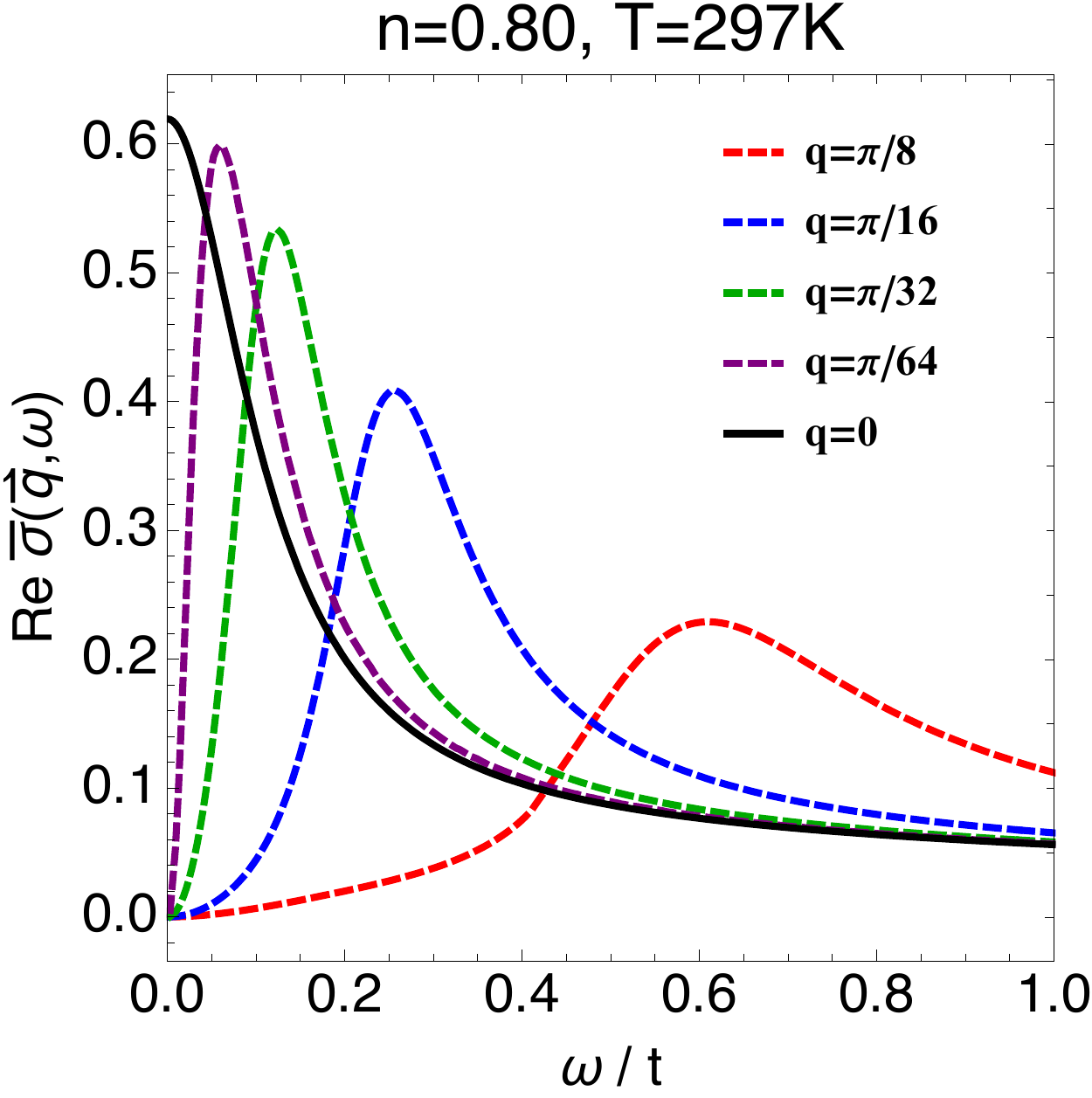}}
 \subfigure[\;\;]{\includegraphics[width=0.27\columnwidth]{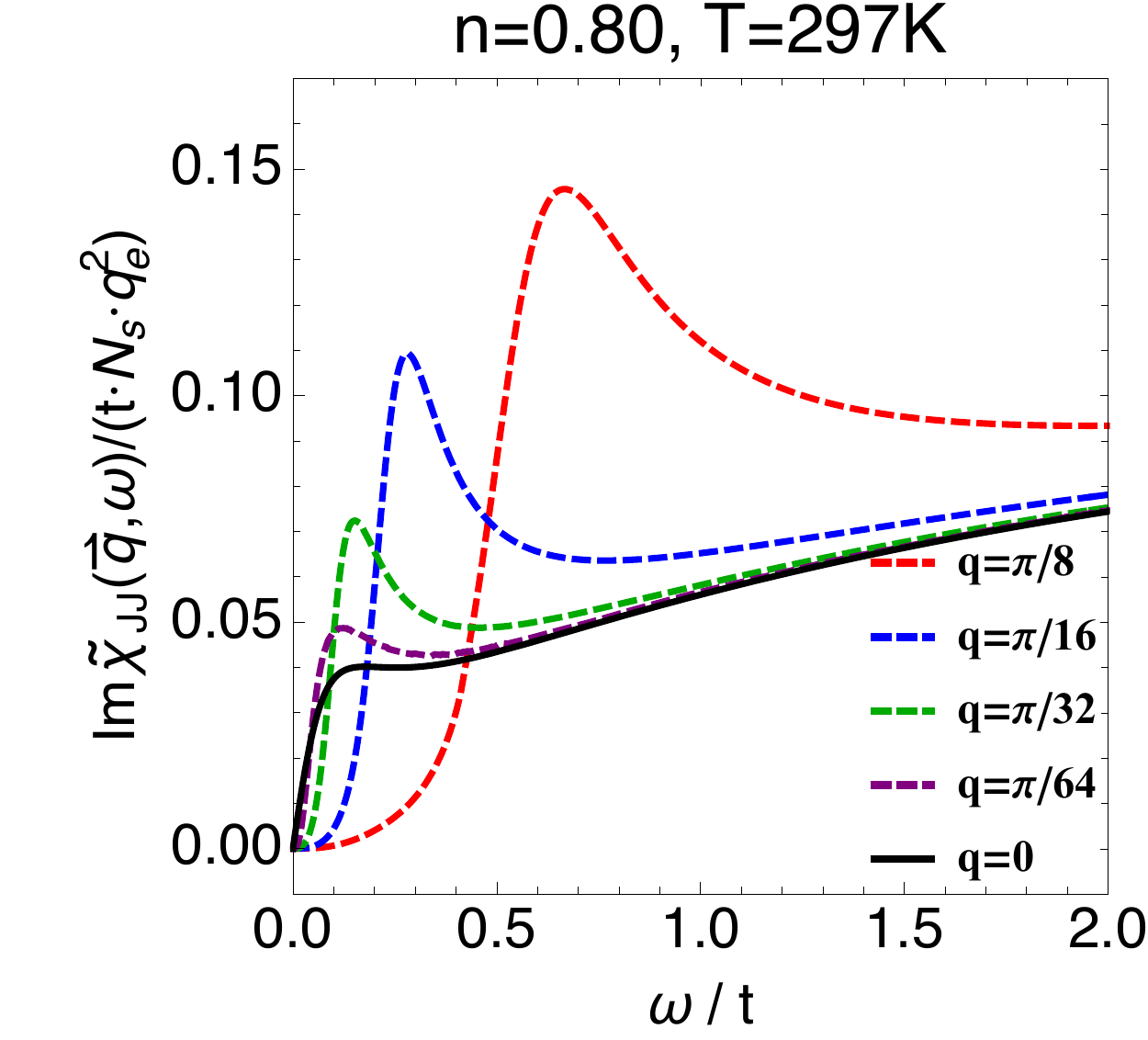}}
 \caption{\footnotesize Three  variables $\Im \hatchi^{(II)}_{\rho\rho}(\vq,\omega)$, $\Re \, \bar{\sigma}(\vq,\omega)$ and $\Im \hatchi_{JJ}(\vq,\omega)$,  closely  interrelated through \disp{rel-all}, each exhibiting   peaks as functions of $\omega$, are compared  at  $T=297$K, and     $n=$$0.85$ and $n=$$0.80$.   The wavevector $\vec{q}=(q,q)$ lies  along $\Gamma \to X$. This computation used $L_x \times L_y = 128\times128$. Panels ({\bf a,d}) display    the density susceptibility  $\Im \, \hatchi^{(II)}_{\rho\rho}(\vq,\omega) $ (\disp{appxII,briefly-hatchi}), panels ({\bf b,e}) display the dimensionless conductivity $\Re \, \bar{\sigma}$  (\disp{sigmabar,resist-q,cond-chiJJ-2d}), and panels   ({\bf c,f}) display the current susceptibility $\Im \hatchi_{JJ}$ (\disp{WtoJ,rel-J-rho}) with the displayed prefactors.  Temporarily ignoring constants $t,N_s,q_e$, the variable  in panel (b) is obtained from the variable in panel (a) by multiplying with $\omega/q^2$, and the variable in panel (c) is obtained from that in panel (b) by multiplying with $\omega$. Similar considerations hold for panels (d,e,f).  The flattening of the curves for $ \Im \, \hatchi_{JJ}$ for all $\vq$ beyond the peak imply that $\Im \, \hatchi_{\rho\rho}$ falls off as $1/\omega^2$ in that region. Such  a feature  was already noted in current experiments \cite{MEELS-1,MEELS-2,MEELS-3}.  The solid black lines for $\vq=0$ in panels ({\bf b},{\bf e})  and ({\bf c},{\bf f}) are separately computed using the current vertex as defined by Eq.~(2) of \refdisp{ECFL-Raman}.   }
 \label{chiJJT297K}
\end{figure*}

\begin{figure*}[h]
 \centering
 \includegraphics[width=.3\columnwidth]{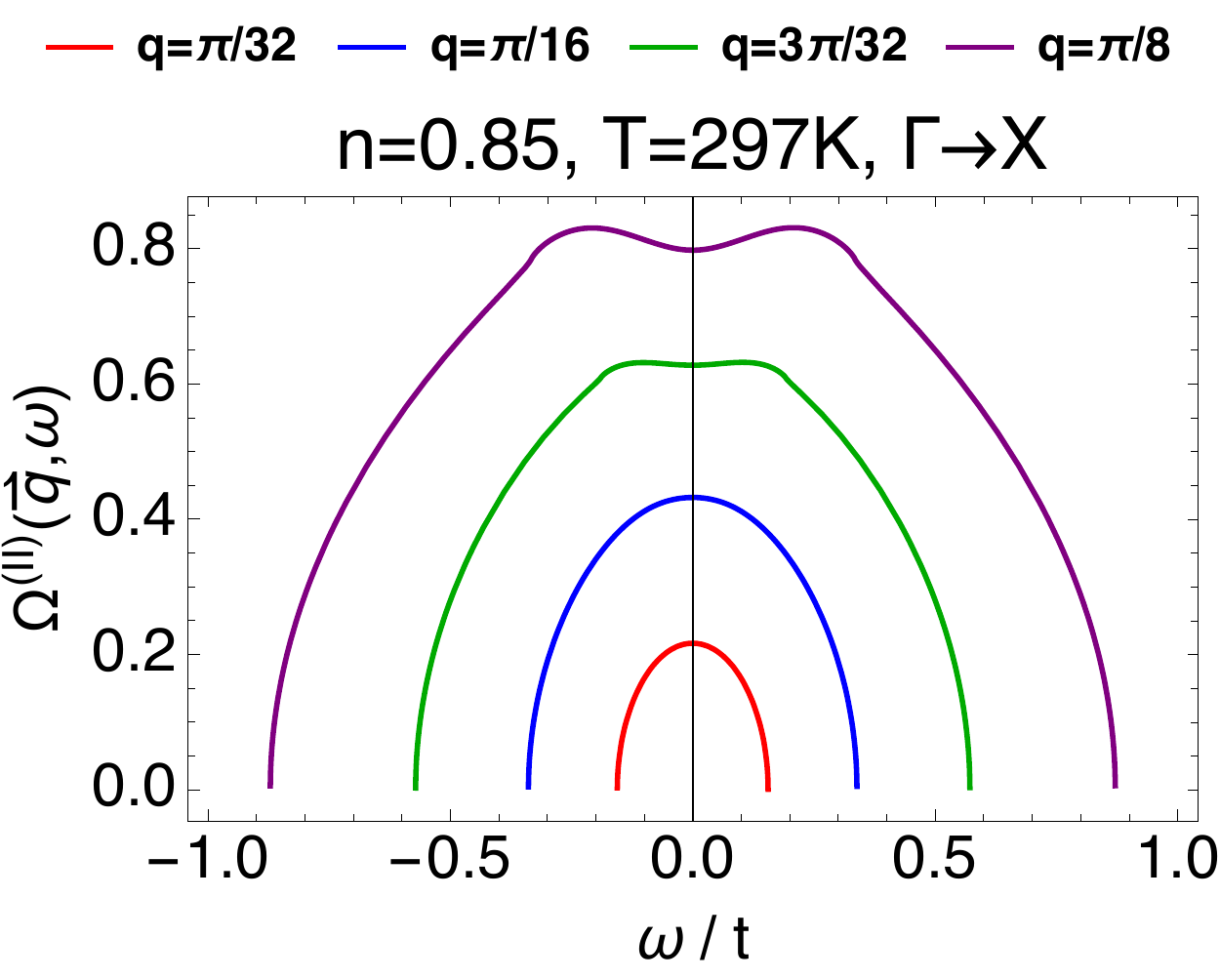}
 \caption{\footnotesize The characteristic energy scale $\Omega(\vq,\omega)$ \disp{Omega} in units of $t$. Here $\Omega^{(II)}$  is found from the peak frequency using \disp{appxII}. Here $n=0.85$ and $T=297$K and $\vec{q}=(q,q)$. The peaks in $\Im \, \hatchi_{\rho\rho}(\vq,\omega)$ are  found from \disp{Omega-p-exact}, or approximately at the  energy $\Omega_p(\vq) \sim \Omega(\vq,0)$, i.e. the $\omega=0$ intercept in the above curves. The intercepts therefore represents the peak energy scale observed in \figdisp{chiGAMMAtoX}. }
 \label{BigOmega}
\end{figure*}
 
\begin{figure*}[h]
 \centering
 \subfigure[\;\;]{\includegraphics[width=0.36\columnwidth]{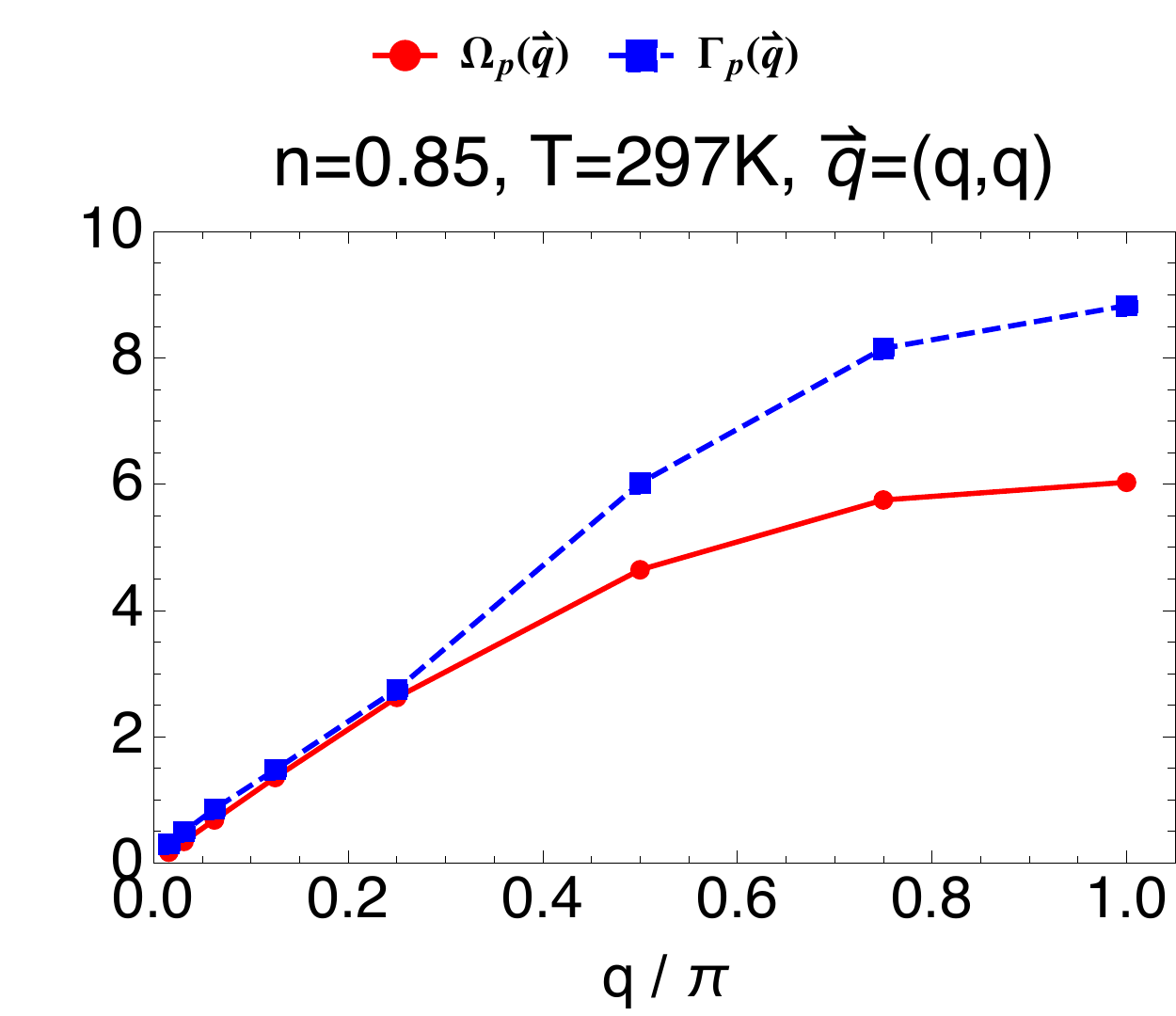}}
 \caption{\footnotesize   The approximate  theoretical peak energy scale $\Omega_p(\vq)$ (\disp{Omega-p,emergentOmega-2}) (obtained by setting  $\Omega(\vq,\omega\to0)$ ) and the width of the peaks $\Gamma_p(\vq)$ (\disp{Gamma-p}). These two scales are enough to reconstruct the peak in  the irreducible susceptibility $\Im \, \hatchi_{\rho\rho}$  using \disp{im-chi-2}. This plot indicates a  peak structure  for small $\vq$. For  higher $q$  the breadth exceeds the peak frequency,  as seen explicitly in \figdisp{chiGAMMAtoX,chiJJT297K}.
  \sr{ $\Omega_p$ are $\Gamma_p$ are  calculated using $\wt{\w}^{(1)} = \wt{\w}^{(1)}_B$ , the thermodynamic variable $\frac{dn}{d\mu}$ (\figdisp{fig-compressibility}). $\Gamma_p$, and $\Psi = \Psi_B + \delta \Psi_{QP}$, where the self energy is defined from the susceptibility $\hatchi_{\rho\rho}$ in \disp{new-rep-2,Omega,gamma}.} These computations use $L_x \times L_y = 128 \times 128$, $n=0.85$, $T=297$K and $\vec{q}=(q,q)$ along $\Gamma \to X$.
 } 
 \label{Omega-p-comp}
\end{figure*}

\begin{figure*}[h]
 \centering
 \subfigure[\;\;]{\includegraphics[width=0.36\columnwidth]{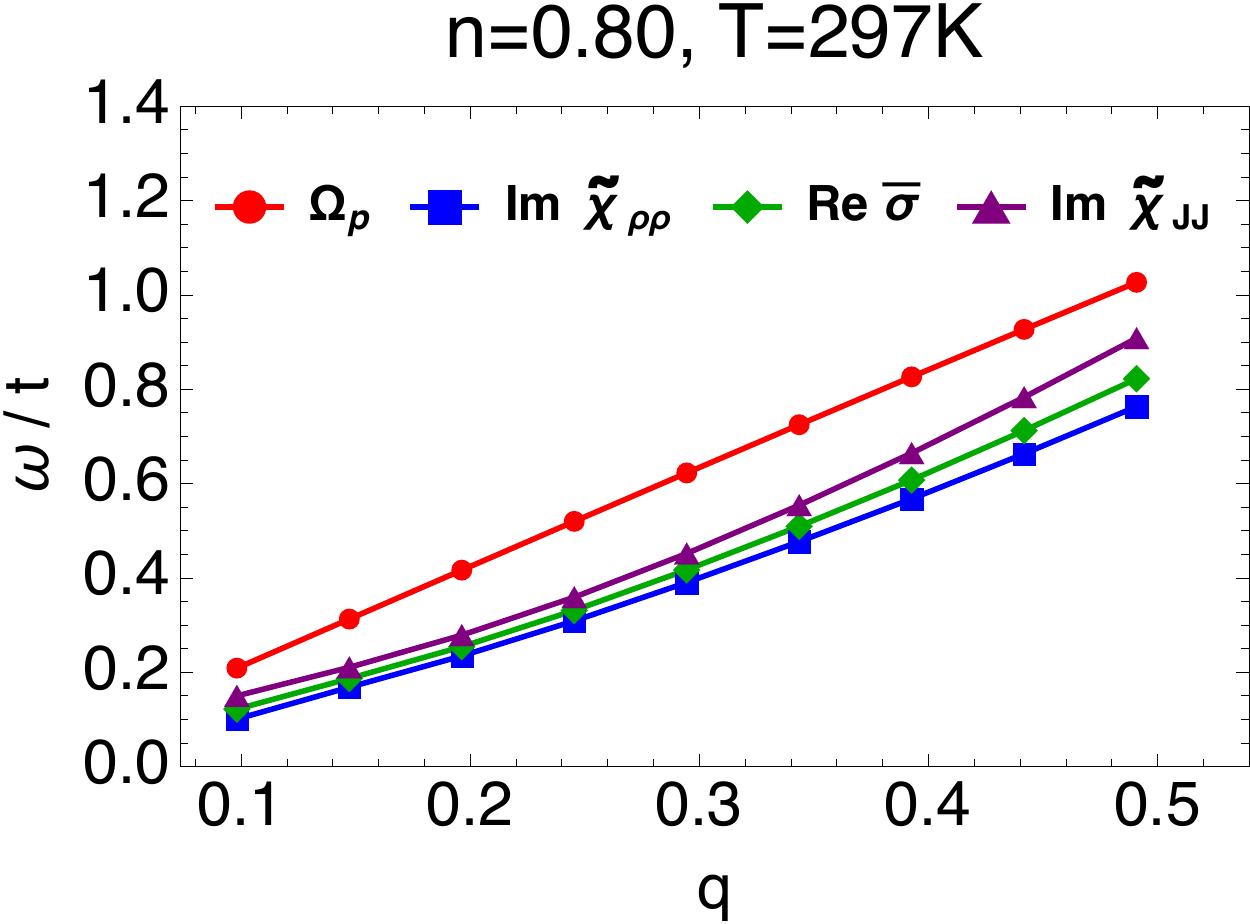}}
 \subfigure[\;\;]{\includegraphics[width=0.36\columnwidth]{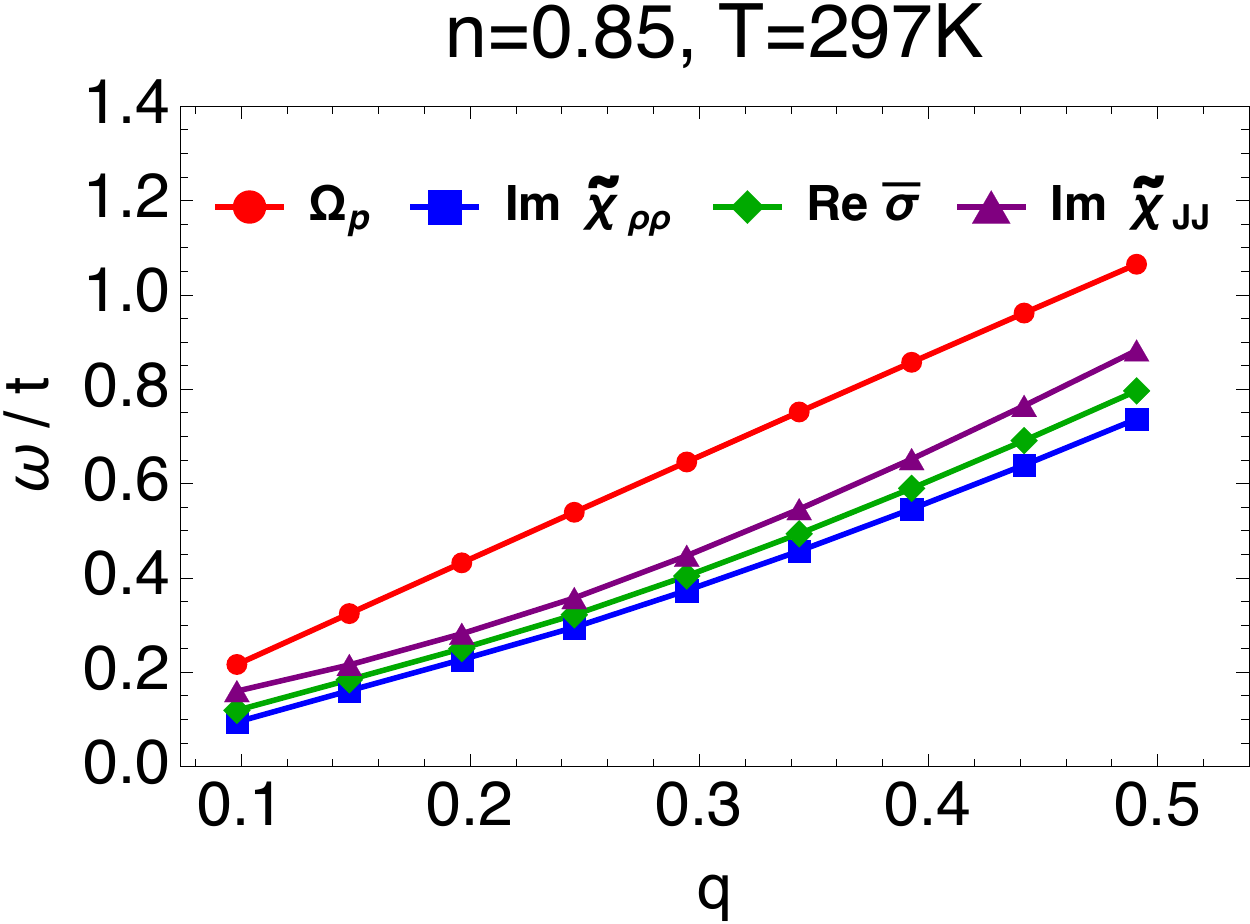}}
 \subfigure[\;\;]{\includegraphics[width=0.36\columnwidth]{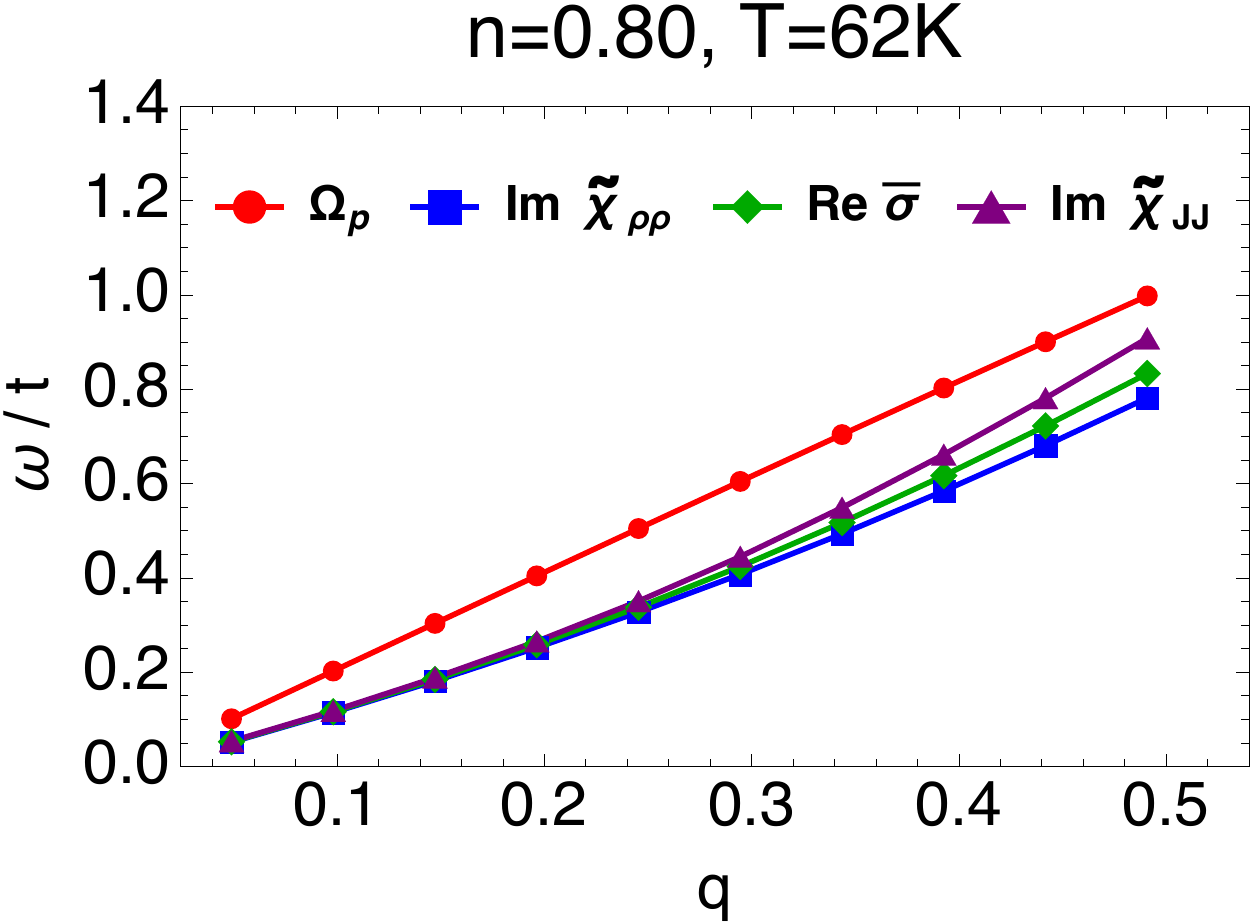}}
 \subfigure[\;\;]{\includegraphics[width=0.36\columnwidth]{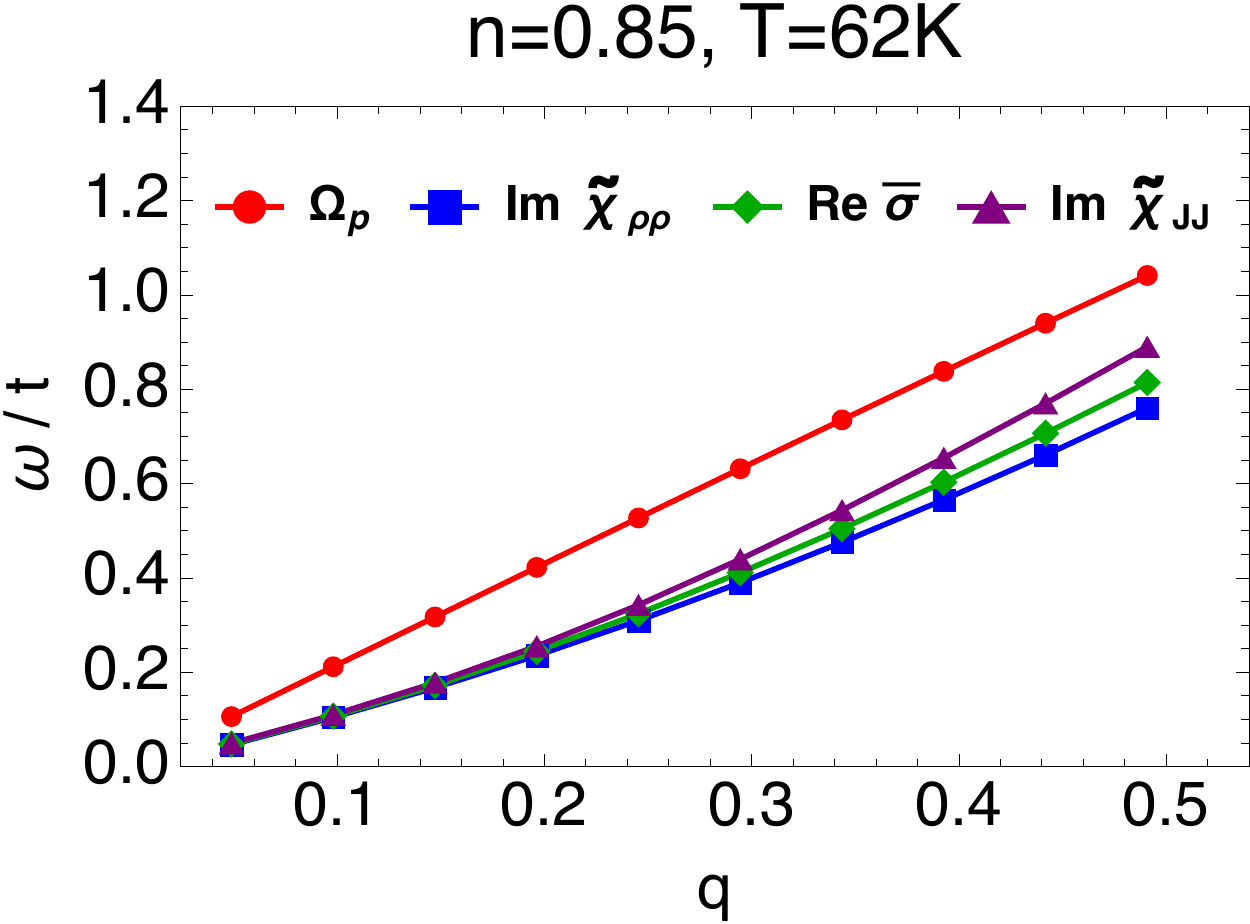}}
 \caption{\footnotesize The approximate  theoretical peak frequency  $\Omega_p(\vq)$ (red) \disp{Omega-p,emergentOmega-2} and the exact peak frequencies  extracted from \figdisp{chiJJT297K} for $\Im \,\hatchi_{\rho\rho}$ (blue), $\Re \, \bar{\sigma}$ (green) and $ \Im \, \hatchi_{JJ}$ (purple),
 with $\vq=\pi \{q,q\}$. This computation used $L_x \times L_y = 128 \times 128$.  The lowest frequency is at $q=\pi/64$ and found to be $0.031$t, $0.027$t, $0.052$t, $0.046$t for panels 
(a,b,c,d) respectively. If we choose $t=0.45$eV, the lowest values of the peak energies lie
between 12meV and 21meV.
 }
 \label{zeroes}
\end{figure*}

 \section{Calculations, Results and Discussion \label{sec4.2}}
 We first summarize the parameters used in our calculations.
We calculate the Green's functions using the set of formulas summarized in \appdisp{App-ECFL} \disprange{eq:ECFL1}{inverseg2nd}, employing the set of band and model parameters
\beq
t=0.45eV, \; t'=-0.20\, t , \; J=0.17\, t,\label{paras-1}.
\eeq
The system sizes used in most of the presented calculations are
\beq
N_\omega=2^{14}, \; L_x\times L_y=64\times 64 \nonumber \mbox{ (correlated  model)} \\
  \label{paras-2}
\eeq
where $N_\omega$ is the number of $\omega$ points in the frequency grid and $L_x,L_y$ are the dimensions of the 2-d lattice. For calculations of the reference uncorrelated model, we use bigger spatial grids $L_x\times L_y=128\times 128$. We present results at a few representative temperatures, and focus on two densities $n=0.80$ and $n=0.85$, corresponding  to the well studied  over-doped regime and optimally doped cases in the family of cuprate superconductors. We now present the results from this formalism, and provide some discussion of these.

 \subsection{ Basic results from ECFL on spectral function, momentum distribution and compressibility \label{Sec6.1}}
 We begin by illustrating the basic results of the ECFL theory for $A(\vk,\omega)$ the single electron spectral function, and $m_k$  the momentum distribution function, which display the strong redistribution of spectral weight from a Fermi gas due to correlations. This is followed by showing the compressibility within ECFL. The compressibility is  reduced considerably from  the Fermi gas  due to Gutzwiller type correlations, as argued originally in\cite{Rice-Brinkman,Vollhardt}.
 
\begin{itemize}
\item{ \figdisp{spectrum} shows  the electronic spectral function $A(\vec{k},\omega)$ obtained by solving for the ECFL Green's function by methods that are  elaborated upon in Appendix (A) . The quasiparticle weight $Z_{k_F}$ (abbreviated as $Z$ is seen to be very small $ Z= 0.06, 0.09$ for $T=99$K and $T=297$K respectively. The area sum-rule for the lower Hubbard band spectral function reads as: $\int d\omega A(\vk,\omega)= 1- \frac{n}{2}$, it is satisfied by depleting the quasiparticle peak, and smearing it over a wide background. This redistribution of weight accounts for the broad and featureless background seen in the spectral functions, it is a reflection of the strong local correlations. The insets  show the spectral function at $k_F$ against $\omega/t$,  over a  wide  energy scale. 
They show that    the small area  under the quasiparticle peak at $\omega\sim0$, due to a tiny  $Z_{k_F}$,  is  compensated by broad features at very high excitation energies $\sim10$t.  Analogous figures for the spectral function at other densities and temperatures over an wider energy window for this theory can be found in \cite{ECFL-2d-Mai-a} (Figs.~(1,2))
    The spectral width depends sensitively on T. This thermal sensitivity is a characteristic of the ECFL theory, where the effective Fermi temperature is much suppressed from the band value. }
  
\item  {  In \figdisp{occ-num} we display the momentum distribution function $m_k = \langle \wt{C}^{\dagger}_{k \uparrow} \wt{C}_{k \uparrow}\rangle$ found from \disp{mk}, together with the analogous   $n_k$ for uncorrelated electrons.  The reduced quasiparticle weight is also reflected in a small (Migdal) jump\cite{Nozieres,AGD,Brinkman-Rice} in the momentum distribution function $m_k$.  The ECFL theory satisfies the Luttinger-Ward theorem and hence the Fermi surface (FS) is unshifted by interactions. In this figure 
       a drastic reduction of the quasiparticle weight $Z_{k_F}$ is evident from the flattening of the correlated distribution $m_k$ in this figure. Certain weighted averages of $m_k$ are required for computing the    function $\frac{1}{t q_e^2}\kappa(\vq)$ (\disp{kappa}), or upon using \disp{moment-2},  the first moment $\widetilde{\omega}^{(1)}(\vq)/t$. These are tabulated in Table \ref{Table1}.
   } 

\item { In \figdisp{{fig-compressibility}} we plot the compressibility using the thermodynamic result  \disp{compressibility-2}, as a function of hole doping $\delta = 1 - n$. Correlations are seen to suppress the compressibility as $\delta$ decreases towards the insulating limit, relative to the compressibility of the free Fermi gas.  We also show the static uniform limit of the susceptibility $\frac{1}{q_e^2 N_s}\lim_{\vec{q}\to0}\hatchi^{\mbox{\tiny Bub}}_{\rho\rho}(\vec{q},0)$ (\disp{Inter-1}). If an exact calculation, going beyond the bubble approximation  were possible, the two curves would coincide, thanks to the sumrule \disp{compress-1}.   We see  that the discrepancy   is quite small  at low  $|\vq|$.   } 
\end{itemize}

\subsection{Static susceptibility and the first moment $\widetilde{\omega}^{(1)}(\vq)$ or $\kappa(\vq)$ \label{Sec6.2}}

\begin{itemize}
\item {  In \figdisp{static} we display the wave vector dependence of the static charge susceptibility  $\hatchi_{\rho \rho}(q,0)$  and compare with  the bare (uncorrelated) static susceptibility using \disp{RPA-1}. Correlations are seen to suppress the magnitudes of the susceptibilities as expected. Somewhat unexpectedly, the relative locations of the three curves for the correlated system, corresponding to different direction in the k-space \sr{undergoes a reshuffle relative to the band susceptibilities. At small q, the $\Gamma \to M$ and $M \to X$ curves are inverted, and the $\Gamma \to X$ also flips. }
} 

\item {  In \figdisp{chiBubWWstatic}, we compare the dimensionless functions $\frac{1}{t q_e^2}\kappa(\vec{q})$ from \disp{kappa} and $\frac{1}{t q_e^2 N_s} \hatchi^{\mbox{\tiny Bub}}_{WW}(\vec{q},0)$ from \disp{Inter-2} plotted over the Brillouin zone. 
 \sr{In \disp{constraint-chiWW} we noted that  the  identity of these objects is required  in an exact theory,  \figdisp{chiBubWWstatic} verifies that the present calculation satisfies this identity exactly at small $\vq$, and   fairly well  over the entire zone zone. }
     }
 
 
 \item  In Fig.~(\ref{sumrule-dispersion}) we displays $\frac{1}{t q_e^2}\kappa(\vq)$  \disp{kappa}, which is equivalent to the first moment $\widetilde{\omega}^{(1)}(\vq)/t$, and also  the 2-d plasmon spectrum( \disp{plasmons-4,3d-plasmon,Coulomb-2d}). The plasmon displays the expected acoustic $\sqrt{|\vq|}$ behavior at low $\vq$ of 2-dimensional plasmons. This feature is followed by a broad continuum at an energy  scale $\omega_p\sim 1.50$t, which is considerably lower than the energy scale without interactions.  

\end{itemize}

\subsection{Irreducible susceptibility $\Im \hatchi_{\rho\rho}(\vq,\omega)$ \label{Sec6.3}}
\begin{itemize}
\item \sr{ We next display in \figdisp{Psi} results for the two successive approximations to the irreducible  susceptibility $\hatchi^{(I)}$ in \disp{appxI}    and $\hatchi^{(II)}$ in \disp{appxII}. These  are constructed using three building blocks (i)
the static susceptibility $\hatchi_A(\vq,0)$ (ii) the plasma frequency $\widetilde{\omega}^{(1)}$ and  (iii) the self energy $\Psi(\vq,\omega)$. The first two  are common, while the third, i.e. the self energy, distinguishes between the two approximations; $\hatchi^{(I)}$ uses the self energy  $\Psi_B(\vq,\omega)$
while $\hatchi^{(II)}$ use  self energy     $\Psi_B(\vq,\omega)+ \delta \Psi_{QP}(\vq,\omega) $. 
Panel (a) shows the imaginary part of these two self energies
  From  these objects  we 
compute its real part using the causality condition \disp{redeltaPsi}. The real parts of these three susceptibilities are  shown in panel (b,e,f). }
 In comparing panels (c) and (e) we clearly see the linear in $\omega$ regime near the origin due to the quasiparticle contribution, which in turn creates the double minimum in the real part seen in  panels (b) and (f).


\item 

  In \figdisp{Psi-scaling} panel (a) we display $\Im \,\Psi_B$  (multiplied by a scale factor $\widetilde{\omega}^{(1)}(\vq)$) at different $\vq$ as functions of $\omega$. We observe that  these collapse to a single curve over the Brillouin zone, when multiplied by $\widetilde{\omega}^{(1)}(\vq)$ (\disp{moment-expansion-1} and in Fig.~(\ref{sumrule-dispersion}.b). The other self energy,  $\Im \{\Psi_B + \delta\Psi_{QP}\}$ at different $\vq$ differ in the low $\omega$ region, due to the presence of the quasi-particle contributions, but do collapse to a single curve at higher frequencies, as seen in panel (b)


\item  { In \figdisp{chiIvschiII} we compare two approximations for the imaginary part of the irreducible (screened) susceptibilities $\Im \, \hatchi^{(I)}$ (solid red line) and $\Im \, \hatchi^{(II)}$ (blue dashed lines), i.e., \disp{appxI,appxII}. As expected the quasiparticle contribution at low frequencies is roughly linear in $\omega$. If we neglect that regime, 
the two approximations lead to similar results, as seen  for $\omega\gssim 0.40t$. The inset shows that the  corresponding non-interacting  complex susceptibility (see \disp{RPA-1}) for  the same  parameters, extends to  much higher frequencies $\omega/t$, and have  different vertical scales and shapes.
    }

\item   In \figdisp{chiGAMMAtoX}
 we show  the $\vq,\omega$ variation of the  imaginary part of the irreducible  susceptibility $\Im \, \hatchi^{(II)}$ \disp{appxII}. \sr{ We show the density and temperature evolutions of the screened susceptibility approximations $\hatchi^{(I)}$ (dashed) and $\hatchi^{(II)}$ (solid) over the ranges $n=0.8,0.85$ and $T=99,198,297$K in the  direction $\Gamma \to X$ (see \figdisp{chiGAMMAtoX}). In all cases we observe that the high $\omega$ fall off of $\Im \, \hatchi$ is $\sim \frac{1}{\omega^2}$, while the curves turn-around at low frequencies to vanish as $\omega \to 0$.}
 
  The significant features from the  $\Im \, \hatchi^{(I)}$ (in \disp{appxI}) are qualitatively similar.
  Each  curves exhibit a $\vq$ dependent peak at an energy  $\Omega_p(\vq) $ from \disp{Omega-p}. The peak shifts towards lower energies as $q$ is reduced, and for a fixed $q$ the intensity drops rapidly with a modest increase of  $T$. The peak energy is a  (measurable) characteristic energy scale, and discussed further in \figdisp{chiJJT297K,BigOmega}. We also note the approximately $\sim \frac{1}{\omega^2}$ falloff of this function beyond the peak frequency. This falloff is similar to that seen in experiments \cite{MEELS-1,MEELS-2,MEELS-3},
  and we correlate this behavior with that seen in optical conductivity and the current-current susceptibility  in \figdisp{chiJJT297K}. We discuss the connection with a related feature observed  Raman scattering  below in Sec.(\ref{Sec6.6}).
    
\end{itemize}
\subsection{Dielectric function $\varepsilon(\vq,\omega)$ \label{Sec6.4}}
\begin{itemize}
\item 
In \figdisp{epsIepsII} we display the approximate dielectric functions, computed from \disp{app-dielectric,gc}, in the form of $-\Im \frac{1}{\varepsilon(\vq,\omega)}$ at two different values of the Coulomb coupling $g_c$. The effective Coulomb coupling constant $g_c$  \disp{gc} involves a combination of material parameters $t,a_0,\varepsilon_\infty$. In the BSCCO material used in the experiments of \cite{MEELS-1,MEELS-2}, using  $t\sim 0.45$ eV, $a_0$$\sim3.81$$\AA$ and   $\varepsilon_\infty$$\sim$$4.5$, we find $g_c$$\sim 11.5$, while  using $t\sim 0.16$eV  gives $g_c\sim$$32.0$.  We provide a results for a few typical values of this parameter, since the basic parameters  vary for different  materials. The variable $-\Im \frac{1}{\varepsilon(\vq,\omega)}$ is directly measured in (inelastic) electron loss type experiments in typical metallic systems. The significance of this variable is that  any peaks signify plasmons. 
We also show the calculated    $\Re \, \varepsilon(\vq,\omega) $, which is also   used to identify plasmons through its zero crossing in  certain optical experiments. 
 From this figure we note that unlike in the RPA calculation\cite{MEELS-2,Fetter-Walecka} for uncorrelated materials, $\Re\{\varepsilon\}$ crosses the zero line only for large $g_c$.

 \item In \figdisp{epsilon,epsilongc}, we show the imaginary part of inverse dielectric function (\disp{app-dielectric}) at $n=0.85$ and $T=297$K   over the ranges $q =\pi/16,\pi/8,3\pi/16$ with $\vq=\{q,q\}$. We also illustrate the dependence on $g_c$ using $g_c =10,50,100$. We note  that when $g_c$ is large, the peaks  are broadened out considerably and pushed out to higher energies, as compared to smaller $g_c$.

\end{itemize}

\subsection{Reducible susceptibility $\bigchi_{\rho\rho}(\vq,\omega)$  \label{Sec6.5}}
\begin{itemize}

\item \sr{  In \figdisp{reducible-chi} we show the   reducible susceptibility $ \Im \, \bigchi_{\rho\rho}$. From \disp{structure-direct} we  note that $\Im \, \bigchi_{\rho\rho}$ is the most directly accessible (i.e. raw) object in inelastic electron scattering experiments, and therefore of considerable interest. 
It is 
 related  to $-\Im \, \{ 1/\varepsilon \}$ plotted in \figdisp{epsIepsII,epsilon,epsilongc} via the relation 
$ \Im \, \bigchi_{\rho\rho} = -\frac{q_e^2}{V_{\vq}} \, \Im \, \{ 1/\varepsilon \} $ from \disp{dielectric-1}. The peaks are located at 
the same frequencies, since the factor connecting them is independent of $\omega$. We note that its connection with the irreducible susceptibility,
\beq\Im \, \bigchi_{\rho\rho}(\vq,\omega) = \frac{ \Im \, \hatchi_{\rho\rho}(\vq,\omega)}{\{ 1+\frac{V_{\vq}}{q_e^2} \Re \, \hatchi_{\rho\rho}(\vq,\omega)\}^2+\{\frac{V_{\vq}}{q_e^2} \Im \, \hatchi_{\rho\rho}(\vq,\omega)\}^2 }, \label{connect} \eeq 
involves an $\omega$ and  (Coulomb constant) $g_c$ (\disp{gc}) dependent denominator. This term
causes the peaks of $\Im \, \bigchi_{\rho\rho}$ to be shifted upwards substantially with respect to those of $ \Im \, \hatchi_{\rho\rho}$. The shift is also strongly dependent on the Coulomb constant $g_c$ \disp{gc}, as seen in \figdisp{epsilongc}. The peaks in the reducible susceptibility \figdisp{reducible-chi} or of \figdisp{epsIepsII,epsilon,epsilongc} are at $\omega$$\sim t$ while those of the irreducible susceptibility in 
 \figdisp{chiGAMMAtoX} are at about $\omega$$\sim 0.15$$ t$ at similar wave vectors. Here we also note a decrease in magnitude of the peak height as $q\to0$. This  is a direct  consequence of the conservation of charge, and serves as a  constraint on  experiments.}

\sr{ The theoretical calculation  of either approximation to the screened susceptibility  $\hatchi_{\rho\rho}$ {\em does not depend} on $g_c$, while the unscreened   $\bigchi_{\rho\rho}$ (inferred from \disp{relations-0} or  \disp{connect}) does so. This implies that uncertainties in the theory or in $g_c$ are magnified in   $\bigchi_{\rho\rho}$. In this sense we might say that $\hatchi_{\rho\rho}$ is the raw theoretical variable.}
 
 \sr{ It is amusing to note that experiments face a converse of the theoretical problem. The measured scattering intensity yields the reducible susceptibility  $\bigchi_{\rho\rho}(\vq,\omega)$, and the extraction of the screened susceptibility  $\hatchi_{\rho\rho}(\vq,\omega)$ requires amongst other assumptions, an estimate of the material dependent Coulomb coupling $g_c$ (from \disp{app-dielectric,gc}). This observation motivates our exploration of a varying the  values of $g_c$. In \figdisp{reducible-chi} we  observe that when $g_c$ is large, the peaks in $ \Im \, \bigchi_{\rho\rho}$ are broadened out considerably and pushed out to higher energies. }

\end{itemize}
\subsection{The variables $\Im \, \hatchi_{\rho\rho}$, $\Re \, \bar{\sigma}$ and $\Im \,\hatchi_{JJ}$ \label{Sec6.6} }
\begin{itemize}
\item    In \figdisp{chiJJT297K} we display these closely related triad of variables, $\Im \, \hatchi_{\rho\rho}(\vq,\omega)$, $\Re \, \bar{\sigma}(\vq,\omega)$ and $\Im \, \hatchi_{JJ}(\vq,\omega)$, which are related through \disp{rel-all}.
 Panels ({\bf a,d}) display    the density susceptibility  $\Im \, \hatchi^{(II)}_{\rho\rho}(\vq,\omega) $ (\disp{appxII,briefly-hatchi}), panels ({\bf b,e}) display the dimensionless conductivity $\Re \, \bar{\sigma}$  (\disp{sigmabar,resist-q,cond-chiJJ-2d}), and panels   ({\bf c,f}) display the current susceptibility $\Im \hatchi_{JJ}$ (\disp{WtoJ,rel-J-rho}) with the displayed prefactors.  Temporarily ignoring constants $t,N_s,q_e$, the variable  in panel (b) is obtained from the variable in panel (a) by multiplying with $\omega/q^2$, and the variable in panel (c) is obtained from that in panel (b) by multiplying with $\omega$. Similar considerations hold for panels (d,e,f).  The  multiplicative factor changes the low $\omega$ behavior of the three variables, and it also affects the location of  the peak frequencies  are slightly shifted from  the theoretical $\Omega_p(\vq)$ \disp{Omega-p}. We explore this shift more closely in the following section.
 
   The evolution of the theoretically calculated $\Im \hatchi$ is quite  complex at low $\vq,\omega$. On the other hand  the theoretical     conductivity $\Re \, \bar{\sigma}$ and
 the current-current susceptibility  $\Im \hatchi_{JJ}$  evolve relatively more smoothly 
 with $\vq$. From this observation we expect that  these curves might serve as  guides  for interpolation  in $\vq$.

\sr{ The above observation suggests  that   Raman scattering experiments and optical conductivity experiments, which probe small $\vq$, could be useful. Assuming smoothness in $\vq$, these experiments can be perhaps useful in constraining the inelastic electron scattering data.
 Firstly we note note that  non-resonant inelastic Raman scattering data in the $B_{2g}$ geometry (e.g. see Fig.6  of \cite{Sugai})  shows a close correspondence with optical experiments  \cite{Sugai,Girsh}. Assuming this relation  one can obtain a rough estimate of $\hatchi_{JJ}$  from Raman measurements \cite{Sugai,Sugai2,Girsh}. The flattening of the theoretical curves for $ \Im \, \hatchi_{JJ}$ for all $\vq$ beyond the peak, also seen in Raman data of \cite{Sugai} at low $\vq$,  imply that $\Im \, \hatchi_{\rho\rho}$ falls off as $1/\omega^2$ beyond any features.   This    is consistent with the observations in  current experiments \cite{MEELS-1,MEELS-2,MEELS-3},  at essentially any $\vq$.  }

\sr{ Another interesting variable is the   (independently measurable) optical conductivity at $\vq=0$, which in turns  evolves continuously from transport measurements at $\omega=0$. The present theoretical calculations show a smooth evolution with $\vq$  above the  peak at $\Omega_p(\vq)$ \cite{FN2}.  Therefore a systematic comparison at a set of $\vq$  of the  $\Re \, \bar{\sigma}(\vq,\omega)$ deduced from $\Im \hatchi(\vq,\omega)$ (by multiplying with $\omega/|\vq|^2$),  with the optical conductivity $\Re \, \bar{\sigma}(0,\omega)$ could be most helpful. One advantage is that  the deduced $\Re \, \bar{\sigma}(\vq,\omega)$ is  expected to be  more stable than  $\Im \, \hatchi_{\rho \rho}$ against low $\omega$ excitations or  noise. }

 \end{itemize}

\subsection{The  energy scale $\Omega_p(\vq)$ and peak frequencies  in  $\Im \, \hatchi_{\rho\rho}$, $\Re \, \bar{\sigma}$ and $\Im \,\hatchi_{JJ}$  \label{Sec6.7}}

\begin{itemize}

\item  {  In \figdisp{BigOmega} We display the  energy scale $\Omega(\vq,\omega)$ \disp{Omega} in units of $t$ using \disp{appxII}. The peaks in $\Im \, \hatchi_{\rho\rho}(\vq,\omega)$, denoted by $\Omega_p(\vq)$ (\disp{Omega-p}), are roughly given by   $\Omega_p(\vq) \sim \Omega(\vq,0)$ i.e. the $\omega=0$ intercept in the above curves. The intercepts therefore represents the peak energy scale observed in \figdisp{chiGAMMAtoX}. Experimentally $\Omega_p(\vq)$ can be inferred from a turn-around feature observed in the plots of $\Im \, \hatchi_{\rho \rho}(\vq,\omega)$, and potentially also in experiments. An explicit expression for the important energy scale $\Omega_p(\vq)$ in the limit of small $\vq$ is given  in \secdisp{sec-4.2.3} and \disp{emergentOmega}. In those sections we also provide an alternate and direct argument that leads to this scale, starting from the normalized spectral function of density fluctuations $\varphi(\vq,\omega)$ \disp{emergentPhi}. }

\item  {In \figdisp{Omega-p-comp} we show the approximate  theoretical peak energy scale $\Omega_p(\vq)$ (\disp{Omega-p,emergentOmega-2}) and the width of the peaks $\Gamma_p(\vq)$ (\disp{Gamma-p}). These two scales are enough to   reconstruct the peak in  the irreducible susceptibility $\Im \, \hatchi_{\rho\rho}$  using \disp{im-chi-2}, at least roughly. This plot indicates a  peak structure  for small $\vq$. For  higher $q$  the breadth exceeds the peak frequency,  as seen explicitly in \figdisp{chiGAMMAtoX,chiJJT297K}.
 } 

\item  {In \figdisp{zeroes}, we show the approximate theoretical peak frequency  $\Omega_p(\vq)$ (red) \disp{Omega-p,emergentOmega-2} and the exact peak frequencies  extracted numerically from \figdisp{chiJJT297K} for $\Im \,\hatchi_{\rho\rho}$ (blue), $\Re \, \bar{\sigma}$ (green) and $ \Im \, \hatchi_{JJ}$ (purple),
 with $\vq=\pi \{q,q\}$.  At the lowest $\vq=\{\pi/64,\pi/64\}$, for $n=0.85$ and T=297K  the exact peak energy (blue) is $\sim 0.027$t, i.e.  about a half of the approximate result (red). With $t=0.45$eV this gives a peak energy $\sim$12 meV, which seems to be at the threshold of  currently available  resolution.}


\end{itemize}



\section{Conclusions and Discussion \label{sec5}}

    We have presented results from our calculation of the dynamics of electron fluctuations in the \tJV model of \disp{model}. We see that the small quasiparticle weight in the normal state gives rise to a broad background in the electron spectral weight \figdisp{spectrum}. This in turn leads to a smearing of sharp features in the dynamical correlations, as we see in \figdisp{chiIvschiII}. The small $Z_{K_F}$ also reflects in the flattening of the momentum distribution, as seen in \figdisp{occ-num}.
 
  The plasmon energy $\omega_{p}(\vq)$ can be extracted in several distinct ways. We have discussed two methods already, from the peaks in $-\Im (\frac{1}{\varepsilon(\vq,\omega)}) $, or from the zeroes of $\Re \, \varepsilon(\vq,\omega)$ as seen in \figdisp{epsIepsII,epsilon}. There is yet another possibility, namely from a measurement of the first frequency moment of the structure function $S(\vq,\omega)$ as in  \disp{equality-1,moment-2}. Here the frequency integration must be large enough to contain all the weight from the primary band containing the Fermi level, but small enough to exclude interband effects. This balance is familiar from studies of optical conductivity in cuprates \cite{optics}, where satisfying the various versions of the $f$-sumrule involves parallel issues.
 
    The results for $\Im \, \hatchi_{\rho\rho}(\vq,\omega)$ presented in \figdisp{chiGAMMAtoX} display a slow fall off for $\omega>\Omega_p$ over a substantial range. This behaviour is similar to the fall off seen experimentally \cite{MEELS-3,MEELS-2}. From \disp{rel-all} this implies that the current susceptibility $\Im \, \hatchi_{JJ }(\vq,\omega)$ should flatten out in the same $\omega$ range. This is indeed seen in \figdisp{chiJJT297K} in panels (a,c). We should note that in the panels (b,d) of this figure, the conductivity shows a related sluggish fall off with $\omega$, consistent with \disp{rel-all}. 
 
In the region $|\omega|\leq \Omega_p(\vq)$, our calculations show that the quasiparticle contribution to $\Im \, \hatchi_{\rho\rho}(\vq,\omega)$ leads to a linear in $\omega$ behavior, as seen in the contrast between the two plots in \figdisp{chiIvschiII}, and in all the low $\vq$ plots of in \figdisp{chiGAMMAtoX}. A low magnitude of $Z_{K_F}$, as in the ECFL theory makes the linear regime small, but remain non-zero, and hence worth looking for in data. 
 
    Finally we believe that extracting systematically the energy scale $\Omega_p(\vq)$ for a range of small $\vq$ values is an important task for future experimental studies. In addition to tracking the peaks of the imaginary parts of the various susceptibilities noted in \disp{rel-all}, as well as \figdisp{chiJJT297K} and related figures, approximately evaluating the formula for the first frequency moment \disp{emergentOmega-2,emergentOmega-3} using data could provide a useful alternative. It is possibly a difficult task if the $\Omega_p(\vq)$ is not sufficiently larger than the experimental resolution, and if other sources such as phonons contribute strongly to the scattering intensity. Such a study would provide insight into the nature of the metallic state in the cuprates.




\section{Acknowledgements: \label{sec6}} 

    We thank P. Abbamonte for a helpful communication.  The work at UCSC was supported by the US Department of Energy (DOE), Office of Science, Basic Energy Sciences (BES), under Award No. DE-FG02-06ER46319. The computation was done on the comet in XSEDE \cite{xsede} (TG-DMR170044) supported by National Science Foundation grant number ACI-1053575.



\appendix

\numberwithin{equation}{section}

\setcounter{equation}{0}

\section{Summary of ECFL  Green's function $\G$ \label{App-ECFL}}
    The ${\cal O}(\lambda^2)$ approximation of the ECFL equations determining the Green's function for the \tJ model has been discussed earlier in our papers \refdisp{ECFL,ECFL-2d,ECFL-2d-Mai-a,ECFL-2d-Mai-b}, so we provide a very short summary of the equations used. In the ECFL theory, the one-electron Green's function $\G$ is found using the Schwinger method\cite{ECFL}, and expressed as a product of an auxiliary Green's function $\GH$ and a ``caparison'' function $\widetilde{\mu}$:
\beq \label{eq:product}
    \G(k) = \GH (k) \times \widetilde{\mu}(k)
\eeq
where $k \equiv (\vec{k}, i\omega_{k})$, and $\omega_{k} = (2k+1) \pi k_B T$ is the Fermionic Matsubara frequency and subscript $k$ is an integer. The auxiliary $\GH(k)$ is a Fermi-liquid type Green's function. The Schwinger equation of motion for the physical Green's function can be symbolically written as\cite{ECFL,ECFL-2d,ECFL-2d-Mai-a,ECFL-2d-Mai-b}
\beq
\left ( \GHI_{0} - \lambda \hat{X} - \lambda  {Y_1}\right ). ~\G = \delta \ ( \iden - \lambda  {\gamma} ). \label{Min-1}
\eeq
where $\hat{X}$ represents a functional derivative and $Y_1$ describes a Hartree-type energy. Here $\lambda$ is an expansion parameter and  set equal to unity after retaining all second order terms. The non-canonical nature of the Gutzwiller projected operators leads to the term $(\iden - \lambda  {\gamma} )$ on the right hand side, this would be just $\iden$ for canonical electrons. The decomposition in \disp{eq:product} circumvents this problem since $\GH$ is constructed so as to satisfy a canonical equation\cite{ECFL}.

To second order {(in $\lambda$)} the ECFL equations\cite{ECFL-2d,ECFL-2d-Mai-a,ECFL-2d-Mai-b} are found to be
\begin{align} 
    \widetilde{\mu}(k) &= 1 - \lambda \frac{n}{2} + {\lambda} \psi(k) \label{eq:ECFL1} \\
    \GH^{-1}(k) &= i\omega_{k} + \chem - \epsilon_{\vec{k}} + {\lambda} \frac{n}{2}\epsilon_{\vec{k}} - \lambda \phi(k)
    \label{eq:ECFL2}
\end{align}
where $\chem$ is the chemical potential  and $\epsilon_{\vec{k}}$ is the bare band energy \disp{dispersion} and  $\psi(k)$ is the second self-energy. The self-energy $\phi(k)$ factors out as $\phi(k) = \chi(k) +
\epsilon_{\vec{k}}'\psi(k)$ where $\chi(k)$ is another function defined below, $\epsilon'_{\vec{k}} =\epsilon_{\vec{k}} - u_0 /
2$, where $u_0$ is  a Lagrange multiplier. Both $\chem$ and $u_0$ are determined by constraining the number of electrons defined respectively using $\G$ and $\GH$ on \disp{eq:numsumrules}. The two self-energies functions $\psi$ and $\chi$ expanded formally in $\lambda$ to second order approximation $\mathcal{O}(\lambda^2)$ are $\psi =
 \psi_{[0]} + \lambda \psi_{[1]} + \ldots$ and $\chi = \chi_{[0]} + \lambda\chi_{[1]} + \ldots $. The expression for these self-energies in the expansion are 
\beq
 \label{eq:selfenergy1}
\psi_{[0]}(k) = 0, \;\;\;\; {\chi_{[0]}(k) = - \frac{1}{N_s}\sum_{p} \left
(\epsilon'_{\vec{p}}+\frac{1}{2}J_{\vec{k}-\vec{p}} \right )\GH(p)} e^{ i \omega_p 0^+}
\eeq
where we used the abbreviation
\beq
\sum_k = \frac{1}{ \beta} \sum_{\vk, \omega_k}, 
\eeq
with $N_s$ the number of lattice sites, and 
\begin{align}\label{eq:selfenergy2}
    \psi_{[1]}(k) &= -\frac{1}{N^2_s} \sum_{pq}\left (\epsilon'_{\vec{p}} + \epsilon'_{\vec{q}} + J_{\vec{k}-\vec{p}} \right ) \GH(p)\GH(q)\GH(p + q - k) \\
    \chi_{[1]}(k) &= - \frac{1}{N^2_s}\sum_{pq} \left (\epsilon'_{\vec{p}} + \epsilon'_{\vec{q}} + J_{\vec{k}-\vec{q}} \right )
    \left (\epsilon'_{\vec{p}+\vec{q}-\vec{k}} + J_{\vec{k}-\vec{p}} \right )\nn \\ &\quad \times\GH(p)\GH(q)\GH(p + q - k) 
\end{align}
where  $J_{\vec{q}}$ is the Fourier transform of $J_{ij}$. With  $\lambda\to 1$, the  expressions for the  $\mathcal{O}(\lambda^2)$ ECFL equations are
\begin{align}
 \widetilde{\mu}(k) &= 1 - \frac{n}{2} + \psi(k) \label{caprison2nd} \\ 
    \GH^{-1}(k) &= i\omega_{k} + \chem - \epsilon_{\vec{k}} + \frac{n}{2}\epsilon_{\vec{k}} - \chi_{[0]}(k)  \label{inverseg2nd} \\
                &\quad\quad-\chi_{[1]}(k) - \epsilon'_{\vec{p}} \psi_{[1]}(k)\;.  \nn 
\end{align}
We can determine the two chemical potentials $\chem$ and $u_0$ by satisfying the following number sum rules
\beq \label{eq:numsumrules}
   \frac{1}{N_s} \sum_k \GH(k)e^{i\omega_{k} 0^{+}} = \frac{n}{2} =\frac{1}{N_s}  \sum_k\G(k)e^{i\omega_{k} 0^{+}}\;,
\eeq
where $n$ is the particle density. The momentum distribution function $m_{\vk}$ is found from $\G$ using
\beq
m_{\vk}= \langle \wt{C}^\dagger_{\vk} \wt{C}_{\vk} \rangle= \frac{1}{\beta}\sum_{i \omega_k} \G(\vk,i\omega_k) e^{i \omega_k 0^+} \label{mk}
\eeq

 We find the spectral function $A(\vec{k},\omega) =
-1/\pi \Im \, \G(k)$ by analytically continuing (i.e., {$i\omega_{k} \to \omega + i\eta$}) and by solving \disp{eq:product} and \disprange{eq:selfenergy1}{eq:numsumrules} {iteratively}. We also note the useful spectral representation expressing $\G$ in terms of $A$:
\beq
\G(\vec{k},  i \omega_n) = \int_{-\infty}^\infty  d\nu \; \frac{A(k , \nu)}{i \omega_n - \nu } \label{spectral}.
\eeq


\section{Susceptibilities and the Structure function \label{App-A}}

    Our focus is on the charge susceptibility and the related structure function, and hence we first summarize some standard results \cite{Nozieres,Kadanoff,AGD,Fetter-Walecka}. Let us define the susceptibility of any pair of operators $A,B$ as
\beq
  \bigchi_{AB}(\omega+i \eta ) = i \int_0^\infty \; dt \; e^{i \omega t- \eta t} \langle [A(t),B(0)]\rangle \label{susceptibility}
\eeq
 where $\eta=0^+$ is a positive infinitesimal, $A(t)= e^{i H t} A e^{- i Ht}$, and the brackets denote the usual thermal average. Its causal nature allows us to write a spectral representation
 \beq
 \bigchi_{AB}(\omega+i \eta ) = -\frac{1}{\pi} \int_{-\infty}^\infty  d\nu \frac{\bigchi''_{AB}(\nu)}{ \omega-\nu+ i \eta} \label{susceptibility-causal}.
 \eeq
  By integration over $t$  we find the usual expression for the structure function
 \beq
 S_{AB}(\omega)= \int_{-\infty}^\infty \frac{d t}{2 \pi} e^{i \omega t} \langle A(t) B(0) \rangle, 
 \eeq
 and
 \beq
 S_{AB}(\omega)= \frac{1}{\pi} \frac{\bigchi''_{AB}(\omega)}{1-e^{-\beta \omega}}. \label{structure}
 \eeq
 In order to obtain the charge density structure function  $S_{\rho \rho}(\vec{q},\omega)$,  we must calculate the charge susceptibility $\bigchi_{\rho \rho}$ defined  from \disp{susceptibility} as
 \beq
 A= \rho_{\vq} = q_e \sum_{\vk \si} \wt{C}^\dagger_{\vk \si} \wt{C}_{\vk+\vq \si},  \;\; \mbox{and}  \;\; B= \rho_{-\vq}=A^\dagger,
 \eeq
where  $q_e= -|e|$ is the electron  charge.   $S_{\rho \rho}(\vec{q},\omega)$ is a very important object since it is   obtained directly from experimentally determined electron scattering intensity, with energy transfer $\hbar \omega$ and momentum transfer $\hbar \vec{q}$. From this object,  the  reducible susceptibility $\bigchi''_{\rho\rho}(\vec{q},\omega)$ can be obtained using the fact that it is an odd function of $\omega$. Hence
\beq
\bigchi''_{\rho\rho}(\vec{q},\omega)= \pi \left(S_{\rho\rho}(\vec{q},\omega) - S_{\rho\rho}(\vec{q},-\omega)\right). \label{structure-direct}
\eeq

 In real space we write the  local charge density $\rho_m$ at site $m$ 
as
\beq
\rho_m\equiv q_e n_m \label{rhoi}, \;\;\mbox{and}\;\; \rho_m= \frac{1}{N_s} \sum_q e^{i \vec{q}.\vec{r}_m} \rho_\vq,  
\eeq
where $N_s$ is the number of lattice sites.
 For our calculations it is more convenient to evaluate the imaginary time object and its Fourier transform
\beq
\bigchi_{AB}(\tau) = \langle T_\tau A(\tau) B(0) \rangle, \;\;\mbox{and} \;\; \bigchi_{AB}(i \Omega_\nu)=
\half \int_{-\beta}^{\beta} d \tau e^{i \Omega_\nu \tau} \bigchi_{AB}(\tau), \label{Matsubara}
\eeq
 where $\Omega_\nu =  \frac{2 \pi}{\beta}  \nu$ and $\nu = 0, \pm1, \pm 2, \ldots$. We can use analytic continuation $i \Omega_\nu \to \omega + i 0^+$ to obtain the physical susceptibility $\chi_{AB}(\omega+ i \eta)$ \disp{susceptibility} from \disp{Matsubara}.

\section{\label{App-B} Reducible susceptibility $\bigchi$ from $\G$ }

    We next turn to calculation of the susceptibilities from the electronic Green's functions. For this purpose we need to calculate the Green's functions in the presence of external potentials, and taking the derivatives we can find the susceptibilities. Although this procedure might be familiar to most readers, we summarize the steps below for completeness. In order to calculate the Green's functions for this model, we add an imaginary time $\tau$ dependent external potential (or source term) ${\cal A}$ to the definition of thermal averages. The expectation of an arbitrary observable $Q(\tau_1,\ldots)$, composed e.g. of a product of several (imaginary) time ordered Heisenberg picture operators, is written in the notation 
\beq
\bra Q(\tau_1,\ldots) \ket=  Tr\; P_\beta \; T_\tau \{ e^{-{\cal A}} Q(\tau_1,\ldots) \}. \label{Bracket}
\eeq
Here $T_\tau$ is the time-ordering operator, an external potential term ${\cal A}= \int_0^\beta d\tau \A(\tau)$, and $P_\beta = e^{-\beta H} /{Tr \left( e^{-\beta H} T_\tau e^{-{\cal A}}\right)}$ is the Boltzmann weight factor including ${\cal A}$. Here $\A(\tau)$ is a sum of two terms, ${\cal A}_{{\cal V}}(\tau)$ involving a density-spin dependent external potential ${\cal V}$, and $ {\cal A}_{u v}(\tau)$ involving external potentials $u_m(\tau),v_m(\tau)$ coupling to the charge and the W variables of \disp{W1,W2}. These are given by
\beq
{\cal A}_{{\cal V}}(\tau)& =& \sum_i   {\cal V}_i^{\si_i \si_j}(\tau) \wt{C}^\dagger_{i \si_i}(\tau) \wt{C}_{i \si_j}(\tau) \nn \\ 
{\cal A}_{u v}(\tau)& =& \sum_{m } \left( u_{m }(\tau) \rho_m(\tau)  +   v_{m}(\tau) W_m(\tau)  \right).
 \label{Sources}
\eeq
At the end of the calculations, the external potentials ${\cal V}, u, v$ are switched off, so that the average in \disp{Bracket} reduces to the standard thermal average. We can find the equation of motion for the electron Green's function
\beq
\G_{i \si_i j \si_j}(\tau,\tau')= - \bra  \wt{C}_{i \si_i}(\tau)  \wt{C}^\dagger_{j \si_j}(\tau') \ket 
\eeq
by standard methods described in literature. In particular by using the identity valid for any operator $Q$ and external potential taken to be $v_i$ for illustration:
\beq
Tr P_\beta  T_\tau \{ e^{-{\cal A}} Q_i(\tau') W_j(\tau) \}=\bra  Q_i(\tau')\ket \; \bra W_j(\tau) \ket - \frac{\delta}{\delta v_i(\tau)} \bra  Q_i(\tau')\ket  \label{TS}
\eeq
we can reduce higher order Green's functions to functional derivatives of the lower order ones. A straightforward calculation using the method described in \cite{ECFL} gives the exact functional differential equation satisfied by $\G$. Let us define
\beq
\gamma_{\si_i \si_j}(i, \tau)&=& \si_i \si_j \bra  \wt{C}^\dagger_{i \sib_i}(\tau) \wt{C}_{i \sib_j}(\tau) \ket \nn \\
 {\cal D}_{\si_i \si_j}(i, \tau))&=&\si_i \si_j \frac{\delta}{\delta {\cal V}_i^{\sib_i \sib_j}(\tau)}, \label{Def-gamma}
\eeq
the non-interacting Green's function $G_0$ including all the external potentials:
\beq
G^{-1}_{0 i \si_i j \si_j}&=&\delta_{ij} \delta_{\si_i \si_j} \left(\chem - \partial_{\tau_i} \right) + t_{ij}  \delta_{\si_i \si_j} - \delta_{ij} {\cal V}_i^{\si_i \si_j}\nn \\
&&- q_e u_i \delta_{ij} - i q_e (v_i-v_j) t_{ij},\label{G0}
\eeq
the standard Hartree type $Y$ variables from \cite{ECFL} 
\beq
Y_{i \si_i j\si_j}&=&t_{ij} \gamma_{\si_i \si_j}(i, \tau_i) - \delta_{ij} \half \sum_k J_{ik}  \gamma_{\si_i \si_j}(k, \tau_i) \nn \\
&& + \delta_{ij} \sum_l V_{il} \langle \langle n_l(\tau_i)\rangle\rangle \label{Y}
\eeq
and the $X$ type functional derivative terms
\beq
X_{i \si_i j\si_j}&=& - t_{ij} {\cal D}_{\si_i \si_j}(i) + \delta_{ij} \half \sum_k J_{ik}  {\cal D}_{\si_i \si_j}(k, \tau_i) \nn\\
&&-{q_e}  \delta_{ij} \sum_l V_{il} \frac{\delta}{\delta u_l(\tau_i)}. \label{X}
\eeq
In the equations \disp{G0,Y,X} a factor of $\delta(\tau_i-\tau_j)$ right-multiplying all the terms has been suppressed for brevity. We find the exact equation for $\G$ in a compact form by using a repeated spin index summation notation as:
\beq
 (G^{-1}_{0 i \si_i  j \si_j} - &Y_{i \si_i j\si_j} - & X_{i \si_i j\si_j}) \G_{j \si_j f \si_f}(\tau_i,\tau_f) \nn \\&& = \delta(\tau_i-\tau_f) \delta_{if} (\delta_{\si_i \si_f} - \gamma_{\si_i \si_f}(i,\tau_i )).\nn \\ \label{EOM}
\eeq

    The expressions for $Y$ in \disp{Y} and $X$ in \disp{X} reduce to the corresponding equations for the pure \tJ model in \cite{ECFL-2d,ECFL-2d-Mai-a,ECFL-2d-Mai-b}, if we drop the Coulomb terms in the last lines, i.e. $V_{il}\to 0$, and also drop the source terms with $u$ and $v$ in the last line of \disp{G0}. Following standard practice for Coulomb interactions \cite{Nozieres}, an implicit neutralizing background term 
cancels the divergence of the $q=0$ component of the last Hartree-type term in $Y$ in \disp{Y}.

    In terms of the Green's function, the expectation value of the density and the W-variables are found as
\beq
\langle \langle \rho_m(\tau) \rangle \rangle &=&  \sum_{if \si} \gamma_\rho (i,f;m) \G_{\si \si}( i \tau , f \tau^+) \label{exrho}\\
\langle \langle W_m(\tau) \rangle \rangle &=&  \sum_{if \si} \gamma_W (i,f;m) \G_{\si \si}( i \tau , f \tau^+)\label{exW}
\eeq
where we introduced the bare vertices for the charge $\rho$ and the divergence of current $W$:
\beq
 \gamma_\rho(i,f;m)\equiv q_e \delta_{i,m} \delta_{f,m}=- \frac{\delta}{\delta u_m(\tau)} G^{-1}_{0 i \si_i  f \si_f} \nn
   \\
   \gamma_W(i,f;m)\equiv i q_e t_{if}  (\delta_{i,m}- \delta_{f,m})= - \frac{\delta}{\delta v_m(\tau)} G^{-1}_{0 i \si_i  f \si_f}. \label{bare-vertex}
\eeq
Using \disp{TS} we write down the four relevant susceptibilities in real space:
\beq
 \bigchi_{\rho \rho}(i \tau_i  j \tau_j) &=& - \frac{\delta}{\delta u_j(\tau_j) } \sum_{l m \si} \gamma_\rho (l,m;i) \G_{\si \si}( l \tau_i, m \tau_i^+) \nn \\
 \bigchi_{W W}(i \tau_i  j \tau_j) &=& - \frac{\delta}{\delta v_j(\tau_j) } \sum_{l m \si} \gamma_W (l,m;i) \G_{\si \si}( l \tau_i, m \tau_i^+) \label{chiWW} \nn \\
 \bigchi_{\rho W}(i \tau_i  j \tau_j) &=& - \frac{\delta}{\delta v_j(\tau_j) } \sum_{l m \si} \gamma_\rho (l,m;i) \G_{\si \si}( l \tau_i, m \tau_i^+) \label{chirhoW}\nn \\
 \bigchi_{W \rho}(i \tau_i  j \tau_j) &=& - \frac{\delta}{\delta u_j(\tau_j) } \sum_{l m \si} \gamma_W (l,m;i) \G_{\si \si}( l \tau_i, m \tau_i^+) \label{chis-1}
\eeq
    To compress the notation we introduce Greek symbols $\mu,\nu$ taking two values, with $\mu=\{\rho,W\}$, with $\rho$ denoting charge and $W$ denoting the W-variable (divergence of current). The two bare vertices $\gamma_\rho$ and $\gamma_W$ in \disp{bare-vertex} can now be represented by $\gamma_\mu$, and the external potentials by $w_\mu$ with $w_\rho(i \tau_i)=u_i(\tau_i)$ and $w_W(i \tau_i)=v_i(\tau_i)$. The four relations in \disp{chis-1} can then be compactly written as
\beq
\bigchi_{\mu \nu}(i \tau_i j \tau_j)= - \frac{\delta}{\delta w_{\nu }(j \tau_j) } \sum_{l m \si} \gamma_\mu (l,m;i) \G_{\si \si}( l \tau_i, m \tau_i^+). \label{chis-2}
\eeq

\section{Irreducible susceptibility $\hatchi$ from $\G$ \label{App-C}} 
    In order to treat the most important effect of long-ranged Coulomb interactions, we must first account for screening. In the case of the electron gas this is achieved by introducing screened vertices and their Feynman diagram definitions in the enlightening discussion in Nozi\`eres book \cite{Nozieres} and useful summaries in \cite{Nambu,Rajagopal}. The projected electrons lack Feynman diagrams and require an alternate treatment. More fundamentally the non canonical nature of the projected electrons creates an obstacle for defining reasonable vertex operators \cite{ECFL}, which tend to free vertices at high frequencies. This situation prevents us from borrowing Nozi\`eres treatment of screening, and an adaptation is necessary. For this purpose a more general discussion is provided here, working directly with the susceptibilities instead of the vertices.

    The main qualitative idea behind our treatment of screening, is to eliminate the long-ranged Hartree-type Coulomb term in the self energy $Y$ appearing on the last line of \disp{Y}. This term is absorbed into the redefined  external potential term $q_e \tilde{u}_i$ in the non-interacting Green's function \disp{G0}. We define a screened external potential 
\beq
q_e \tilde{u}_i(\tau)= q_e u_i(\tau) + \sum_l V_{il} \langle \langle n_l(\tau) \rangle \rangle. \label{newu}
\eeq
The Green's function is unchanged since we merely shifted the location of the Hartree-type term in \disp{EOM}. We may now regard the Green's function as a functional of $\tilde{u}_i$ rather than $u_i$. With this modification, we can use a chain rule for taking derivatives  
\beq
 \frac{\delta}{\delta u_i(\tau_i) }&=& \frac{\delta}{\delta \tilde{u}_i(\tau_i) } +\sum_j   \int_0^\beta d\tau_j       \frac{\delta \tilde{u}_j(\tau_j)}{\delta {u}_i(\tau_i) }
 \frac{\delta}{\delta \tilde{u}_j(\tau_j) } \nn \\
 &=&\frac{\delta}{\delta \tilde{u}_i(\tau_i) } - \frac{1}{q_e^2}\sum_j\int_0^\beta d\tau_j \;  V_{ij}   \bigchi_{\rho \rho}(j \tau_j, i \tau_i)  
 \frac{\delta}{\delta \tilde{u}_j(\tau_j) }.\nn \\ \label{screenu}
\eeq
Here the partial derivative $\frac{\delta}{\delta \tilde{u}_j(\tau_j)}$ is taken at fixed values of $\tilde{u}_i$, where $i\neq j$. 

    In order to take the derivatives $\frac{\delta}{\delta v_i}$ in \disp{chis-1}, we should note that a variation of $v_i$ also induces a variation in $\tilde{u}_i$, which depend on it through the second term in \disp{newu}. We can account for this dependence by defining a screened set of potentials $\{\tilde{v}_j\}$, which are independent of $\tilde{u}_i$.

    The derivatives with respect to ${v}_i$ are relatable to the derivatives with respect to $\tilde{v}_i$ and $\tilde{u}_i$ through the chain rule:
\beq
 \frac{\delta}{\delta v_i(\tau_i) }&=& \frac{\delta}{\delta \tilde{v}_i(\tau_i) }+\sum_j  \int_0^\beta d\tau_j       \frac{\delta \tilde{u}_j(\tau_j)}{\delta {v}_i(\tau_i) }
 \frac{\delta}{\delta \tilde{u}_j(\tau_j) } \nn \\
 &=& \frac{\delta}{\delta \tilde{v}_i(\tau_i) } - \frac{1}{q_e^2}\sum_j\int_0^\beta d\tau_j \;  V_{ij}   \bigchi_{\rho W}(j \tau_j, i \tau_i)  
 \frac{\delta}{\delta \tilde{u}_j(\tau_j) }.\nn \\ \label{screenv}
\eeq
The second term captures the non-local variation of the $\tilde{u}_j$ by changing $v_i$ that is evident in \disp{newu}. Therefore for computing the susceptibilities in \disp{chis-1} and \disp{chis-2}, we can replace the derivatives with respect to the independent sets of external potentials $\{ u_j, v_j\}$ by another independent set of potentials $\{\tilde{u}_j,\tilde{v}_j\}$ related by \disp{screenv}. 

    Combining \disp{screenu} and \disp{screenv} we write
\beq
 \frac{\delta}{\delta w_\nu(i \tau_i) }
 &=& \frac{\delta}{\delta \tilde{w}_\nu(i\tau_i) } - \frac{1}{q_e^2}\sum_j\int_0^\beta d\tau_j \;  V_{ij}   \bigchi_{\rho \nu}(j \tau_j, i \tau_i)  
 \frac{\delta}{\delta \tilde{u}_j(\tau_j) }.\nn \\ \label{screenw}
\eeq

    To summarize the above discussion, the Green's functions of the theory, while \disp{EOM} is unchanged, \disp{G0,Y,X} are now functionals of the variables $\tilde{u}_i,\tilde{v}_i$, 
\beq
G^{-1}_{0 i \si_i j \si_j}&=&\delta_{ij} \delta_{\si_i \si_j} \left(\chem - \partial_{\tau_i} \right) + t_{ij}  \delta_{\si_i \si_j} - \delta_{ij} {\cal V}_i^{\si_i \si_j}\nn \\
&&- q_e \tilde{u}_i - i q_e (\tilde{v}_i-\tilde{v}_j) t_{ij}, \label{G0-2} \\
Y_{i \si_i j\si_j}&=&t_{ij} \gamma_{\si_i \si_j}(i \tau_i) - \delta_{ij} \half \sum_k J_{ik}  \gamma_{\si_i \si_j}(k \tau_i) \nn \\
X_{i \si_i j\si_j}&=& - t_{ij} {\cal D}_{\si_i \si_j}(i) + \delta_{ij} \half \sum_k J_{ik}  {\cal D}_{\si_i \si_j}(k \tau_i) \nn\\
&&-{q_e}  \delta_{ij} \sum_l V_{il} \frac{\delta}{\delta u_l(\tau_i)}, \label{X2}
\eeq
where the derivative $\frac{\delta}{\delta u_l(\tau_i)}$ in the last term, can be eliminated using \disp{screenu}. The Hartree type approximations  made below throws out this last term completely, and hence we skip the details.

    We now denote the set of four screened susceptibilities $\hatchi_{\mu \nu}$ in the form of \disp{chis-2}
\beq
\hatchi_{\mu \nu}(i \tau_i, j \tau_j)= - \frac{\delta}{\delta \tilde{w}_{\nu }(j \tau_j) } \sum_{l m \si} \gamma_\mu (l,m;i) \G_{\si \si}( l \tau_i, m \tau_i^+)\label{chis-22}
\eeq
where $\tilde{w}_\mu$ is either $\tilde{u}$ or $\tilde{v}$. Using the chain rules \disp{screenw} we find the important result connecting the unscreened and screened susceptibilities
\beq
&&\bigchi_{\mu \nu}(i \tau_i, j \tau_j)=\hatchi_{\mu \nu}(i \tau_i, j \tau_j) \nn \\ &&- \frac{1}{q_e^2}\sum_m \int_0^\beta d\tau_m \;  V_{im}   \hatchi_{\mu \rho}( i \tau_i, m \tau_m)  \bigchi_{\rho \nu}( m \tau_m, j \tau_j) \label{connect-a}
\eeq
Upon switching off the external potentials we recover translation invariance, and on  taking  the Fourier transform of this equation, we  find an algebraic equation at each $q\equiv \{\vec{q}, i \Omega\}$
\beq
&&\bigchi_{\mu \nu}(q)= \hatchi_{\mu \nu}(q) - \frac{1}{q_e^2} V(\vq)  \hatchi_{\mu \rho}(q) \bigchi_{\rho \nu}(q). \label{hatunhat-a}
\eeq
This can be solved for all the components and displays the   screened nature of the resulting susceptibilities. The density-density response $\bigchi_{\rho \rho}$ is simplest since all terms on the right have the same subscripts. Gathering terms $\bigchi_{\mu \nu}(q)$ on the left, we find
\beq
&&\bigchi_{\rho \rho}(q)= \frac{\hatchi_{\rho \rho}(q)}{ \varepsilon(q) }, \label{relations-0-a} 
\eeq
where dielectric function is given (exactly) by
\beq
\varepsilon(q) \equiv \varepsilon(\vec{q}, \omega) = 1+ \frac{1}{q_e^2} V(\vq) \hatchi_{\rho \rho}(\vec{q}, \omega), \label{dielectric-constant-1-a}
\eeq
with the Coulomb potential given by \disp{Coulomb-3d,Coulomb-2d}.  
Proceeding similarly we find the other three susceptibilities in terms of their screened counterparts as
\beq
&& \bigchi_{\rho W}(q)= \frac{\hatchi_{\rho W}(q)}{ \varepsilon(q) },\label{relations-1-a} \\
&&  \bigchi_{W \rho }(q)= \frac{\hatchi_{W \rho }(q)}{ \varepsilon(q) },\;\;\; \label{relations-2-a} \\
&&\bigchi_{W W}(q)= \hatchi_{W W}(q)- \frac{V(\vq)}{q_e^2 \varepsilon(q)}\hatchi_{W \rho}(q)\; \hatchi_{\rho W}(q). \label{relations-3-a}
\eeq

\section{Low and high $\omega$ limits of  $\varepsilon(\vq,\omega)$ \label{App-Limits}}

\subsection{Low $\omega$: Static Screening and Compressibility \label{sec3.5}}

    At low frequencies $\omega\to0$ and in the long-wavelength limit $|\vec{q}|\ll 1 $, the screened susceptibility $\hatchi_{\rho \rho}$ defined in \disp{chis-2} equals the thermodynamic derivative
\beq
\lim_{q\to0} \lim_{\omega\to0}\hatchi_{\rho_q \rho_{-q}}(\vq,\omega)= q_e^2 \frac{d n}{d \mu} N_s.    \label{compress-1} 
\eeq
In view of the connection with the compressibility \disp{compressibility-2}, this is often called the {\em compressibility sum-rule}. To see this we note that a space independent $-q_e \tilde{u}$ is {\em additive} to the chemical potential $\mu$ in \disp{X2}, and since the nominally divergent Hartree term is removed in defining $\tilde{u}$ the uniform limit is safely taken. This gives the compressibility sum-rule, i.e., the screening limit of the dielectric constant \cite{Nozieres,Rice-Brinkman,Vollhardt}
\beq
\lim_{q\to0} \lim_{\omega\to0}\varepsilon(\vec{q}, \omega)&=& 1+ V(\vq) N_s \frac{d n}{d \mu} \label{compress-2} 
\eeq
Thus in 3-d and 2-d we get the exact result:
\beq
\varepsilon &\to& 1+ \frac{q^2_s}{|\vq|^2}, \;\; \mbox{(3-d) with } q^2_s= \frac{4 \pi q_e^2}{\varepsilon_\infty} \frac{d n}{d \mu} \label{static-screening-3d}\\
\varepsilon &\to& 1+ \frac{q_s}{ |\vq |}, \;\; \mbox{(2-d) with } q_s=  \frac{2 \pi q_e^2}{\varepsilon_\infty} \frac{d n}{d \mu} \label{static-screening-2d}
\eeq
Using the thermodynamic relation for compressibility $\chi_{comp}$
\beq
\chi_{comp}= \frac{1}{n^2} \frac{d n}{d \mu}, \label{compressibility-2}
\eeq
the screening length $\lambda_s= 2\pi/q_s$  can thus be related to the  compressibility $\chi_{comp}$.

    Strongly correlated systems  near half filling display a reduced compressibility, and are therefore expected to show very poor screening, i.e., $\lambda_s  \gg 1$ (we set the lattice constant $a_0=1$).

\subsection{High $\omega$: Plasmon Dispersion in $\varepsilon(q)$ \label{sec3.6}} 

    In the limit $\omega \gg t$ the behavior of the dielectric function is easily read off from \disp{dielectric-constant-2}. Neglecting  $\frac{ \hatchi_{W W}(\vq,\omega)}{ \hatchi_{W W}(\vq,0)}$ compared to unity,  we get
\beq
\lim_{\omega\gg t}\varepsilon(\vec{q}, \omega)= 1- \frac{\omega_p^2(\vq)}{\omega^2}. \label{plasma-limit}
\eeq
In both 3-d and 2-d,  the plasma frequency is given in terms of $\kappa$ by
\beq
 \omega^2_p(\vq)=\frac{N_s }{q_e^2} V(\vq) \kappa(\vq). \label{3d-plasmon}
\eeq 
In 3-d the plasma frequency can be written using \disp{kappa} and \disp{Coulomb-3d}   as
\beq
\omega^2_p(\vq)&=& \frac{8 \pi q_e^2}{\varepsilon_\infty |\vq|^2} \frac{1}{N_s} \sum_{k \si} (\varepsilon_{\vk+\vq}-\varepsilon_\vk) \langle \wt{C}^\dagger_{\vk \si} \wt{C}_{\vk+\vq \si}\rangle \label{plasmons-1}.
\eeq
In the long wavelength limit we find
\beq
\lim_{q\to0}\omega^2_p(\vq)&=&\frac{4 \pi q_e^2}{\varepsilon_\infty  } \frac{1}{N_s} \sum_{k \si} (\frac{d^2\varepsilon_\vk}{d k_x^2}) \langle \wt{C}^\dagger_{\vk \si} \wt{C}_{\vk \si}\rangle = \frac{4 \pi}{\varepsilon_\infty} {\cal T}, \label{plasmons-2}
\eeq
where we used \disp{kappa-smallq} in the last line. For quadratic dispersion $\varepsilon_k= |\vk|^2/(2m)$, we get the familiar expression $\omega^2_p=\frac{4 \pi n q_e^2}{m \varepsilon_\infty}$. The  f-sumrule \disp{f-sumrule} is expressible in terms of the plasma frequency as
\beq
\int_{-\infty}^\infty \frac{ d \omega}{\pi} \Re \, \sigma(\omega) = \frac{\varepsilon_\infty}{4 \pi} {\omega^2_p(0)}. \label{f-sumrule-2}
\eeq

    In 2-d  using \disp{kappa} and \disp{Coulomb-2d}  we obtain the acoustic plasmon energy 
\beq
\omega^2_p(\vq)&=& \frac{4 \pi q_e^2}{\varepsilon_\infty  |\vq|} \frac{1}{N_s} \sum_{k \si} (\varepsilon_{\vk+\vq}-\varepsilon_\vk) \langle \wt{C}^\dagger_{\vk \si} \wt{C}_{\vk+\vq \si}\rangle \label{plasmons-3}\\
\lim_{q\to0}\omega^2_p(\vq)&=&|\vq| \times \frac{2 \pi q_e^2}{\varepsilon_\infty  } \frac{1}{N_s} \sum_{k \si} (\frac{d^2\varepsilon_\vk}{d k_x^2}) \langle \wt{C}^\dagger_{\vk \si} \wt{C}_{\vk \si}\rangle = |\vq| \times \frac{2 \pi }{\varepsilon_\infty  } {\cal T}. \label{plasmons-4}
\eeq
For quadratic dispersion this reduces to $\omega^2_p=|\vec{q}| \times \frac{2 \pi n q_e^2}{m \varepsilon_\infty}$. This implies that the plasmon mode, found as the zero of the dielectric function is gapless in 2-d with a dispersion $\omega_q \propto \sqrt{ q}$, as opposed to the usual gapless mode in 3-d.

    Let us note that the effect of Gutzwiller type short range correlations is seen most directly in expressions for ${\cal T}$ in \disp{Big-Tau} and in \figdisp{sumrule-dispersion}. We discuss in \secdisp{moments-irreducible-chi} the connection of this result with the first frequency sum rule for  the electron structure function.

\subsection{The Resistivity Formula:} 
    We note that the formula in \disp{dielectric-constant-2} also gives the correct resistivity formula used in studies of the \tJ model. Let us first examine the 3-dimensional case with a cubic unit cell, and assume that the electric field polarization is longitudinal, i.e. the current is along $\vq$. From the usual relation between the induced current and the polarization 
$\vec{J}_{ind}=  \dot{\vec{P}}$, and $\vec{P}=\frac{1}{4 \pi}( \vec{D}-\vec{E})$
combined with the constitutive relations $\vec{J}_{ind}= \sigma \vec{E}$ and $\vec{D}= \varepsilon \vec{E}$ we obtain $\sigma(q)=\frac{ \omega}{4 \pi  i} (\varepsilon(q)-1)$ and on using \disp{dielectric-constant-2} 
\beq
\sigma(\vec{q},\omega)= \frac{i}{ |\vec{q}|^2 \, \omega }
\left(  \kappa(q) - \frac{1}{N_s} \hatchi_{W W}(q)  \right).
\label{resist-q}
\eeq
In the  uniform limit  $q\to0$ we note from \disp{OPWtoJ} that  $W_q \to -i \vec{q}.\vec{J}_q$ and $ W_{-q} \to i \vec{q}.\vec{J}_{-q} $; therefore \beq
\mbox{For}\;\;  |\vq| a_0 \ll 1, \;\;
 \hatchi_{J J}(\vq,\omega) = \frac{1}{|\vq|^2 }\hatchi_{W W}(\vq,\omega). \label{WtoJ}
\eeq
This is the screened analog of \disp{chiWtoJ}. In the limit $\vec{q}=0$, there is no distinction between longitudinal and transverse response, and hence using \disp{kappa-smallq} we get the conductivity accessible in optical experiments
\beq
\sigma(\omega)&=& \frac{i}{\omega}\frac{1}{N_s} \left(q_e^2 \sum_{k\si}  \left( \frac{d^2 \varepsilon_{k}}{d k_x^2}  \right) \langle  \wt{C}^\dagger_{k \si} \wt{C}_{k \si} 
 \rangle   - \hatchi_{J J}(\omega) \right), \nn \\
 &=& \frac{i}{\omega} \left( {\cal T} - \frac{1}{N_s}  \hatchi_{J J}(\omega) \right)   \label{resistivity}
\eeq
with  $\omega\equiv \omega+i 0^+$. 
Let us note an important consequence of \disp{resist-q}: 
\beq
\Re \,  \sigma (\vq,\omega) = \frac{1}{\omega N_s} \Im \, \hatchi_{J J}(\vq,\omega),  \label{cond-chiJJ}
\eeq
thus relating the dissipative part of conductivity with $\Im \, \hatchi_{J J}(\vq, \omega)/\omega$. In \disp{cond-chiJJ} we have suppressed an implicit prefactor $\frac{1}{a_0}$, which needs modification for quasi 2-dimensional system such as the cuprate materials analyzed in \cite{ECFL-Resistivity,ECFL-2d-Mai-a,ECFL-2d-Mai-b}. Here the theory proceeds by assuming that the unit cell is body centered tetragonal instead of cubic. Here $a_0$ is replaced by $c_0$, the separation between two copper oxide layers in the simple case of single layer cuprates, so that $c_0\gg a_0$. The different layers are assumed to be decoupled as far as electron hopping is concerned, while their polarizations add up. We then obtain an appropriate generalization of \disp{cond-chiJJ} 
\beq
\Re \, \sigma (\vq,\omega) = \frac{q_e^2}{c_0 h} \left( \frac{ h}{q_e^2  \omega N_s} \Im \, \hatchi_{J J}(\vq,\omega)\right),  \label{cond-chiJJ-2d}
\eeq
where the object in parentheses is ${\cal O}(1)$ and dimensionless. We note that \disp{resistivity} is almost identical to the standard formula for the optical conductivity $\sigma(\omega)$ obtained from the Kubo formula for Hubbard model or \tJ model type systems without the long ranged Coulomb interaction, e.g. see Eq.~(A1-A5) in \cite{Shastry-rho}. The only change is that the screened current susceptibility $\hatchi_{JJ}$ replaces the unscreened $\bigchi_{JJ}$. This object can be obtained from \disp{relations-3} in the limit of small $\vq$. Physically the tilde means that the calculation of the current-current correlators must discard direct contributions from the Coulomb potential. The f-sumrule for the conductivity $\int_{-\infty}^\infty \frac{d \omega}{\pi} \; \Re \, \sigma(\omega) = {\cal T}$ given in \disp{f-sumrule}, follows by first writing the Kramers-Kronig relation
\beq
\Im \,  \sigma(\omega)= \frac{1}{\pi} \int_{-\infty}^\infty  \, d \nu \frac{\Re \, \sigma(\nu) }{\omega-\nu} \label{KK},
\eeq
 taking the limit $\omega\gg 0$, and finally comparing the expression with the coefficient of $1/\omega$ in \disp{resistivity}.


\section{Structure Function Frequency Moments  \label{App-X}} 

    The recent momentum dependent electron energy loss experiments (M-EELS) \cite{MEELS-1,MEELS-2,MEELS-3} probe charge response inferred from the inelastic momentum resolved scattering of electrons from the surface of the high $T_c$ superconductor Bi2212 $Bi_2Sr_2CaCu_2O_{8+x}$. Making various simplifying assumptions that are argued for in the important work of Mills \cite{Mills}, the experiment gives a readout of the structure function
\beq
 S_{\rho \rho}(\vec{q}, \omega)= \int_{-\infty}^\infty \frac{dt}{2 \pi} e^{i \omega t} \langle \rho_{\vec{q}}(t) \rho_{-\vec{q}}(0)\rangle = \frac{1}{\pi} \frac{\bigchi''_{\rho \rho}(\vq,\omega)}{1-e^{-\beta \omega}},
\eeq
over a substantial portion of the $\vec{q},\omega$ region with remarkably high precision. The energy resolution $\Delta \omega \sim 2$meV. Here $\vec{q}$ is taken to be 2-dimensional. These works present direct information about $\bigchi_{\rho\rho}$, in fact using the odd-ness of $\bigchi''_{\rho\rho}$ we can extract this object by combining energy loss and energy gain data:
\beq
 \bigchi''_{\rho\rho}(\vq, \omega)=\pi \left( S_{\rho\rho}(\vq,\omega)- S_{\rho\rho}(\vq,-\omega) \right) \label{chi-structure}
\eeq

    The work of \cite{MEELS-1,MEELS-2,MEELS-3} presents data for the $\bigchi''_{\rho\rho}(\omega)$ as well as the inferred screened susceptibility $\hatchi_{\rho \rho}$.

\subsection{High frequency moments: reducible susceptibility \label{moments-reducible-chi}}
    Using the familiar analyticity of $\bigchi_{\rho\rho}(\vq,\omega)$ in the upper half of the complex $\omega$ plane, we can write a spectral representation 
\beq
\bigchi_{\rho\rho}(\vq,\omega) = - \frac{1}{\pi} \int_{-\infty}^\infty d\nu \,  \frac{\bigchi''_{\rho\rho}(\vq,\nu)}{\omega-\nu+ i 0^+}. \label{dispersion-relation-1}
\eeq
We note that $\bigchi''_{\rho\rho}(\vq,\nu)$ is odd in $\omega$ and hence as $\omega \gg 0$ we get a moment expansion with even terms \cite{Pathak}
\beq
\lim_{\omega \gg 0} \bigchi_{\rho\rho}(\vq,\omega) =- q_e^2 N_s \left( \frac{\omega^{(1)}(\vq)}{\omega^2} + \frac{\omega^{(3)}(\vq)}{\omega^4}+\ldots \right), \label{moment-expansion-1}
\eeq
where the frequency moments $\omega^{(2j+1)}(\vq)$ are given by
\beq
\omega^{(2j+1)}(\vq)= \frac{1}{q_e^2 N_s} \int_{-\infty}^\infty \frac{d \omega}{\pi} \omega^{2 j +1} \bigchi''_{\rho\rho}(\vq,\omega), 
\eeq
or upon using \disp{chi-structure}
\beq
\omega^{(2j+1)}(\vq)= \frac{2}{q_e^2 N_s} \int_{-\infty}^\infty {d \omega} \omega^{2 j +1}  S(\vq,\omega).
\eeq
\subsection{High frequency moments: irreducible susceptibility \label{moments-irreducible-chi}}

    In the presence of long-ranged Coulomb interactions it is necessary  \cite{Nozieres} to distinguish between reducible susceptibility (or polarization) $\bigchi_{\rho \rho}$ and the irreducible susceptibility (or polarization) $\hatchi_{\rho \rho}$. The irreducible susceptibility $\hatchi_{\rho\rho}$ can be shown to satisfy a spectral representation 
\beq
\hatchi_{\rho\rho}(\vq,\omega) = - \frac{1}{\pi} \int_{-\infty}^\infty d\nu \,  \frac{\hatchi''_{\rho\rho}(\vq,\nu)}{\omega-\nu+ i 0^+}. \label{dispersion-relation-11}
\eeq

    This is completely analogous to \disp{dispersion-relation-1}, and using a moment expansion analogous to \disp{moment-expansion-1} we get
\beq
\lim_{\omega \gg 0} \hatchi_{\rho\rho}(\vq,\omega) =- q_e^2 N_s \left( \frac{\widetilde{\omega}^{(1)}(\vq)}{\omega^2} + \frac{\widetilde{\omega}^{(3)}(\vq)}{\omega^4}+\ldots \right), \label{moment-expansion-2}
\eeq
In order to determine the moments $\widetilde{\omega}^{(2l+1)}(\vq)$, we recast \disp{relations-0} in the form
\beq
 \hatchi_{\rho\rho}(\vq,\omega)= \frac{\bigchi_{\rho\rho}(\vq,\omega)}{1- \frac{V(\vq)}{N_s} \bigchi_{\rho\rho}(\vq,\omega)} \label{polarization-Dyson-1}.
\eeq
We next plug into this expression the high frequency expansion \disp{moment-expansion-1} giving an infinite series in $\frac{1}{\omega^2}$. Comparing with \disp{moment-expansion-2}, the moments $\widetilde{\omega}^{(2j+1)}(\vq)$ can be determined in terms of ${\omega}^{(2j+1)}(\vq)$. For our purpose we only need the first moment:
\beq
 \widetilde{\omega}^{(1)}(\vq)={\omega}^{(1)}(\vq) \label{equality-1}.
\eeq
    We make extensive use of the first moment $\omega^{(1)}(\vq)$ below, let us note that it is in frequency units and provides a very important scale in the problem. We now relate this frequency to $\kappa(\vq)$. From \disp{dielectric-constant-1} we note that
 \beq
\lim_{\omega \gg0} \varepsilon(\vq,\omega)  \to 1- V(\vq) N_s  \left( \frac{\widetilde{\omega}^{(1)}(\vq)}{\omega^2} + \frac{\widetilde{\omega}^{(3)}(\vq)}{\omega^4}+\ldots \right). \label{dielectric-constant-22}
\eeq
Comparing the leading term with the expression in \disp{plasma-limit,3d-plasmon}, we get
\beq
\widetilde{\omega}^{(1)}(\vec{q})&=& \frac{a_0 \hbar}{q_e^2} \kappa(\vec{q}), \label{moment-2}
\eeq 
where we temporarily reintroduced the lattice constant $a_0$ and $\hbar$ to emphasize
that $\widetilde{\omega}^{(1)}$ is in frequency units, while $\kappa$ is the square of a frequency \cite{dimension-kappa}.

    Using \disp{3d-plasmon}, the first moment also determines the plasmon energy as $\omega_p(\vq)=\sqrt{\frac{N_s }{q_e^2}V(\vq) \kappa(\vq)} $. Proceeding further we can express $\kappa(\vq)$ in 2-d  explicitly in terms of $\vq$, the band hopping parameters and the averages over the momentum distribution function $\langle \wt{C}^\dagger_{k } \wt{C}_{k }\rangle $  of the type $\langle \cos k_x\rangle_{ave} \equiv \frac{1}{N_s}\sum_k \cos k_x  \langle \wt{C}^\dagger_{k } \wt{C}_{k }\rangle$. Using \disp{kappa} and the band dispersion parameters $t,t'$ representing the nearest and next nearest neighbor hops on the square lattice: 
\beq \varepsilon_k= - 2 t (\cos k_x +\cos k_y) - 4 t' \cos k_x \cos k_y. \label{dispersion}
\eeq  
We find 
\beq
\widetilde{\omega}^{(1)}(\vq) &=& 8 t \langle \cos k_x\rangle_{ave} (2- \cos q_x -\cos q_y)  
+ 16 t' \langle \cos k_x \cos k_y\rangle_{ave} (1- \cos q_x \cos q_y) \label{mu-dispersion}.
\eeq
For small $\vq$ we find
\beq
\lim_{\vq \to 0} \widetilde{\omega}^{(1)}(\vq) \to  |\vq|^2 \frac{\cal T}{q_e^2}, \label{low-q-w1}
\eeq 
where we utilized \disp{low-q-kappa}, and 
\beq
{\cal T}= q_e^2 \left( 4 t \langle \cos k_x\rangle_{ave} +8 t' \langle \cos k_x \cos k_y\rangle_{ave} \right) \label{Big-Tau}
\eeq
We see from \disp{f-sumrule} that ${\cal T}$ determines the total weight of the optical conductivity. The relevant averages of the cosines are tabulated in \tbldisp{Table1}, where we see the enormous reduction from uncorrelated values brought about by the strong correlations.

    For completeness we note that our notation for the reducible $\bigchi_{\rho\rho}$ and irreducible $\hatchi_{\rho\rho}$ polarizations can be mapped into that used in \cite{MEELS-1,MEELS-2,MEELS-3} by setting 
\beq
\bigchi_{\rho\rho}\to-\bigchi \nn \\ \hatchi_{\rho\rho}\to - \Pi \nn \\ \varepsilon \to \varepsilon/\varepsilon_\infty.
\eeq






\begin{thebibliography}{99}
\bibitem{FN1}  We  denote by   $U$ and $t$ the standard interaction and hopping parameters of the Hubbard model with $N_s$ sites. Here the lattice constant is $a_0$,    and we  denote by $q_e$ the electron charge, i.e.  $q_e=-|e|$.
\bibitem{dimension-kappa}  A comment on the dimensions of these variables may be useful. The variables $\kappa(\vq)$ and ${\cal T}$  have dimensions of  frequency squared. This feature  is submerged in the notation since we have set the lattice constant $a_0\to1$ and also set $\hbar \to 1$. To restore the dimensions we should read \disp{kappa} as
\beq
\kappa(\vq)=  \frac{2 q_e^2}{a_0 \hbar^2 N_s}\sum_{k \si} \left( \varepsilon_{\vk+\vq}-\varepsilon_{\vk}  \right) \langle  \wt{C}^\dagger_{\vk \si} \wt{C}_{\vk \si} 
 \rangle. \label{kappa-2}
\eeq
and
\disp{kappa-smallq} as
\beq
{\cal T}= \frac{q_e^2}{a_0 N_s \hbar^2} \sum_{k \si} \left( \frac{d^2 \varepsilon_{\vk}}{d (k_xa_0)^2}  \right) \langle  \wt{C}^\dagger_{\vk \si} \wt{C}_{\vk \si} 
 \rangle, \label{kappa-smallq-3}
 \eeq
\bibitem{FN3} For this purpose it may be more useful to rewrite \disp{emergentOmega-2}
in the form
\beq
\Omega_p^2(\vq) = \frac{\int_{-\infty}^\infty d\omega \, \omega \,  \Im \, \hatchi_{\rho\rho}(\vq,\omega)  }{\int_{-\infty}^\infty d\omega \, \frac{1}{\omega}\, \Im \, \hatchi_{\rho\rho}(\vq,\omega)  }. \label{emergentOmega-3}
\eeq
For the purpose of estimation, one may perform both integrations over the finite range of  available frequencies. The estimate can be quite reasonable if the range is not too small.
\bibitem{FN2} This is also true theoretically, as seen in \figdisp{chiJJT297K}. (We checked the continuity in $\vq$  by independent calculations of the two sets of variables using the current vertex defined by Eq~(2) of \cite{ECFL-Raman}.
\end{thebibliography}
\end{document}